\documentclass[12pt, letterpaper]{article}
\usepackage{amsfonts}

\usepackage[utf8]{inputenc}
\usepackage{amsmath}
\usepackage{bbm}
\usepackage{graphicx}
\usepackage{setspace}
\usepackage{natbib}
\usepackage{verbatim}
\usepackage{clipboard}

\usepackage[pdftex, 
				pdfauthor={Dominic Smith and Sergio Ocampo},
                pdftitle={The Evolution of U.S. Retail Concentration},
                pdfsubject= {Retail Concentration},
                pdfkeywords = {Retail} {Walmart} {Target} {Concentration}]{hyperref}
\usepackage{xcolor}
\usepackage{listings}

\usepackage[]{threeparttable}
\usepackage{multirow}
\usepackage{tikz}
\usepackage{siunitx}
\usepackage{csvsimple}
\usepackage{breqn}
\usepackage{enumitem}
\usepackage{caption}
\usepackage{subcaption}
\usepackage[section]{placeins}
\usepackage{nicefrac}
\usepackage{lipsum}
\usepackage{mathtools}
\usepackage{chngcntr}
\usepackage{minitoc}

\usetikzlibrary{arrows,positioning} 

\newcommand\Tstrut{\rule{0pt}{2.6ex}}         %

\gdef\notesoff{\gdef\note##1{}
	\usepackage[top=1.05in, bottom=1.05in, left=1.05in, right=1.05in, heightrounded]{geometry}
}
\notesoff
\setlist[itemize]{noitemsep}
\setlist[enumerate]{noitemsep}
\setlist[description]{noitemsep}
\hypersetup{
    colorlinks=true,
    linkcolor={cite_blue}, %
    citecolor={cite_blue}, %
    urlcolor={cite_blue} %
}

\usepackage[toc,page,header]{appendix}
\setcitestyle{authoryear, open={((},close={)}}
\emergencystretch=\maxdimen
\usepackage[normalem]{ulem}

\usepackage{amsthm}

\definecolor{mDarkBrown}{HTML}{604c38} 
\definecolor{mDarkTeal}{HTML}{23373b}
\definecolor{mLightBrown}{HTML}{EB811B} 
\definecolor{mLightGreen}{HTML}{14B03D}
\definecolor{red}{RGB}{252,141,89}
\definecolor{blue}{RGB}{101,134,153}%
\definecolor{cite_blue}{RGB}{51,102,153}


\renewcommand \thepart{}
\renewcommand \partname{}
\renewcommand{\tabcolsep}{20pt}
\usepackage{titletoc}

\title{The Evolution of U.S. Retail Concentration\thanks{ This paper is conditionally accepted at American Economic Journal: Macroeconomics.
    Any views expressed are those of the authors and not those of the U.S. Census Bureau. 
    The Census Bureau's Disclosure Review Board and Disclosure Avoidance Officers have reviewed this information product for unauthorized disclosure of confidential information and have approved the disclosure avoidance practices applied to this release. This research was performed at a Federal Statistical Research Data Center under FSRDC Project Numbers 1179 and 1975 (CBDRB-FY19-P1179-R7207, CBDRB-FY20-P1975-R8604 and CBDRB-FY23-P1975-R10585). \\ 
    We gratefully acknowledge advice from Thomas Holmes, Teresa Fort, Amil Petrin, Joel Waldfogel, Emek Basker, Brian Adams, Fatih Guvenen, Juan Herreño, Gueorgui Kambourov, Burhan Kuruscu, Rory McGee, Emily Moschini, Martin O'Connell, Jane Olmstead-Rumsey, Pascual Restrepo, Jeff Thurk, and Nicholas Trachter. 
    We are also indebted to an anonymous referee for constructive comments.
    We also thank participants of seminars at various institutions as well as participants at the 2021 NBER SI CRIW, 2021 CEA, 2021 IIOC, and 2019 SED meetings and the 2018 FSRDC conference.
    This paper is based upon work supported by the Doctoral Dissertation Fellowship at the University of Minnesota.
    }}

\author{Dominic A. Smith\thanks{Email: smith.dominic@bls.gov; Web: \url{https://www.bls.gov/pir/authors/smith.htm}.}\\ Bureau of Labor Statistics \and Sergio Ocampo\thanks{Email: socampod@uwo.ca; Web: \url{https://sites.google.com/site/sergiocampod/}.}\\
University of Western Ontario}

\date{\today }

\begin{document}
\doparttoc %
\faketableofcontents %

\maketitle
\thispagestyle{empty}

\vspace{-1cm}

\begin{abstract}
\singlespacing 
Increases in national concentration have been a salient feature of industry dynamics in the U.S. and have contributed to concerns about increasing market power.
Yet, local trends may be more informative about market power, particularly in the retail sector where consumers have traditionally shopped at nearby stores.
We find that local concentration has increased almost in parallel with national concentration using novel Census data on product-level revenue for all U.S. retail stores between 1992 and 2012.
The increases in concentration are broad based, affecting most markets, products, and retail industries.
We show that the expansion of multi-market firms into new markets explains most of the increase in national retail concentration, with consolidation via increases in local market shares increasing in importance between 1997 and 2007, and single-market firms playing a negligible role.
Finally, we find that increases in local concentration can explain one-quarter to one-third of the observed rise in retail gross margins. 



\end{abstract}

\setcounter{page}{0}
\clearpage
\newclipboard{results}

\Copy{local_prod_cz_increase}{2.2}
\Copy{local_prod_cz_lvl_1992}{6.4}
\Copy{local_prod_cz_lvl_2012}{8.6}
\Copy{nat_prod_lvl_1992}{1.3}
\Copy{nat_prod_lvl_2012}{4.3}
\Copy{nat_prod_lvl_1992_raw}{0.013}
\Copy{nat_prod_lvl_2012_raw}{0.043}
\Copy{cz_collocation}{2}
\Copy{local_ind_cz_increase}{12.6}
\Copy{markup_prod_increase}{2.1}
\Copy{local_chg_min}{-1.5}
\Copy{local_chg_max}{12.6}
\Copy{pct_local_increase_1992}{69}
\Copy{pct_local_increase_1992_weighted}{72}
\Copy{pct_local_increase_2002}{60}
\Copy{pct_local_increase_2002_weighted}{}
\Copy{ARTS_Markup_9312}{6.0}

\section{Introduction}

There is an economy-wide trend toward greater ownership concentration and an increase in the dominance of large, established firms.
These trends have been accompanied by rising markups, which raises concerns about increasing market power.\footnote{
    See \citet*{Autor2017} for evidence of increased concentration in retail and other sectors, and \citet*{Decker2014,Decker2020} for the dominance of large firms. 
    \citet*{DeLoecker2017} and \citet*{Hall2018} document increasing markups.
    }
The increase in concentration has been particularly strong in the retail sector, which accounts for 11 percent of U.S. employment and 6 percent of U.S. GDP. %
Both the share of sales going to the largest firms and the national Herfindahl-Hirschman Index (HHI) have been increasing for decades across retail industries \citep*{Autor2017,Hortacsu2015}.
These changes potentially reflect a decrease in retail competition.

However, local concentration is more informative than national concentration about the degree of competition and the evolution of retail markups, because consumers in the retail sector primarily choose among local stores.
Retail firms also compete within and across industries, as firms from different industries often sell identical products. %
This raises the need for new measures of retail firm concentration that reflect the evolution of local retail markets at the product and industry levels.

In this paper, we use novel U.S. Census data covering all retail establishments to show that both national and local firm concentration have increased.
The data come from the Census of Retail Trade (CRT) and span 1992 to 2012, allowing us to measure changes in local and national concentration over 20 years.
Our data allow us not only to measure industry-based concentration, but also to construct sales by product for individual retail stores with which we compute new measures of concentration for local product markets, handling retailers that sell multiple products by assigning their sales to the appropriate markets.
We consistently find increases across these measures.

Our data show that the national and local HHI increased almost in parallel between 1992 and 2012.
We show that the HHI measures the probability that two dollars spent at random are spent at the same firm. 
We use this fact to interpret changes in the HHI.
The national product HHI increased from \Paste{nat_prod_lvl_1992} to \Paste{nat_prod_lvl_2012} between 1992 and 2012, indicating that the probability that two random dollars spent on a product anywhere in the U.S. are spent at the same firm has increased by 3 percentage points.
Local (commuting zone) concentration in product markets increased by \Paste{local_prod_cz_increase} percentage points, from \Paste{local_prod_cz_lvl_1992}  to \Paste{local_prod_cz_lvl_2012}. 
Moreover, we find that the increases in local retail concentration hold and are often larger when looking at an extended sample dating back to 1982, when changing the geographical definition of local markets, or when concentration is measured using the sales share of the largest firms.

We find that the increases in local concentration were widespread, with a majority of markets and product categories experiencing increasing concentration.
The local HHI increased in 72 percent of commuting zones between 1992 and 2012, %
with markets with increasing concentration accounting for 66 percent of retail sales in 2012. %
Moreover, the increases in concentration were substantial, with 40 percent of markets presenting increases of more than 5 percentage points (well above the thresholds in the merger guidelines of the \citealp{Justice2010}). 
Concentration also increased for seven of the eight major product categories in retail between 1992 and 2012, with Clothing being the exception.\footnote{
    The eight major product categories are Clothing, Furniture, Sporting Goods, Electronics \& Appliances, Health Goods, Toys, Home Goods, and Groceries. 
    These categories account for 94 percent of retail sales.
    For comparison, the eight largest (out of 61) six-digit NAICS codes in retail account for two-thirds of sales.
    } 

We examine how online and other non-store retailers affect local concentration and find  they have a small effect because they account for less than 10 percent of CRT sales throughout our sample.
Establishing the exact effect of non-store retailers on local concentration is challenging because the CRT does not contain the location of sales for non-store retailers. %
Nevertheless, we obtain bounds for the effect of non-store retailers by assigning their national sales to local markets using a range of assumptions on how concentrated their local sales are. 
Including non-store retailers implies smaller increases in national and local concentration under most assumptions. %

We also measure local and national retail concentration in six-digit NAICS industries and compare them to our product-based results.
We find that industry-based measures exhibit a stronger increase in concentration than product-based measures. 
Commuting zone concentration increased by \Paste{local_ind_cz_increase} percentage points between 1992 and 2012, an increase six times greater than the increase in local product concentration.
Local concentration increased within all eight three-digit NAICS subsectors, led by a 28 percentage point increase in the industries that make up the General Merchandising subsector.

The main difference between product- and industry-based measures of concentration is the type of competition that they emphasize. 
Product-based measures emphasize competition in the sale of goods, while industry-based measures emphasize competition in retail services.
This difference is made clear in the treatment of general merchandisers and other multi-product retailers, which, by definition, sell the same products as retailers in other industries, but offer a different service precisely by offering a wider range of products.\footnote{
    For example, Walmart is in the general merchandising subsector (three-digit NAICS 452) but competes with grocery, clothing, and toy stores. 
    However, a retailer in a clothing industry is likely to carry a large number of clothing items, while a general merchandiser like Walmart carries other products in addition to a smaller selection of clothing. 
    Walmart reports SIC code 5331 to the Security and Exchange Commission, which corresponds to NAICS 452990 \citep{Commission2020}. 
    References to specific firms are based on public information and do not imply the company is present in the confidential data.
    } 
In fact, general merchandisers account for more than 20 percent of sales in Electronics \& Appliances, Groceries, and Clothing, and their expansion has been linked to the closure of grocery stores \citep*{Arcidiacono2016}, showing that competition across industries is a relevant feature of retail markets.
The same pattern arises in more detailed industries, as we show in Section \ref{sec:data} and Appendix \ref{app:crt}.

Having established the increase in both national and local retail concentration, we investigate the relationship between these two trends.
We find that the two trends are linked by the expansion and consolidation of multi-market retailers, with the expansion of large retailers across markets accounting for 89 percent of the increase in national retail concentration between 1992 and 2012.
In this way, concentration has increased as consumers in different markets increasingly buy from the same firms, 
adding to the findings of previous papers such as \citet*{Cao2019}, \citet*{Hsieh2019}, and \citet*{Rossi-Hansberg2018} on the role of the expansion of large firms in explaining changes in the U.S. economy.\footnote{
    The expansion of large retail firms has led to the closing of small stores \citep*{Jia2008,Haltiwanger2010} and grocery chains \citep{Arcidiacono2016}, as well as higher retail employment in local labor markets \citep{Basker2005} and increased welfare for consumers \citep*{LeungBigBox, LeungRetailConcentration}.
    }

Our findings also reveal that single-market firms operating in local product markets have a negligible effect in the evolution of national concentration. 
This is because the distribution of retail sales across locations in the U.S. implies that even the largest retail markets in the U.S. are too small to affect national trends.
By contrast, the national market share of retail firms present in at least 50 commuting zones increased from 34 to 58 percent between 1992 and 2012, despite there being fewer than 350 of these firms in the U.S. (or less than 0.1 percent of retail firms).
All this while their average local market share barely increased from 3.2 to 3.4 percent.

We end the paper with a discussion of the implications of our findings. 
Under Cournot competition there is a direct link between local concentration and firms' margins.
This link implies increases in retailers' margins of 1.6 percentage points from the overall increase in local product market concentration.
Taking into account differences in concentration trends and retailers' margins across products increases the estimate to 2.1 percentage points.\footnote{
    Many of the concerns about concentration leading to higher markups (\citealp*{Hall2018,Traina2018,Edmond2018}; \citealp{DeLoecker2017}) would operate through local markets, particularly in labor and retail markets. 
    For instance, higher local employment concentration has been shown to negatively impact wages \citep*{Jarosch2019,Azar2019,Rinz2018,Berger_Herkenhoff_Mongey_2022}, and has been associated with the presence of large publicly traded firms that are increasingly owned by common investors \citep*{Sojourner_Common_Ownership_2022}. 
    }
These changes in margins are meaningful, accounting for one-fourth to one-third of %
the \Paste{ARTS_Markup_9312} percentage points increase in retailers gross margins between 1993 and 2012 from the Annual Retail Trade Survey (ARTS).
We also discuss the importance of multi-market pricing and the growth of online retailers. 
These trends imply that retailers may increasingly set prices based on their average market power across markets, attenuating the impacts of increasing local concentration on retailer margins.

\paragraph{Comparison to Previous Concentration Results}
Our finding of parallel increases in local and national concentration complements work documenting increasing concentration across the U.S. economy.\footnote{ 
    See \citet*{Basker2012,Foster2015,Hortacsu2015,Grullon2019,Ganapati2018a}.
    }
In particular, our product-based measures complement work finding increasing national concentration at the industry level using the Census of Retail Trade \citep{Autor2017}.
The increases in local concentration in the retail sector that we document are in line with the findings of \citet{Rinz2018} and \citet{Lipsius2018}, who study local labor markets using the Longitudinal Business Database (LBD), another U.S. Census dataset, and contrast with the decreasing trends in local concentration that they and \citet{Rossi-Hansberg2018} find outside of retail. 
We provide more evidence that retail is the \textit{only} sector with consistently increasing local concentration.

We provide new series of concentration by product categories, which better reflect the nature of competition in retail, as well as series by industry at different levels of geographic aggregation.
Our results differ from the industry-based results in \citet{Rossi-Hansberg2018} and studies of consumer purchasing decisions by \citet*{Neiman2019} and \citet*{Benkard2021}, that find decreasing concentration.

\citet{Rossi-Hansberg2018} base their results on data from the National Establishment Time Series (NETS), a private dataset, that does not contain sales by product categories, preventing them from addressing cross-industry competition. 
Moreover, the NETS includes restaurants in the retail sector while the CRT does not, which explains part of the difference between the two studies \citep{Trachter2021}.\footnote{ 
    The NETS also has issues tracking establishments and imputing sales of retailers \citep[see,][]{Crane2020,Decker2020_Discussion}.
    }
There are also methodological differences between our studies.
We consider a range of methods to calculate the local HHI to make our studies more comparable. 
We find increases in local concentration with all but one, with changes in local industry concentration between \Paste{local_chg_min} and \Paste{local_chg_max} percentage points.
The baseline estimate in \citeauthor{Rossi-Hansberg2018} that local retail concentration decreased by 17 percentage points falls significantly outside this range. 
Half of the remaining difference is due to data source as we show in Section \ref{sec:industry} and Appendix \ref{app:rh}.

On the other hand, \citet{Benkard2021} study the ownership of the brands of goods and services that consumers purchase, finding that both national and local concentration among the producers of these products decrease over time, while \citet{Neiman2019} use scanner data to show that aggregate concentration in household UPC purchases of groceries has fallen.
We calculate firm-level concentration among the retailers that sell those (and other) goods using data covering all retail sales. %
Taken together, our results are complementary as they speak to different aspects of the retail sector.
On aggregate, consumers are simultaneously purchasing a wider variety of brands as they buy those products from a smaller set of retail firms.
In this way, increasing retail concentration could cause retail firms to have both more market power with consumers and better negotiating power with suppliers.

Finally, we show how the consolidation and expansion of single- and multi-market firms has shaped the relationship between local and national concentration trends. 
Our results extend previous work documenting the expansion of individual retail firms \citep{Basker2005,BaskerJEP,Holmes_Walmart_2011} and national chains \citep*{Miranda,Foster2006Market,Foster2015} in the 1980s and 1990s.
We further contribute by showing that this expansion was followed by within consolidation of large retailers that explains 40 percent of the increase in national concentration between 1997 and 2007.

The rest of the paper proceeds as follows.
Section \ref{sec:data} describes the data and how we construct store-level sales by product. 
Section \ref{sec:concen} measures retail concentration and documents its evolution. 
Section \ref{sec:decomposition} decomposes national concentration into local and cross-market concentration.
Section \ref{sec: Discussion} discusses implications for retailers' market power.

\section{Data: Retailer Revenue for All U.S. Stores} \label{sec:data}

This section describes the creation of new data on store-level revenue for 18 product categories for all stores with at least one employee in the U.S. retail sector.  These data allow us to construct detailed measures of concentration that take into account competition between stores selling similar products in specific geographical areas.

\subsection{Data Description}

We use confidential U.S. Census Bureau microdata that cover 1992 to 2012 \citep{CRT}.  
The data source is the Census of Retail Trade (CRT), which provides revenue by product type for retail stores (establishments) in years ending in 2 and 7. 
We compile CRT data on product-level revenue and information on each store's location to define which stores compete with each other. 
Importantly, a store’s local competition will include stores in many different industries inside the retail sector because stores of different industries can sell similar products. 
This is particularly relevant for stores in the general merchandising subsector, but it also affects stores across more detailed industries (e.g., family clothing stores, women's clothing stores, and men's clothing stores). 
The data we create here are uniquely equipped to deal with cross-industry competition.

We combine the CRT data with the Longitudinal Business Database (LBD) \citep{Jarmin2002Data}, which contains data on each store's employment and allows us to track stores over time.
The LBD's firm identifiers allow us to determine which stores are owned or controlled by the same entity.\footnote{
    See \url{https://www.census.gov/econ/esp/definitions.html}
    } 
This definition of firms combines stores with different names if they are subsidiaries of the same firm. 
It also combines the stores of entities that merge into one firm.\footnote{  
    The effect of mergers on concentration depends on whether the stores that merge sell the same products, are in the same industries, and are located in the same geographical markets.
    }
We calculate all concentration measures at the firm level by combining store sales of a firm in each market. 

\subsection{Sample Construction}

The retail sector is defined based on the North American Industrial Classification System (NAICS) as stores with a two-digit code of 44 or 45.  
As such, it includes stores that sell final goods to consumers without performing any transformation of materials. 
We use the NAICS codes available from the CRT as the industry of each store. 
The sample includes all stores with positive sales and valid geographic information that appear in official CRT and County Business Patterns (CBP) statistics that sell one of the product categories used in this study.\footnote{
    We exclude sales of gasoline and other fuels, autos and automotive parts, and non-retail products because franchising makes it difficult to identify firms.  
    In our main results we exclude non-store retailers because sales from these stores are typically shipped to different markets than their physical location. 
    We explore the implications of this assumption in Section \ref{sec:online}.
    }

Table \ref{tab:summary} shows summary statistics for our sample. 
Even though the number of establishments and firms fluctuates over time, there is an overall decrease in both counts between 1992 and 2012. 
Notably, the decrease in firms is double the decrease of establishments.
This trend is consistent with the growing importance of multi-market firms in rising cross-market concentration that we show in Section \ref{sec:decomposition}.
Despite these trends, employment increases over time, representing about 9 percent of nonfarm U.S. employment over the whole sample period.\footnote{
    U.S. employment numbers come from Total Nonfarm Employees in the Current Employment Statistics \citep{BLS_EMP}.
    }

\renewcommand{\tabcolsep}{10pt}
\begin{table}[tbh]
\caption{Sample Summary Statistics}
\begin{center}
\begin{threeparttable}
\begin{tabular}{l ccccc }
\hline \hline
\Tstrut
  &  1992 & 1997 & 2002 & 2007 & 2012   \\
 \hline
 \Tstrut
    Establishments	&	908	&	942	&	913	&	912	&	877 \\
    Firms	&	593	&	605	&	589	&	566	&	523 \\
    Sales	&	1,004	&	1,368	&	1,657	&	2,062	&	2,195 \\
    Employment	&	9.91	&	11.60	&	11.89	&	12.78	&	12.31 \\
\hline
\end{tabular}
\begin{tablenotes}
\item {\footnotesize \textit{Notes:} Establishment and firm numbers are expressed in thousands. Sales and employment numbers are expressed in millions. The numbers are based on calculations from the Census of Retail Trade and the Longitudinal Business Database. (CBDRB-FY20-P1975-R8604)}
\end{tablenotes}
\end{threeparttable}
\end{center}
\label{tab:summary}
\end{table}
\renewcommand{\tabcolsep}{20pt}

\subsection{Creation of Product-Level Revenue}\label{sec:Department-Level_Revenue}

We construct product-level revenue data for all U.S. stores, allowing us to assign a store in a given location to markets based on the types of products it carries.
To do this, we exploit the CRT's store-level data on revenue by product line (e.g.,  men's footwear, women's pants, diamond jewelry). 
We then aggregate product line codes into 18 categories such that stores in industries outside of general merchandise and non-store retailers sell primarily one type of product.\footnote{
    Table \ref{app:dept_list} lists all the product categories. 
    Unless otherwise stated, we use data from all products for our aggregate results. 
    In Section \ref{sec:Product_Concentration} we focus on the eight ``main'' product categories that account for 94 percent of store sales in our sample for results for individual product categories.  
    The remaining categories are individually small and have not been released due to disclosure limitations.
    } 
For instance, stores in industries beginning with 448 (clothing and clothing accessory stores) primarily report sales in products such as women's dress pants, men's suits, and footwear, which are grouped into a Clothing category. %

The product-level data we construct opens up the possibility to study product markets covering all of retail. %
Alternative data sources, such as the NielsenIQ Homescan Consumer Panel and the Nielsen Scanner Data, provide more detailed product descriptions but lack the representativeness of our sample, focusing mostly on grocery and health products for a limited sample of people or retailers, making them less useful for computing concentration at the firm level.\footnote{ 
    Firm level concentration is the relevant measure of concentration for antitrust in retail. 
    However, detailed product concentration is informative about consumer behavior and retailers upstream market power with their suppliers \citep{Benkard2021, Neiman2019}.
    }
Our definition of product categories implies slightly higher levels of aggregation than six-digit NAICS industry codes while dealing with cross-industry competition (see Appendix \ref{app:crt}). 
It also addresses the fact that several product lines are consistently sold together, as is the case for men's wear (product line code---plc---20200), women's juniors' and misses' wear (plc 20220), and footwear products (pcl 20260), or for major household appliances (plc 20300) and TVs (plc 20320).

Aggregating product lines into categories allows us to accurately impute revenue by category for stores that do not report product-level data. 
The CRT asks for sales by product lines from all stores of large firms and a sample of stores of small firms.  
For the remainder, store-level revenue estimates are constructed from administrative data using store characteristics (e.g., industry and multi-unit status). 
These revenue estimates are constructed for stores that account for about 20 percent of sales in each year.  
Appendix \ref{app:lines} provides the details of this procedure. 

Our product-level revenue data also accounts for the presence of multi-product stores.
When a store sells products in more than one category, we assign the store's sales in each category to its respective product market. 
Consequently, a given store faces competition from stores in other industries. 
For example, an identical box of cereal can be purchased from Walmart (NAICS 452), the local grocery store (NAICS 448), or online (NAICS 454).\footnote{
    The authors found a 10.8 oz box of Honey Nut Cheerios at Walmart, Giant Eagle, and Amazon.com on June 22, 2020.
    }

Table \ref{tab:multi_prod} shows that cross-industry competition is pervasive in retail.
On average, the main subsector for each product accounts for just over half of the product's sales. 
The remaining sales are accounted for by multi-product stores, particularly from the general merchandise and non-store retailer industries, which are included in the appropriate product markets based on their reported sales.
The high sales shares of these multi-product stores makes industry classifications problematic when studying competition.
Table \ref{tab:multi_prod_9212a} reports the composition of sales for each product category, further distinguishing between general merchandisers and other multi-product retailers. 

Moreover, even stores in detailed industries (six-digit NAICS) sell the same products. 
For example, men's clothes are available at Men's clothing stores (448110), and family clothing stores (448140), in addition to department stores (452111) and discount department stores (452112), see Table \ref{app:product_by_ind} in Appendix \ref{app:crt}. 
Although the shopping experience may differ between stores in different industries a consumer looking to buy a new pair of jeans might consider stores in more than four different NAICS industries.

\renewcommand{\tabcolsep}{20pt}
\begin{table}[thb]
    \caption{Share of Product Category Sales by Main Subsector}
    \begin{center}
    \begin{threeparttable}
    \begin{tabular}{l ccc }
    \hline \hline
    \Tstrut
            & 1992 & 2002 & 2012 \\
    \hline
    \Tstrut
    Avg. Main Subsector Share	& 55.8 & 53.2 &	50.0 \\ 
    Max Main Subsector Share	& 79.8 & 73.1 &	72.4 \\ 
    Min Main Subsector Share	& 30.3 & 27.6 &	22.0 \\ 
    \hline
    \end{tabular}
    \begin{tablenotes}
    \item {\footnotesize\textit{Notes:} The numbers are based on calculations from the Census of Retail Trade. The average is the arithmetic mean across the eight main product categories of the share of sales accounted by establishments in the product's associated subsector. Shares are multiplied by 100.}
    \end{tablenotes}
    \end{threeparttable}
    \end{center}
\label{tab:multi_prod}
\end{table}

\subsection{Definition of Local Markets}

We use the 722 commuting zones that partition the contiguous U.S. as our definition of local markets. 
Commuting zones are defined by the U.S. Department of Agriculture such that the majority of individuals work and live inside the same zone, and they provide a good approximation for the retail markets in which stores compete.
If individuals live and work in a commuting zone, they likely do most of their shopping in that region.

Our results regarding the increasing trends in local concentration and the role of local trends for national concentration are robust to changes in the definition of retail markets. 
Choosing a larger geographical unit when defining retail markets, such as commuting zones, typically increases the contribution of local concentration to national concentration relative to smaller geographical units such as counties or zip codes.
Larger geographical units also tend to have lower levels of concentration than smaller units. 
However, despite differences in levels of concentration, measures at the zip code, county, commuting zone, and Metropolitan Statistical Area (MSA) levels lead to the same conclusions about the trend in local concentration, even in an extended sample dating back to 1982 (see Appendix \ref{app:1982_Sample}).

\FloatBarrier
\section{Changes in Retail Concentration} \label{sec:concen}

In this section, we use the detailed microdata described in Section \ref{sec:data} to measure national and local concentration in the U.S. retail sector. 
We find that local concentration has increased almost in parallel with national concentration. 
The increases in concentration are broad based, affecting most markets, product categories, and retail industries.

Our primary measure of concentration is the firm Herfindahl-Hirschman Index (HHI) for a given product category. %
We denote by $i$ an individual firm and by $j$ a product so that $s_{i}^{jt}$ represents the sales share of firm $i$ in product $j$ at time $t$. 
More generally, we define subscripts and superscripts such that $s_a^b$ is the share OF $a$ IN $b$.
The national HHI in a year is defined as the sum of the product-level HHIs weighted by the share of product $j$'s sales in total retail sales, $s_j^t$:
\begin{align} \label{eqn:National_HHI}
    HHI^t = \sum_{j=1}^J s_j^t   HHI_j^t, \quad \text{with} \quad  HHI_j^t = \sum_{i=1}^N \left(s_i^{jt}\right)^2,
\end{align}
while the HHI of location $\ell$ and product $j$ in year $t$ is calculated as
\begin{align} \label{eqn:Local_HHI}
    HHI_{\ell j}^t = \sum_{i=1}^N \left( s_{i}^{j \ell t} \right)^2.
\end{align}
The HHI for product $j$ measures the probability that two dollars, $x$ and $y$, chosen at random, are spent at the same firm. 
This probabilistic interpretation of the HHI provides a direct way to understand its level and changes.

\begin{figure}[tb!]
\begin{center}
    \caption{National and Local Concentration}
    \includegraphics[]{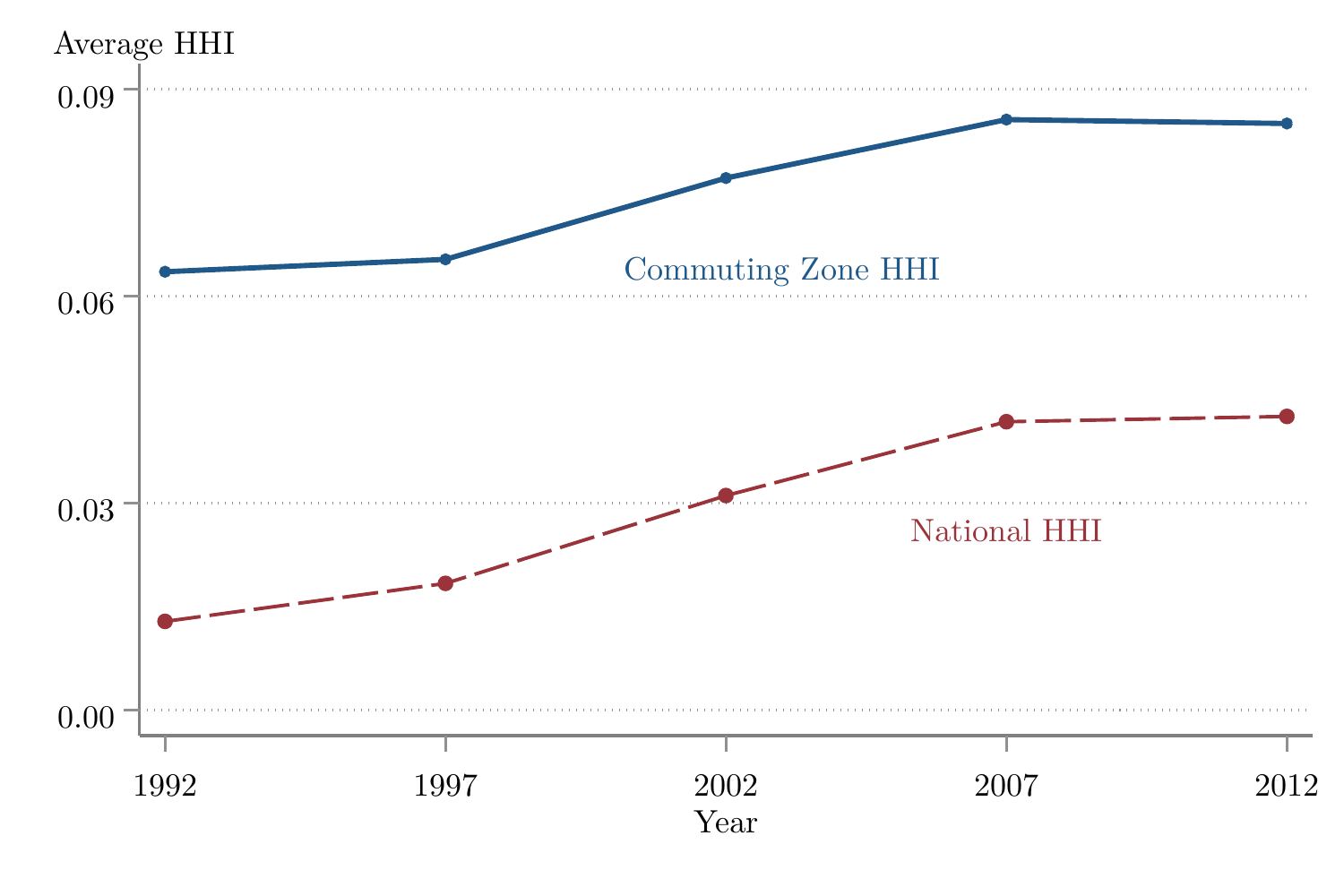}
    
    \caption*{\footnotesize \textit{Notes:} The numbers are based on calculations from the Census of Retail Trade.  The figure plots the Herfindahl-Hirschman Index (HHI) for local markets defined at the commuting zone level and national concentration. The local HHI is aggregated using each location's share of national sales within a product category. The numbers are sales-weighted averages of the corresponding HHI across product categories. (CBDRB-FY20-P1975-R8604)}
    \label{fig:agg_concen}
\end{center}
\end{figure}

Figure \ref{fig:agg_concen} plots national and local concentration in the U.S. retail sector as measured by the HHI. %
Between 1992 and 2012, both national and local concentration increased at a similar pace.
National concentration more than tripled from \Paste{nat_prod_lvl_1992_raw}  to \Paste{nat_prod_lvl_2012_raw}.
That is, the probability that two dollars are spent in the same firm in the U.S. goes from \Paste{nat_prod_lvl_1992} percent in 1992 to \Paste{nat_prod_lvl_2012} percent in 2012.
Local concentration, measured by the commuting zone HHI, increased by 34 percent from \Paste{local_prod_cz_lvl_1992} percent to \Paste{local_prod_cz_lvl_2012} percent, a similar increase  to that of the national HHI. 

We extend these results back to 1982 and consider additional measures of local  concentration measures at the zip code, county, and MSA level (Appendix \ref{app:1982_Sample}).
We find no change in the increasing trends for national and local concentration.
Most of this increase occurred between 1997 and 2007,  after which all concentration measures plateau.
In fact, the national HHI was low and grew at a low rate before 1997. 
National concentration increased by 1 percentage point in the 15 years between 1982 and 1997; by contrast, it increased 2.3 percentage points in the 10 years between 1997 and 2007.
We also show that increases in local concentration are found with other definitions of local markets and that these changes are broad based across products and geographic areas.

The national concentration results are consistent with previous industry-level work using sales and employment for various sectors, including retail  \citep{Basker2012,Foster2015,Lipsius2018,Autor2017,Rinz2018,Rossi-Hansberg2018}. 
The local concentration results are also consistent with studies on labor market concentration that find increasing local concentration in retail but decreasing local concentration in other sectors \citep{Rinz2018,Lipsius2018}.
We show that industry-based measures of sales concentration also rise at both the national and local level in Section \ref{sec:industry} and contrast these results with those of \citet{Rossi-Hansberg2018} who report decreasing local concentration.

\subsection{Changes in Concentration across Markets} \label{sec:hetero}

\begin{figure}[tbh!]
    \begin{center}
    \caption{Changes in Concentration across Markets}
    
    \begin{subfigure}[]{0.49\textwidth}
        \begin{center}
        \caption{Weighted 1992--2012}\label{fig:distr_W}
        \includegraphics[scale=0.825]{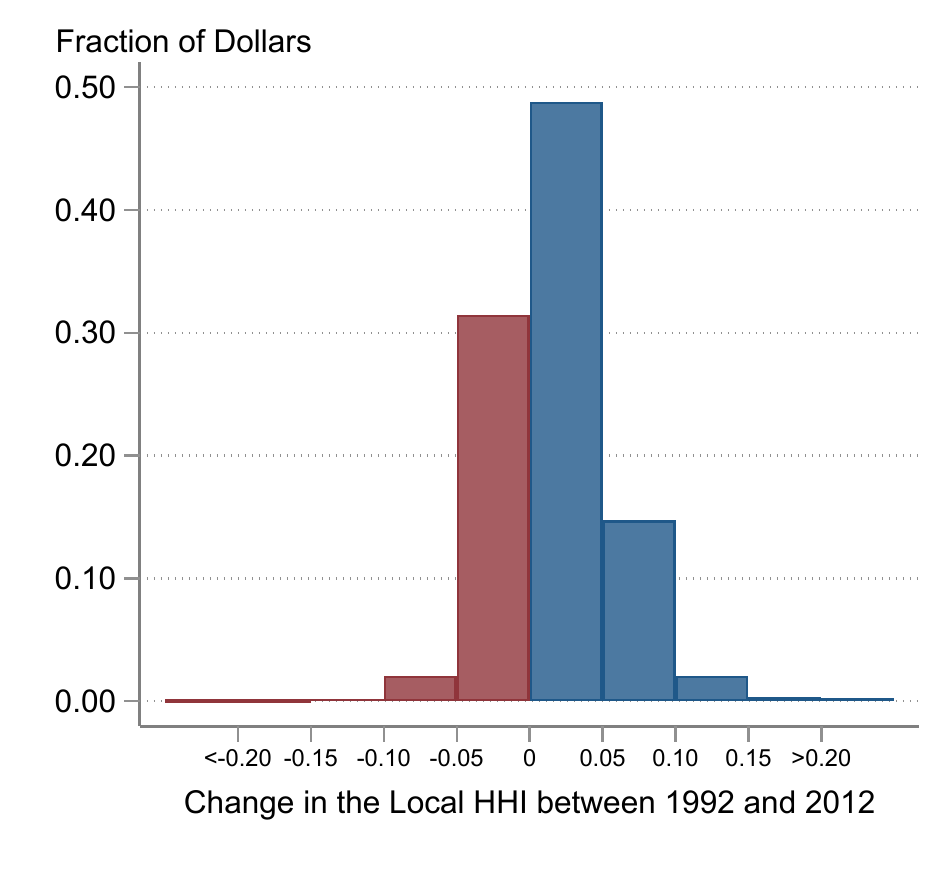}
        \end{center}
    \end{subfigure}
    \begin{subfigure}[]{0.49\textwidth}
        \begin{center}
        \caption{Unweighted 1992--2012}\label{fig:distr_U}
        \includegraphics[scale=0.825]{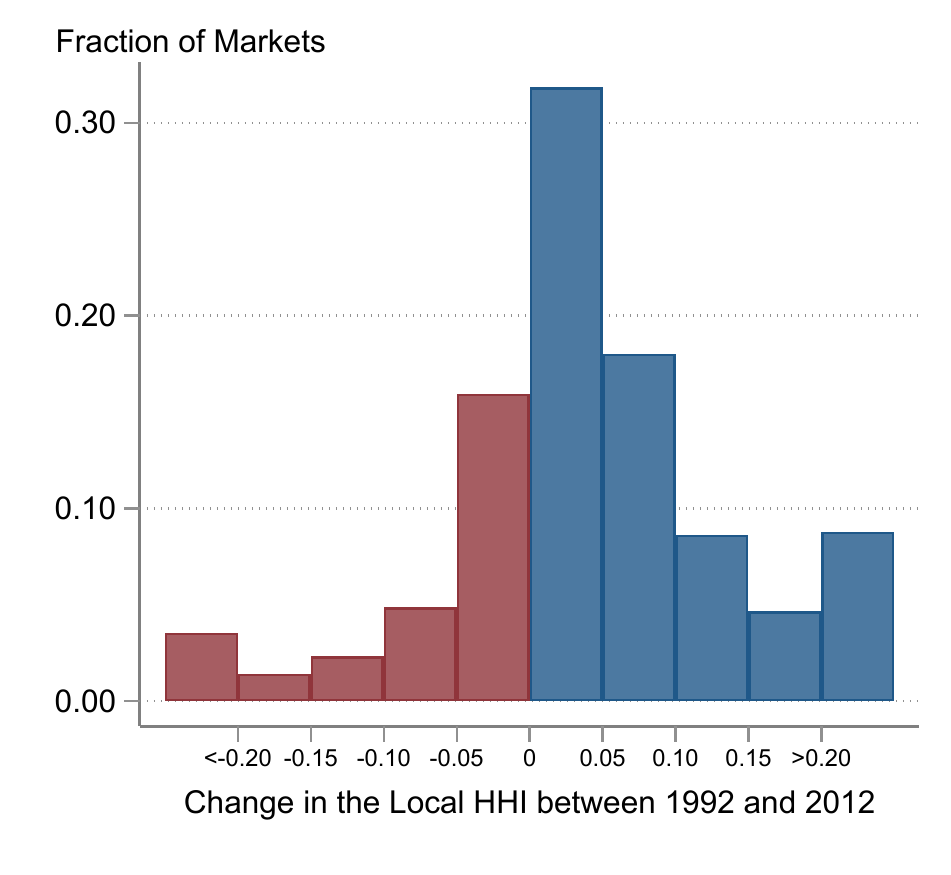}
        \end{center}
    \end{subfigure}%

    \caption*{\footnotesize \textit{Notes:} The numbers are based on calculations from the Census of Retail Trade.  
    The left panel weights markets by the average value of sales in the market in 1992 and 2012. 
    The right panel shows the fraction of markets, commuting zone/product category pairs, with changes in concentration of a given size.  
    (CBDRB-FY23-P1975-R10585)}
    \label{fig:distr}
    \end{center}
\end{figure}

We now turn to the distribution of changes in concentration across markets.
We find that the increases in concentration have been broad based. 
Sixty-six percent of dollars spent in 2012 are spent in markets that have increased concentration since 1992 (Figure \ref{fig:distr_W}). 
In 20 years, 40 percent of markets accounting for 17 percent of spending have increases in concentration of over 5 percentage points (Figure  \ref{fig:distr_U}).   
These changes are significant.  
For comparison, the Department of Justice considers a 2 percentage point increase in the local HHI potential grounds for challenging a proposed merger \citep{Justice2010}. 
On the other side, only 12 percent of markets accounting for 2 percent of spending experienced decreases in the local HHI at at least 5 percentage points.

Even though local retail concentration has sustainably increased between 1992 and 2012, the changes in concentration were more widespread between 1992 and 2002, when concentration increased in almost 70 percent of markets.
The share of markets with increasing concentration was lower in the following decade with just under 60 percent of markets increasing their concentration. 
We explore these patterns in in Appendix \ref{app: 10y_Concentration_Change} where we present changes in concentration by decade.

\subsection{Changes in Concentration across Products}\label{sec:Product_Concentration}

\begin{figure}[tbh!]
    \begin{center}
    \caption{Local and National Concentration across Product Categories}
    
    \begin{subfigure}[th!]{1\textwidth}
        \caption{Local Concentration}\label{fig:local_prod}
        \vspace{-0.24cm}
        \includegraphics[scale=1.05]{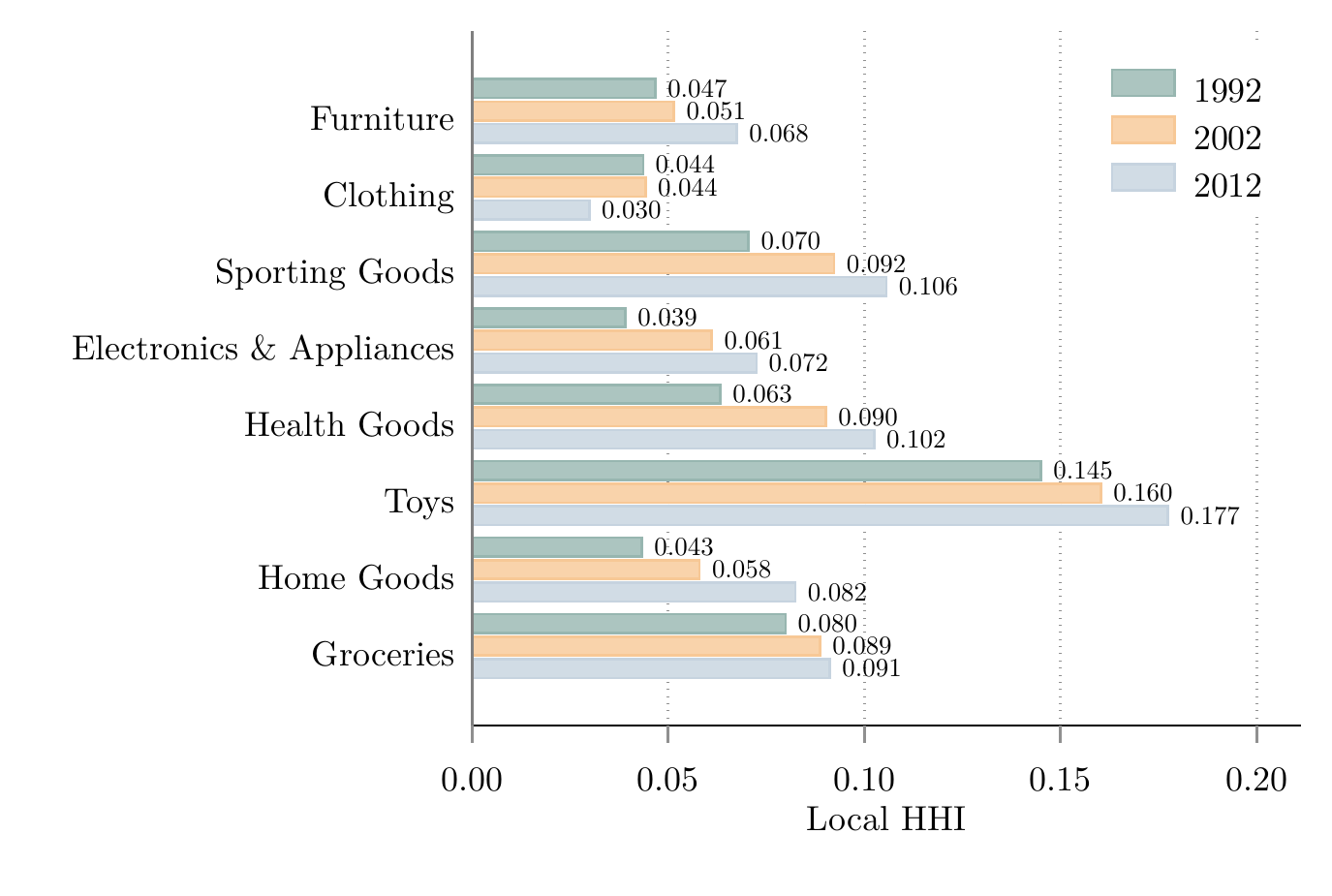}
        \vspace{-0.26cm}
    \end{subfigure}%
    
    \begin{subfigure}[th!]{1\textwidth}
        \vspace{-0.11cm}
        \caption{National Concentration}\label{fig:nat_prod}
        \vspace{-0.28cm}
        \includegraphics[scale=1.05]{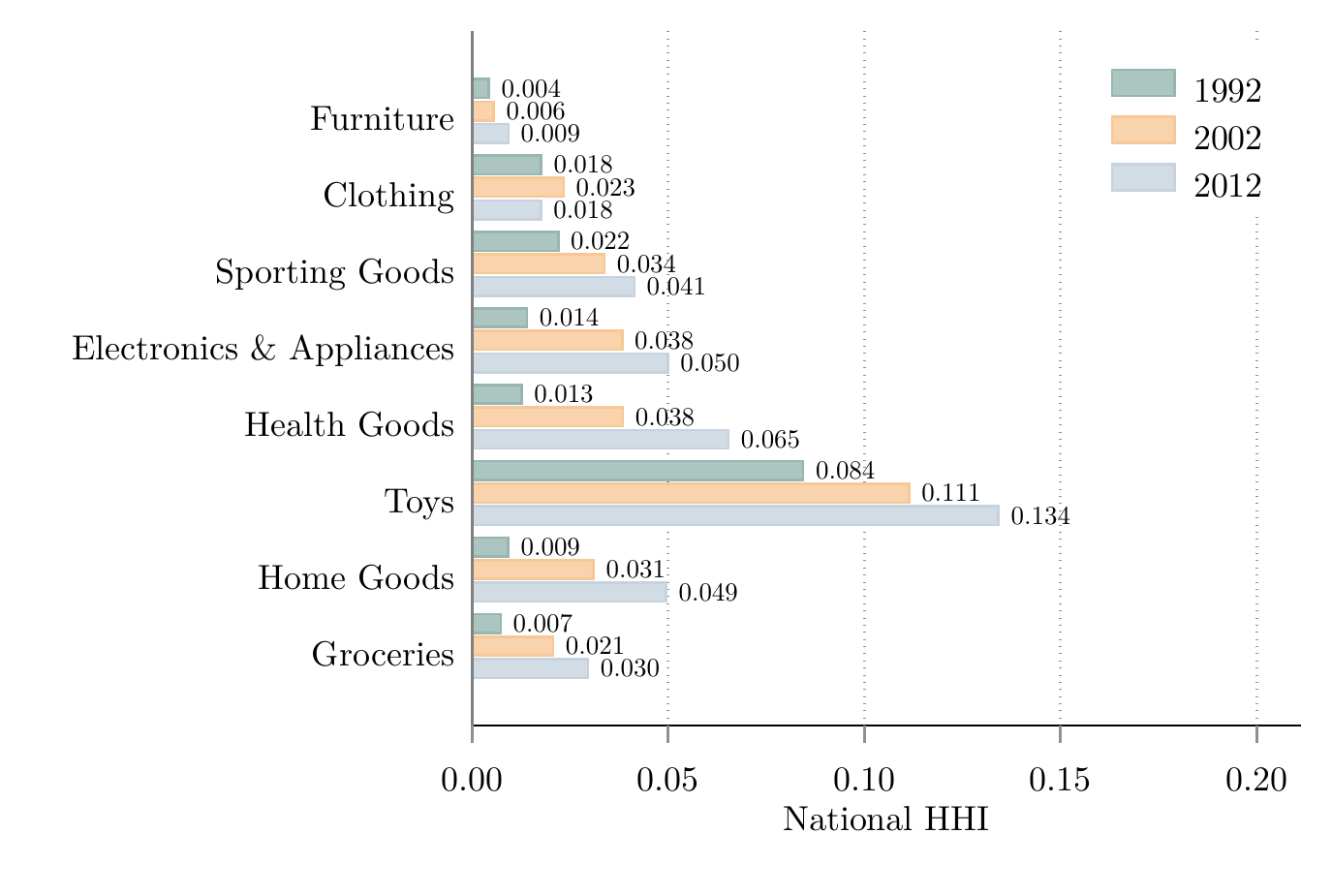}
        \vspace{-0.52cm}
    \end{subfigure}%
    
    \caption*{\footnotesize \textit{Notes:} The numbers are national and local Herfindahl-Hirschman Indexes (HHI) by product weighted by market size from the Census of Retail Trade microdata. (CBDRB-FY20-P1975-R8604)}
    \label{fig:nat_local_prod}
    \end{center}
\end{figure}

Between 1992 and 2012, both local and national concentration increased for seven of the eight major product categories,  Clothing being the exception. 
Figure \ref{fig:local_prod} shows that these increases were large for many products. %
Six of the eight categories had an increase in HHI between 3 and 4 percentage points. %
Despite this common trend, the changes in concentration vary substantially across product categories.  
Local concentration in Groceries increased by only 1.1 percentage points and decreased in Clothing by 2012, while it almost doubled in Home Goods and Electronics \& Appliances.

Figure \ref{fig:nat_prod} shows the levels of national concentration for each product category between 1992 and 2012.
The increases in national concentration are widespread and significant.
Six of the categories have larger absolute changes in national concentration relative to local concentration even though the levels of national concentration are markedly lower than those of local concentration.

Finally, comparing Figures \ref{fig:local_prod} and \ref{fig:nat_prod} shows that not all product markets evolved in the same way between 1992 and 2012.  
The markets for Furniture and Clothing changed very little, and both have relatively low levels of both local and national concentration.  
On the other hand, local markets for Groceries and Health Goods became slightly more concentrated, while at the national level, concentration has increased more than fourfold. %

\subsection{Impact of Online and Other Non-Store Retailers}  \label{sec:online}

The previous results calculated local concentration using only brick-and-mortar retailers. 
In what follows, we consider the potential impact of online and other non-store retailers on local concentration.
The market share of non-store retailers has more than tripled between 1992 and 2012.
However, the overall importance of non-store retailers remained limited through 2012. 
The initial sales share of non-store retailers is low, just 2.7 percent in 1992.
This low share reflects both the absence of online retailers and the limited role of other retailers that rely on mail order and telephone sales. 
The sales share of non-store retailers had risen to 9.5 percent by 2012, driven by an increase in online sales. 
The increase was uneven across product categories. 
Non-store retailers had significant market share in product categories, such as Furniture, Clothing, and Sporting Goods, but have almost no market share in Groceries and Home Goods (see Appendix \ref{app:nonstore_share}).

The effect of online and other non-store retailers on local concentration depends on how their sales are distributed across and within markets. 
Unfortunately, the CRT does not record the location in which non-store retailers sell their products, making it impossible to determine the exact effect of these retailers on local concentration. 
Nevertheless, we can generate bounds for the effect of non-store retailers while being consistent with their behavior at the national level.
To do this, we assume that the share of retail spending that goes to non-store retailers is constant across markets within a product category and is equal to the national sales share of non-store retailers in that category.\footnote{
    It is possible for brick-and-mortar retailers to sell online becoming mixed channel retail firms. 
    The CRT assigns these transactions to the establishment that fulfills them. %
    Transactions shipped from a brick and mortar establishment are typically included in the sales of that establishment and are thus already accounted for in our concentration calculations.
    Transactions shipped from a dedicated fulfillment center are assigned to an establishment of the same firm in a non-store retailer industry. 
    We have studied multiple ways of assigning the sales of mixed channel retailers and found that they have limited effects through 2012 because their sales are relatively low.
    }

Having distributed the sales of non-store retailers across markets, we can construct a lower and upper bound for the local HHI.
The total effect on concentration depends on the total market share of non-store retailers and how concentrated they are. 
The lower bound assumes that non-store retailers are atomistic, with the sales share of each non-store retailer equal to zero.
The lower bound is
\begin{align}
    \underline{HHI}=\left(1-s_{NS}\right)^2 HHI_{BM}, 
\end{align}
where $s_{NS}$ corresponds to the sales share of non-store retailers and $HHI_{BM}$ to the HHI of brick-and-mortar stores.
In this case, non-store retailers decrease concentration by reducing the sales share of brick-and-mortar stores. The size of this decrease depends on the sales shares of non-store retailers in the product category.
The upper bound assumes that all the sales of non-store retailers belong to a single stand-in firm. 
The upper bound is
\begin{align}
    \overline{HHI}=\left(1-s_{NS}\right)^2 HHI_{BM} + s_{NS}^2.
\end{align}
This is an upper bound on concentration under the assumption that firms do not have both brick-and-mortar and non-store establishments, which is consistent with the data.

\begin{figure}[tb!]
\begin{center}
    \caption{Local Concentration and Non-Store Retailers}
    \includegraphics[scale=1]{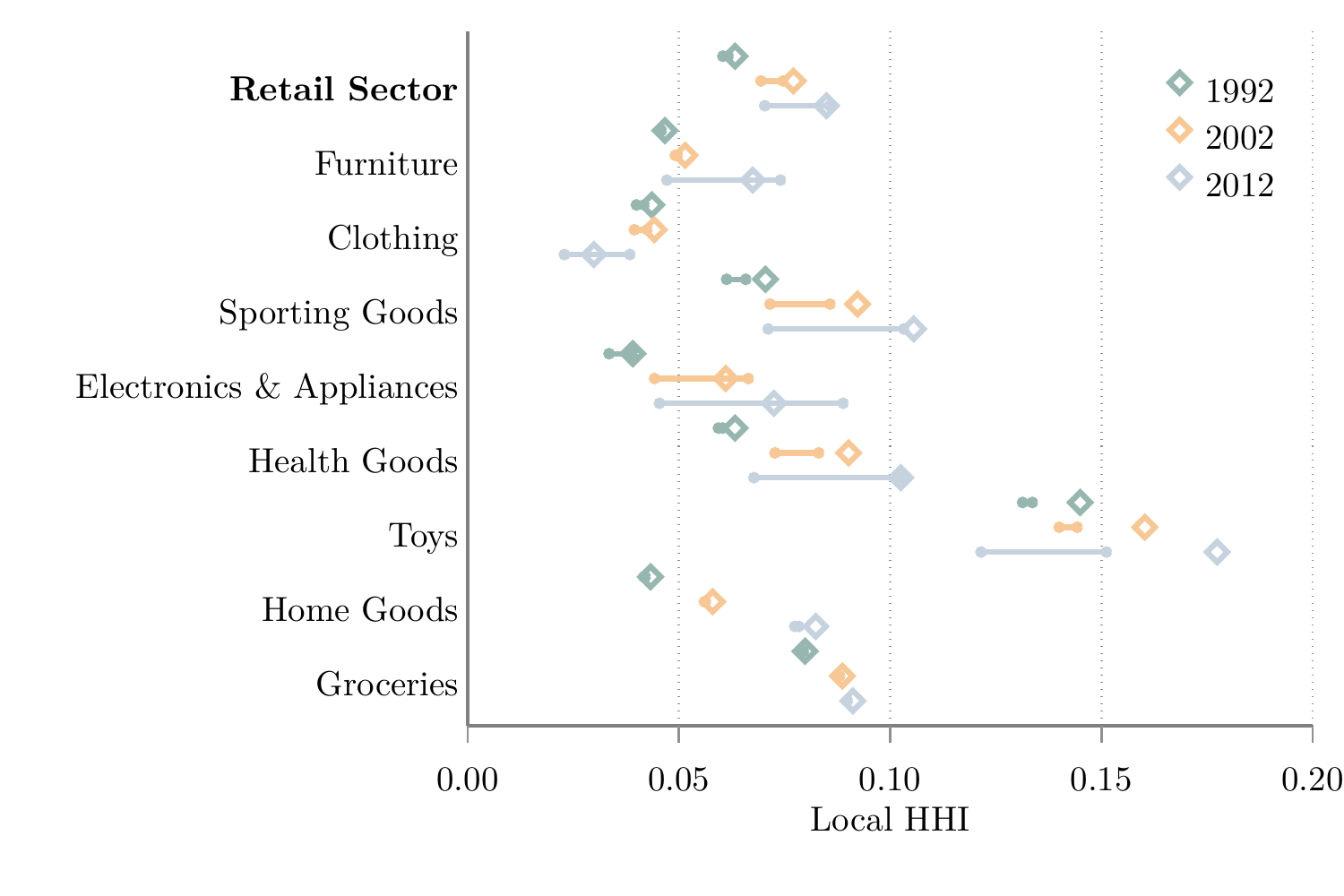}
    
    \caption*{\footnotesize \textit{Notes:} The numbers are based on calculations from the Census of Retail Trade. Diamonds mark local concentration for brick-and-mortar stores as measured by the Herfindahl-Hirschman Index (HHI) at the commuting zone level. The continuous lines cover the bounds on concentration including non-store retailers. We assume that sales of non-store retailers are distributed across local markets proportionately to the sales of brick-and-mortar retailers. The upper bound assigns all the sales of non-store retailers to a single stand-in firm. The lower bound assumes that non-store retailers are atomistic, with the sales share of each individual non-store retailer equal to zero. (CBDRB-FY20-P1975-R8604)}
    \label{fig:online_bounds_prod}
\end{center}
\end{figure}

Figure \ref{fig:online_bounds_prod} shows the bounds we construct for local concentration across product categories in the retail sector.
As expected, including non-store retailers for categories like Home Goods or Groceries hardly affects the level of concentration because the market share of non-store retailers remains low throughout. 
The effects are larger for the other categories, especially for 2012. 
Accounting for non-store retailers reduces concentration in most categories because of the decrease in market share among brick-and-mortar stores. 
For most product categories, the bounds for local concentration lie below the estimated HHI for brick-and-mortar stores (marked by the diamonds in the figure). 
It is only in Electronics \& Appliances, and to a lesser extent in Clothing, that the market share of non-store retailers is large enough for their inclusion to potentially increase concentration. 

When non-store retailers are included, there is still a clear increase in local concentration between 1992 and 2002, although the levels are slightly lower. 
Moving from 2002 to 2012, the story becomes ambiguous, especially for product categories with a significant share of their sales going to non-store retailers. 
In many cases, the bounds for 2012 contain the bounds for 2002, indicating local concentration could either be increasing or decreasing depending on the concentration among non-store sales. 
At a national level, non-store retailers were not highly concentrated during this time period \citep{Hortacsu2015}, and thus the increasing importance of non-store retailers is potentially limiting the increase in local market concentration between 2002 and 2012.
In Table \ref{tab:nonstore_nat} of Appendix \ref{app:nonstore_share}, we show that national concentration is unaffected by non-store retailers until 2002 and does reduce its increase afterwards.

\FloatBarrier
\subsection{Comparing Industry- and Product-Based Results}  \label{sec:industry}

We now document the evolution of industry-based concentration measures. 
These measures capture the variation in retail services offered by different industries.
For instance, general merchandisers offer a variety of products, and consumers value the ability to buy multiple products in one location \citep{Seo2019}. 
In this sense, industry based measures focus on a different dimension of competition between retail stores than the product-based measures we have presented.

We find larger increases in both national and local concentration at the industry level than at the product level.
We study industry-based measures of concentration defined at the six-digit NAICS level, the most disaggregated industries available in the data.
We aggregate these results to the three-digit NAICS level using a sales-weighted average.
Table \ref{tab:industry_vs_prod} shows that national industry-based concentration increases by 8.7 percentage points, 5.7 percentage points more than with product-based measures. 
Commuting zone concentration goes up by 12.6 percentage points between 1992 and 2012 when measured at the industry level, 10.4 percentage points more than the product-based measure. 
The same patterns arise when defining markets at the zip code level and show that the large increase in industry-based local concentration is not a feature of the geographical aggregation of markets.

A significant portion of the increase in industry concentration comes from six-digit industries within the general merchandise subsector (NAICS 452), where local concentration increased by 28.2 percentage points (see Figure \ref{fig:nat_local_ind} in Appendix \ref{app:ind}). 
This change is at least partially due to general merchandisers selling an increasing number of products and may not reflect increasing market power.

\renewcommand{\tabcolsep}{14pt}
\begin{table}[tb]
\caption{Comparison of Product- and Industry-Based Concentration}
\begin{center}
    \begin{threeparttable}
    \begin{tabular}{l lllll}
    \hline
    \hline 
    &      \multicolumn{5}{c}{National Concentration \Tstrut}  \\
    \hline  \Tstrut       

      & 1992 & 1997 & 2002 & 2007 & 2012 \\
    \hline
    \Tstrut
    Product Based & 0.013 & 0.018 & 0.031 & 0.042 & 0.043 \\
    Industry Based & 0.029 & 0.046 & 0.085 & 0.105 & 0.116 \\
    \hline 
    &   \multicolumn{5}{c}{Commuting Zone Concentration \Tstrut}  \\
    \hline \Tstrut
    Product Based & 0.064 & 0.065 & 0.077	& 0.086	& 0.085 \\
    Industry Based & 0.177 & 0.199 & 0.263 &	0.287 & 0.302 \\
    \hline
    &   \multicolumn{5}{c}{Zip Code Concentration \Tstrut}  \\
    \hline \Tstrut
    Product Based & 0.264 & 0.277 & 0.288 & 0.286 & 0.277 \\
    Industry Based & 0.530 & 0.552 & 0.602 & 0.611 & 0.615 \\
    \hline
    \end{tabular}
    \begin{tablenotes}
    \item {\footnotesize\textit{Notes:} The numbers come from the Census of Retail Trade and are the level of concentration in a given year with markets defined according to the noted geography. All local measures of concentration use the establishments included in the sample for the product-based results. Industry concentration uses six-digit NAICS codes. (CBDRB-FY20-P1975-R8604, CBDRB-FY19-P1179-R7207)}
    \end{tablenotes}
    \end{threeparttable}
\end{center}
\label{tab:industry_vs_prod}
\end{table}

\begin{figure}[tb!]
\begin{center}
    \caption{Product vs Industry Concentration}
    \includegraphics[scale=0.40]{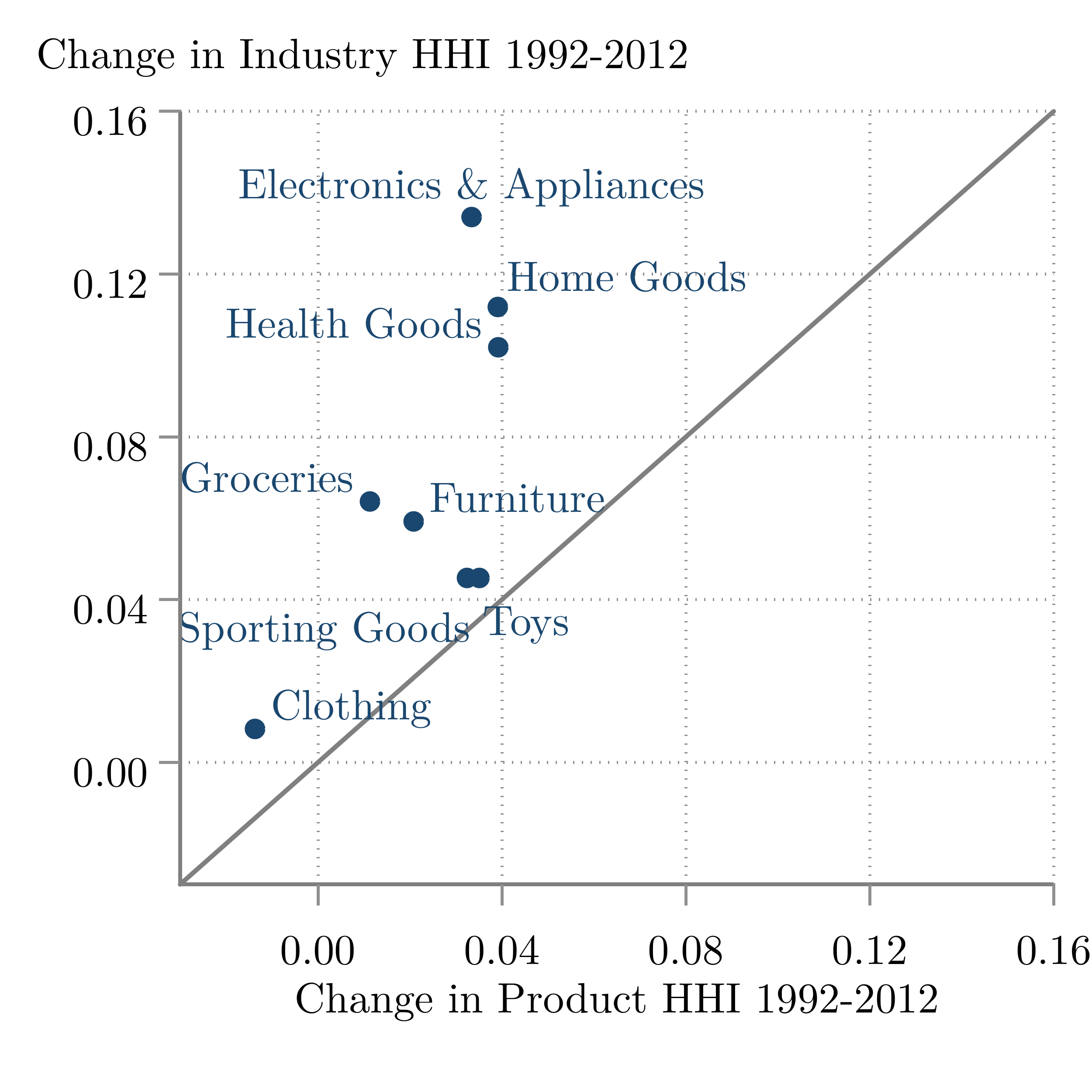}
    
    \caption*{\footnotesize \textit{Notes:} The data are from the Census of Retail Trade.  Each point marks the change in local Herfindahl-Hirschman Index (HHI) of a product category and its main subsector between 1992 and 2012. Markets are defined at the commuting zone level and are aggregated using each market's share of national sales in the relevant industry or product. (CBDRB-FY20-P1975-R8604)}
    \label{fig:prod_vs_ind}
\end{center}
\end{figure}

We match each product to the subsector that primarily sells that product (e.g., Clothing and NAICS 448: Clothing and Clothing Accessory Stores) and plot the changes in concentration in Figure \ref{fig:prod_vs_ind}. 
The figure shows a positive correlation between industry and product concentration despite the differences in market definition. 
However, the increases in concentration are larger when measured at the industry level, which explains the larger increases in overall retail concentration shown in Table \ref{tab:industry_vs_prod}. 
Appendix \ref{app:ind} complements these results by reporting the levels of local and national concentration for the industries corresponding to our main product categories and general merchandisers.

The picture that emerges from our data differs from \cite{Rossi-Hansberg2018}, who find a decrease in industrial retail concentration at the zip code level of 17 percentage points between 1992 and 2012.
Our studies differ in both data and methodology.

\citeauthor{Rossi-Hansberg2018} use U.S. National Establishment Time Series (NETS) data, a private dataset, to calculate concentration using sales and employment for multiple sectors.\footnote{ 
    \citet{Crane2020} show that the NETS has issues tracking establishments over time, making it problematic for measuring trends and \citet{Decker2020_Discussion} argues that NETS sales are typically imputed from employment, making studies with NETS most comparable to studies of employment concentration. 
    } 
The NETS defines the retail sector using Standard Industrial Classification (SIC) codes, while the CRT uses NAICS. 
The main difference is that SIC includes restaurants in retail while the NAICS does not, making it so that our results are not directly comparable \citep{Trachter2021}.

We also differ in the aggregation methodology for local concentration. 
The methodology in \citeauthor{Rossi-Hansberg2018} places more weight on markets with declining concentration because it uses each market's final share of employment and markets become less concentrated as they grow. %
Weighting markets using their average share of employment over time or their initial share always implies increasing local concentration.

We replicate the methodology of \citeauthor{Rossi-Hansberg2018} in our data and find a 1.5 percentage point decrease in local industrial concentration at the zip code level. 
The remaining 15.5 percentage points between our results are equally due to differences in market definition and data sources.
For their baseline result, \citeauthor{Rossi-Hansberg2018} define markets based on eight-digit SIC codes, while we use six-digit NAICS codes in our industry measures.\footnote{
    Eight-digit SIC codes may be overly detailed for retail markets because many retailers will sell multiple types of goods.  
    For example, concentration in eggs and poultry (54999902) would miss the fact that many eggs and poultry are sold by chain grocery stores (54119904) and discount department stores (53119901).
    }
\citeauthor{Rossi-Hansberg2018} show that moving from eight- to four-digit SIC codes in NETS implies a decline in concentration of 8 percent, explaining about half of the difference between our results.
Four-digit SIC codes are comparable to the six-digit NAICS available in the CRT.
The change from NETS data to CRT data explains the other half of the difference.
We provide full details of these exercises in Appendix \ref{app:rh}.

All other concentration measures we calculate for the retail sector---varying the level of geographical aggregation, aggregation methodology, and definition of markets by product or industry---imply an increase in local concentration between 1992 and 2012.
Taken together, we find robust evidence for increases in local retail concentration. 

\section{The Relationship Between National and Local HHIs}\label{sec:decomposition}

We now turn to the relationship between national and local concentration, and the role of single- and multi-market firms in shaping this relationship.
National concentration can increase as single- and multi-market firms consolidate in local markets and increase local concentration.
But concentration can also increase as multi-market firms expand across markets, capturing a larger share of national sales without necessarily increasing local concentration.
This expansion makes it so that consumers in different markets increasingly buy from the same firms.
We refer to this as cross-market concentration.

We disentangle the contribution of consolidation and expansion of single- and multi-market retailers in increasing national concentration in three steps. 
First, we show that the contribution of single-market firms is negligible.
Then, we establish that this is necessarily the case given the distribution of economic activity in the U.S. retail sector.
Finally, we conduct a counterfactual exercise to separate the impact of consolidation and expansion in  the growth of national concentration. 
We find that expansion plays a prominent role between 1992--2002 explaining about 89 percent of the increase in national concentration, with consolidation increasing in importance in the 1997--2007 period.

\subsection{The Contribution of Single-Market Firms to Concentration}

Single-market firms can only affect national trends through local concentration as they do not affect cross-market concentration. 
To quantify their role, we consider a counterfactual where all firms are single-market firms by breaking up multi-market firms into separate firms in each commuting zone.
This counterfactual gives us an upper bound on the role of single-market firms in explaining national trends, while keeping the level and evolution of local concentration unchanged.

If all firms operated as single-market firms, the national HHI would have been 0.06 percent in 1992 and 0.07 percent in 2012.
These counterfactual levels are 95-98 percent smaller than the actual national HHI, showing that the increase in local concentration coming from single-market firms cannot explain the evolution of U.S. retail concentration.\footnote{ 
    Similar patterns arise for individual product markets. 
    The level of the product HHI in the counterfactual varies between 0.03 and 0.14 percent and remain flat between 1992 and 2012. 
    }

This exercise further implies that, in the U.S., national retail trends must be due to changes in cross-market concentration and therefore to the behavior of multi-market firms. 
We make this precise by decomposing the national HHI into the role of local and cross-market concentration. 
Recall that the HHI for product $j$ measures the probability that two dollars, $x$ and $y$, chosen at random, are spent at the same firm. 
Using the Law of Total Probability we decompose the national HHI based on whether the two dollars are spent in the same or different markets:\footnote{
    The same principle also applies in other settings. 
    We can use the law of total probability to decompose the HHI into any set of mutually exclusive components.
    These can be the urban or rural markets, or markets with different demographic characteristics.
    We provide a detailed description of the decomposition in Appendix \ref{app:concen_decomp_details}.
    }   
\begin{align} \label{eqn:decomp}
    \underbrace{P(i_x=i_y)}_{\textrm{National HHI}} = \overbrace{P(\ell_x = \ell_y)}^{\textrm{Collocation}} \underbrace{P(i_x=i_y | \ell_x = \ell_y)}_{\textrm{Local HHI}} + \overbrace{P(\ell_x \neq \ell_y)}^{\textrm{1 -- Collocation}} \underbrace{P(i_x=i_y | \ell_x \ne \ell_y)}_{\textrm{Cross-Market Concentration}},
\end{align}
where $i_x$ is the firm at which dollar $x$ is spent and $\ell_x$ is the location of the market in which dollar $x$ is spent, and likewise for $y$.

Equation \eqref{eqn:decomp} has three components.  
The first component, $P(\ell_x = \ell_y )$, which we term collocation, captures the probability that two dollars are spent in the same location.
The second component, $P(i_x=i_y | \ell_x = \ell_y)$, is an aggregate index of local concentration, with local concentration measured as in equation \eqref{eqn:Local_HHI}.
This captures the extent  to which consumers in a local market shop at the same firm.
The third component, $P(i_x=i_y | \ell_x \ne \ell_y)$, captures cross-market concentration, the probability that a dollar spent in different markets is spent at the same firm.

The decomposition shows that 98 percent of the level of national concentration comes from the cross-market term.
The reason for this is that, in the U.S., even the largest markets represent only a small fraction of total retail sales, resulting in a low collocation term.
This limits the impact that concentration on local markets on its own can have on national trends.\footnote{
    The collocation term is
    $P(\ell_x = \ell_y ) = \sum_{\ell = 1}^L \left( s_\ell \right)^2,$
    where $s_\ell$ is the share of location $\ell$ in national sales.
    Our estimates of the collocation term for the retail sector using confidential data are about 0.012. 
    See figures \ref{fig:decomp} and \ref{fig:coll_prod} in Appendix \ref{app:concen_decomp_details}.
    } 
For national and local concentration to increase as we observe, it must be that the increase in local concentration is accompanied by cross-market concentration, with consumers in different locations increasingly shopping at the same (large) firms.
This can only happen via the consolidation and expansion of multi-market firms.

\subsection{The Contribution of Multi-Market Firms to Concentration}

The previous results show that the increase in national retail concentration reflects the activity of multi-market firms, particularly that of the largest retailers, that affect cross-market concentration through their expansion and consolidation across markets. 
Take, for instance, the behavior of firms present in at least 50 commuting zones.
Table \ref{tab:many-market} shows that their share of national sales increased from 34 percent to 58 percent between 1992 and 2012, despite there being no more than 350 of these firms in the U.S..
By contrast, their average local share increased only slightly from 3.2 to 3.4 percent. 
This suggests that the increase in U.S. retail concentration is mostly due to the expansion of multi-market retailers into new markets, rather than consolidation in the markets they were already present in.\footnote{
    In fact, large retailers went from being present in 139 commuting zones on average in 1992 to 163 commuting zones in 2012, an increase of 27 percent.
    This adds to the findings of \citet{Cao2019}, \citet{Hsieh2019}, and \citet{Rossi-Hansberg2018}, who also highlight the role of firm expansion.
    }

\begin{table}[t]
    \caption{Multi-Market Firms Present in at Least 50 Commuting Zones}
    \begin{center}
    \begin{threeparttable}
    \begin{tabular}{l c ccc}
        \hline \hline
        \Tstrut
         & & 1992 & 2002 & 2012 \\
        \hline
        Commuting Zones per firm & & 131.5 & 154.7 & 169.2 \\
        Establishments per firm  & & 616.2 & 686.7 & 814.9 \\
        National market share    & & 0.388 & 	0.533 & 0.581 \\
        Local market share       & & 0.032 & 	0.033 & 0.034 \\
        \hline
    \end{tabular}
    \begin{tablenotes}
        \item {\footnotesize \textit{Notes:} 
        The numbers come from the Census of Retail Trade and report averages across retailers with establishments in at least 50 commuting zones in a given year.
        National market share is the total sales of establishments of these firms in sales of product categories in our sample divided by total sales of product categories in sample.
        The local market share is a simple average across commuting zones, product categories, and firms of the market share of firms active in a given product category and commuting zone in a given year.  
        (CBDRB-FY23-P1975-R10585)}
    \end{tablenotes}
    \end{threeparttable}
    \end{center}
    \label{tab:many-market}
\end{table}

We find that most of the increase in U.S. retail concentration comes from the expansion of multi-market retailers, with consolidation in markets they are already in playing a minor role.
We do this by taking into account the behavior of all U.S. retailers and conducting an additional counterfactual exercise that allows us to isolate the changes in national concentration coming from higher concentration at the local level (consolidation) from those coming from the expansion of firms across markets, all while being consistent with the observed evolution of local concentration.

To isolate the contribution of within-market consolidation, we generate a counterfactual economy where the rank of firms is preserved across time.\footnote{
    We thank an anonymous referee for suggesting this exercise.
    } 
This means that the identity of the leader in each market is counterfactually preserved, as is the identity of the r\textsuperscript{th} largest firm in the market at time $t$, and they are assigned the sales share of the corresponding r\textsuperscript{th} largest firm in the market at time $t'$.
In this way, the counterfactual economy assigns the increases in local market shares to consolidation of existing firms (the previous leaders) and not to changes induced by the expansion of firms into new markets.
The counterfactual preserves the \textit{local market structure} of time $t$ and replicates the changes in local concentration by construction.

Formally, for each product $j$, we consider all firms $f$ active at some time $t$ in each location $\ell$ and record their rank $r(f,j,\ell,t)\in\left\{1,\ldots,N_{j\ell}^{t}\right\}$. 
These ranks capture the market structure at time $t$ in each local product market. 
We then consider a future time $t'$ and assign to each firm $f$ active at time $t$ the share of the firm with the same rank $r(f,j,\ell,t)$ operating in time $t'$:
$\tilde{s}_{f}^{j \ell t'} = s_{i(r(f,j,\ell,t),j,\ell,t)}^{j \ell t'}$, where $i(r,j,\ell,t')$ gives the identity of the $r^{th}$ largest firm selling product $j$, in location $\ell$ at time $t'$ and $s_{i(r(f,j,\ell,t),j,\ell,t)}^{j \ell t'}=0$ if $r(f,j,\ell,t)>N_{j\ell}^{t'}$.
So that the counterfactual matches the evolution of local concentration, we first assign market shares to the $N_{j\ell}^{t}$ firms operating at time $t$ and then assign any remaining market shares to entrants (in order of rank in $t'$).
Having completed this process, we aggregate the counterfactual shares at the national level, $\tilde{s}_{f}^{j  t'}=\sum_\ell \tilde{s}_{f}^{j \ell t'}$ and compute the national HHI as in \eqref{eqn:National_HHI}.

\begin{figure}[tb]
    \centering
    \caption{National Concentration Without Retailers' Expansion}
    
    \includegraphics{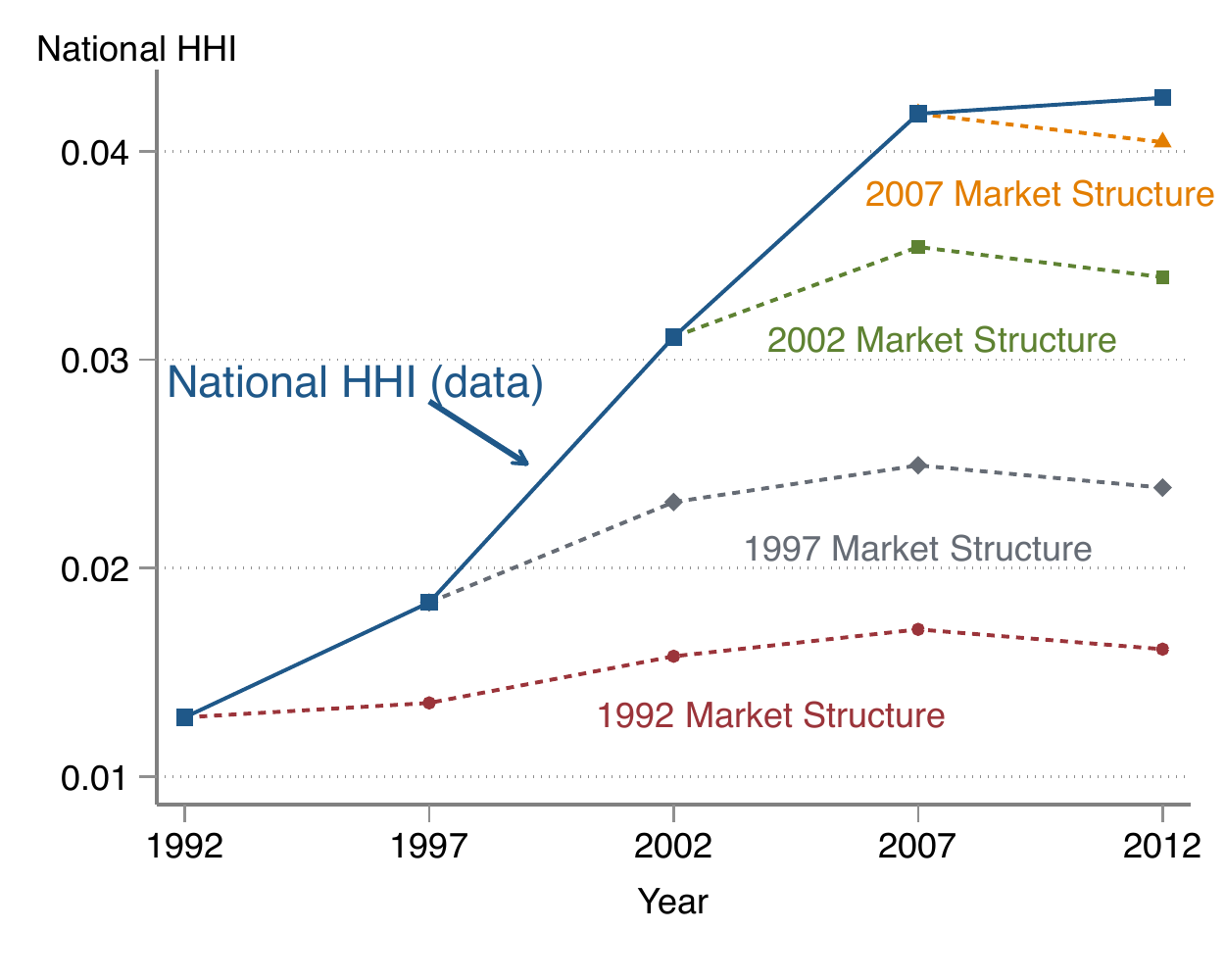}
    
    \caption*{\footnotesize \textit{Notes:} 
    The numbers are based on calculations from the Census of Retail Trade.  
    The figure reports the level of the national HHI as defined in equation \ref{eqn:National_HHI} and its counterfactual levels preserving the market structure of each Census year. 
    The counterfactual values are computed by preserving the rank of firms according to the initial market structure, assigning to the firm with the $r^{th}$ largest shares in the initial year the market share of the $r^{th}$ firm in each subsequent year.
    The numbers for this figure are presented in Table \ref{tab:counterfactual_hhi} of Appendix \ref{app:counterfactuals}.
    (CBDRB-FY23-P1975-R10585)
    }
    \label{fig:counterfactual_hhi}
\end{figure}

We conduct the counterfactual taking the market structure of each census year and present the results in Figure \ref{fig:counterfactual_hhi}.
Without the expansion of retailers across markets the national HHI would have only increased 0.33 percentage points between 1992 and 2012, this is just 11 percent of the observed change in concentration. 
The share of the increase in concentration accounted for within-market consolidation does rise when we start from the 1997 or 2002 market structure.
Consolidation explains up to 40 percent of the increase in the 1997--2007 period, the years when national concentration increased the most.
Nevertheless, the message is clear. 
The majority of the increase in U.S. retail concentration comes from the increase in the market share of multi-market retailers as they expand into new markets.

The same pattern of low but increasing importance of consolidation holds when we look at the evolution of concentration across product categories. 
Take, for example, the largest product category, groceries. 
Where expansion accounts for 92 percent of the change in national concentration between 1992 and 2002, falling to 80 percent between 2002 and 2012. 
These results align with the behavior of large general merchandisers like Walmart, that expanded into new markets during the 1980s and 1990s \citep[][Fig. 2, pg. 255]{Holmes_Walmart_2011} and went from not selling groceries to being the largest grocer in the U.S. \citep{BaskerJEP}. 
We extend these results and show that within-market consolidation plays a larger role in the period after 1997.
Our data indicates that this was part of a broader phenomena across multiple product categories.
However, there are two categories that stand as outliers.
Consolidation captures less than 2 percent of the increase in concentration in Clothing  between 1992 and 2002, while it accounts for 60 percent of the increase in concentration in Sporting Goods between 1992 and 2002, and 84 percent between 2002 and 2012.

\section{Local Retail Concentration and Retailers' Margins}\label{sec: Discussion}

The previous sections document increases in product- and industry-based measures of local retail concentration that are broad based and follow the overall trend of higher national concentration in Retail and other sectors. %
These trends in concentration can imply higher markups and ultimately affect consumer prices. 
However, studying this relationship is challenging because long series on prices or costs for U.S. retailers are unavailable.
We now provide a discussion of the magnitude of these effects in light of the observed increases in retailers' gross margins. %

To get a sense of the potential effect of concentration on markups and deal with data limitations, we use the relationship between market profitability and market concentration as measured by the HHI that arises under Cournot competition \citep[e.g.,][pg. 221-223]{Tirole_1988}.
The gross margins $(\mu)$ in a local product or industry market satisfy\footnote{ 
    The relationship above applies when goods are homogeneous, however, we generalize it to the case  in which goods are differentiated and preferences are homothetic in Appendix \ref{app: HHI_Markups_General}.
    Then, the gross-margin is approximately given by $\mu = \frac{\varepsilon}{\varepsilon-1}\left[1-\text{HHI}\right]^{-1}$.
    This is the same relationship between gross margins and concentration in models of oligopolistic competition like \citet{Atkeson2008Pricing-to-MarketPrices} and \citet{Grassi2017}.
    } 
\begin{align}
    \mu \equiv \frac{\text{Revenue}}{\text{Cost of Goods Sold}} =\left[1-\frac{\text{HHI}}{\varepsilon}\right]^{-1}, \label{eq: Cournot_HHI_Homogeneous}
\end{align}
where $\varepsilon$ is the elasticity of demand.
This relationship gives us a measure of the effect of the change in local concentration on retailers' margins coming from the increased market power of retailers in the local markets they operate in.

\renewcommand{\tabcolsep}{10pt}
\begin{table}[t]
    \caption{Change in Retailers' Margins: 1992-2012}
    \begin{center}
    \begin{threeparttable}
    \begin{tabular}{l ccc c ccc }
        \hline \hline
        \Tstrut
         & \multicolumn{3}{c}{Products $(\Delta\mu)$} & & \multicolumn{3}{c}{Industry $(\Delta\mu)$} \\ 
         & $\varepsilon=1.5$ & $\varepsilon=3$ & $\varepsilon=6$ & & $\varepsilon=1.5$ & $\varepsilon=3$ & $\varepsilon=6$ \\
        \hline
        Commuting Zone & 1.59 & 0.75 & 0.37 & & 11.89 & 4.95 & 2.27 \\
        Zip Code       & 1.27 & 0.52 & 0.23 & & 14.87 & 4.34 & 1.74 \\
        \hline
    \end{tabular}
    \begin{tablenotes}
        \item {\footnotesize \textit{Notes:} The numbers are change in gross margins for the retail sector implied by equation \eqref{eq: Cournot_HHI_Homogeneous} for changes in product and industry HHIs for different geographical definitions of local markets.
        The changes of the HHI are calculated using Table \ref{tab:industry_vs_prod}. (CBDRB-FY20-P1975-R8604)}
    \end{tablenotes}
    \end{threeparttable}
    \end{center}
    \label{tab:Cournot_Margins}
\end{table}

We use our measures of the local HHI for product and industry markets to get the changes in firms' margins between 1992 and 2012 implied by equation \eqref{eq: Cournot_HHI_Homogeneous}. 
We do this for different values of the elasticity of demand based on the range of estimates in \citet[][fig. 3]{Brand2020}.
As shown in Table \ref{tab:Cournot_Margins}, the change in local product concentration implies an increase in margins of 1.6 percentage points for a low value of $\varepsilon$, but the number is cut almost by half for larger values of the elasticity of demand.
The changes implied by industry concentration are much larger, reflecting the findings of Section \ref{sec:industry} documenting that the increases in concentration are larger at the industry than at the product level.
Industry concentration implies increases in markups between 14.9 and 1.7 percentage points. %
However, these magnitudes can be misleading as they do not take into account cross-industry competition taking place in retail. %

The change in margins in product markets implied by local product concentration is meaningful and accounts for about one-fourth of the observed changes in retailers gross margins from the Annual Retail Trade Survey (ARTS).\footnote{ 
    The changes in markups implied by local concentration as well as the changes in retailers gross margins from the ARTS are significantly lower than those found for retail by \citet[][fig. VI]{DeLoecker2017}.
    }
The ARTS data show an increase in retailers' margins of \Paste{ARTS_Markup_9312} percentage points between 1993 and 2012 \citep{ARTS}.
Moreover, taking into account the differences in concentration trends and elasticities of demand across products implies a larger increase in margins. 
In Appendix \ref{app:Model}, we estimate a model of oligopolistic competition in local product markets based on \citet{Atkeson2008Pricing-to-MarketPrices} and find that average product markups increase by \Paste{markup_prod_increase} percentage points between 1992 and 2012, or about one-third of the increase in margins reported in the ARTS.

Looking forward, the growth of multi-market and online retail presents new challenges for understanding retail markets. 
There is evidence that multi-market firms charge uniform prices across locations \citep{Adams2017,Dellavigna2017}. 
This practice initially leads to lower markups because larger (less concentrated) markets have more weight in the pricing decisions of multi-market retailers (Appendix \ref{app:pricing}). 
Simultaneously, the growth of e-commerce has led to new options for consumers in all markets, but much of this growth is due to a few large firms and may have caused the closing of brick-and-mortar retailers.

\section{Conclusion} \label{sec:conclusion}

Consumers have traditionally chosen between nearby stores selling a given product when purchasing goods. 
This fact makes local market conditions relevant for assessing the competitive environment in the retail sector.
Accordingly, we measure concentration on local product markets using novel Census data on all U.S. retailers.
We find increases in concentration covering the majority of markets which hold for product- and industry-based measures of concentration, even after taking into account the role of online and other non-store retailers.
These trends match that of national retail concentration that also rises strongly between 1992 and 2012.

Further, we show that the trends of increasing national and local concentration are linked through the expansion and consolidation of multi-market firms, with single-market firms playing no role in the increase of national concentration. 
We find that the observed increases in national concentration between 1992 and 2002 are due mostly to the expansion of multi-market firms into new markets, with consolidation of these firms in the markets they operate in increasing in importance in the 2002-2007 period.

The increases in local concentration that we document can account for one-quarter to one-third of the rise in retailers' gross margins observed in the retail sector by increasing retailers' local market power.
This can reflect an increase in markups that can potentially hurt consumers. 
However, if increases in concentration are caused by low-cost multi-market firms increasing their market share, prices may fall despite increases in markups \citep{Bresnahan1989}. 
In fact, the 1.6 percentage point increase in markups due to local concentration is small relative to the 34 percent decrease in relative retail prices observed in the same period.
These cost advantages may be due to direct foreign sourcing \citep{Smith2019a}, negotiating power with suppliers \citep{Benkard2021}, or investments in information and communication technologies \citep{Hsieh2019}.
Moreover, increases in e-commerce penetration since 2012 may have tempered the increasing trends in  local concentration. %

\clearpage
\singlespacing
\bibliography{concentration}
\onehalfspacing
\clearpage

\singlespacing
\thispagestyle{empty}
\setcounter{page}{0}
\setcounter{footnote}{0}
\appendix
\counterwithin{figure}{section}
\counterwithin{table}{section}
\counterwithin{equation}{section}
\addcontentsline{toc}{section}{Appendix} %
\part{\centering Online Appendix} %
\begin{center}
    \textbf{\Large The Evolution of U.S. Retail Concentration}\footnote{Any views expressed are those of the authors and not those of the U.S. Census Bureau. 
    The Census Bureau's Disclosure Review Board and Disclosure Avoidance Officers have reviewed this information product for unauthorized disclosure of confidential information and have approved the disclosure avoidance practices applied to this release. 
    This research was performed at a Federal Statistical Research Data Center under FSRDC Project Numbers 1179 and 1975 (CBDRB-FY19-P1179-R7207, CBDRB-FY20-P1975-R8604 and CBDRB-FY23-P1975-R10585).}

    \vspace{2.0mm}
    
    \textbf{\Large Dominic A. Smith\footnote{Bureau of Labor Statistics; Email: smith.dominic@bls.gov; Web: \url{https://www.bls.gov/pir/authors/smith.htm}.} \& Sergio Ocampo\footnote{University of Western Ontario; Email: socampod@uwo.ca; Web: \url{https://sites.google.com/site/sergiocampod/}.}}
    
\end{center}
\parttoc %

\newpage
\onehalfspacing

\setlength{\abovedisplayskip}{5pt}
\setlength{\belowdisplayskip}{5pt}
\newpage
\section{Concentration Decomposition} 
\label{app:concen_decomp_details}

We calculate the HHI for the retail sector at a time $t$, as the sales-weighted average of the product-HHIs:
\begin{align} \label{eqn:national}
HHI^t \equiv \sum_{j=1}^{J} s_j^t HHI_j^t .
\end{align}

The HHI for a given product $j$, $HHI_j^t$,  can be decomposed into the contribution of local and cross-market concentration.
The decomposition starts from the probability that two dollars $(x,y)$ spent on a product during some time period are spent at the same firm $(i)$, which gives the HHI at the national level:
\begin{align} 
    HHI_j^t \equiv P(i_x = i_y; j,t) =  \sum_i \left( s_i^{j t}\right)^2,
\end{align}  
where $s_i^{j t}$ is the share of firm $i$ in product $j$ during period $t$.
This probability can be divided into two terms by conditioning on the dollars being spent in the same location, $\ell_x = \ell_y$:
\begin{align}
    P(i_x = i_y; j,t) &= \underbrace{ \overbrace{P(i_x = i_y | \ell_x = \ell_y; j,t)}^{\text{Local Concentration}} \overbrace{P(\ell_x = \ell_y; j,t)}^{\text{Collocation}} }_{\textbf{Local Term}} \\[10pt]
    & \qquad + \underbrace{ \overbrace{P(i_x=i_y | \ell_x \ne \ell_y; j,t)}^{\text{Cross-Market Concentration}} \overbrace{P(\ell_x \ne \ell_y; j,t)}^{\text{1 - Collocation}} }_{\textbf{Cross-Market Term}}. \nonumber
\end{align}
When we report contribution of local and cross-market concentration for the retail sector, we report the sales-weighted average of these two terms across products.

The collocation probability is calculated as: 
\begin{align} \label{eqn:collocation}
    P(\ell_x = \ell_y;j,t)  &= \sum_{\ell = 1 }^{L} \left(s_\ell^{jt}\right)^2.
\end{align}
When we report the collocation for the  retail sector, we report the sales-weighted average of collocation across products: $\text{Collocation}_t=\sum_j s_j^t P(\ell_x = \ell_y;j,t)$.

The first component, $P(i_x=i_y | \ell_x = \ell_y)$, is an aggregate index of local concentration, with local concentration measured as in equation \eqref{eqn:Local_HHI}.\footnote{
    In the decomposition, each local market is weighted by the conditional probability that the two dollars are spent in location $\ell$ given that they are spent in the same location: $\nicefrac{s_\ell^2}{\left(1-\sum_p s_p^2\right)}$. 
    These weights give more importance to larger markets than the more usual weights $s_\ell$---the share of sales (of product $j$) accounted for by location $\ell$ (at time $t$). 
    We present aggregated series for local concentration in Section \ref{sec:concen} that use the latter weights. 
    Appendix \ref{app:concen_decomp_details} derives these results in detail.
    } 
This captures the extent  to which   consumers in a local market shop at the same firm.
Local concentration is calculated as: 
\begin{align}
P\left(i_x=i_y|\ell_{x}=\ell_{y}; j,t \right) & =\sum_{\ell=1}^{L}\underbrace{P\left(\ell_{x}=\ell|\ell_{x}=\ell_{y};j,t\right)}_{\text{Location Weights}}\overbrace{P\left(i_x=i_y|\ell_{x}=\ell,\ell_{x}=\ell_{y};j,t\right)}^{\text{Local HHI}} \nonumber\\
 & =\sum_{\ell=1}^{L}\frac{(s_{\ell}^{jt})^{2}}{\sum_{n} (s_{n}^{jt})^{2}}\sum_{k=1}^{K}\left(s_{k}^{j \ell t}\right)^{2}.
\end{align}

In the main text, when we report the local HHI for individual product categories we also report the retail sector's average local HHI using sales weights instead of the weights implied by the decomposition to facilitate comparison to other research:  
\begin{align}
\text{HHI}_t^{\text{Local}}=\sum_j^{J} s_j^t\sum_\ell^{L} s_{\ell}^{jt} \sum_i \left(s_{i}^{j \ell t}\right)^{2}.    
\end{align}

The third component, $P(i_x=i_y | \ell_x \ne \ell_y)$, which we call cross-market concentration, captures the probability that a dollar spent in different markets is spent at the same firm:
\begin{align}
    P(i_x=i_y | \ell_x \ne \ell_y) = \underbrace{\sum_\ell \sum_{n \ne \ell}  \frac{s_\ell s_n}{1-\sum_p s_p^2}}_{\textrm{Weights}} \underbrace{\sum_{i=1}^N s_{i}^\ell s_{i}^n}_{\textrm{Cross-Market}}. 
\end{align}
The cross-market concentration between two markets (say $\ell$ and $n$) is given by the product of the shares of the firms in each location (the probability that two dollars spent in different locations are spent in the same firm).
The pairs of markets are then weighted by their share of sales and are summed.  

The cross-market term as a whole is calculated as: 
\begin{align*}
    P\left(\ell_{x}=\ell_{y};j,t\right)P\left(i_x=i_y|\ell_{x}\ne \ell_{y};j,t\right)  &= (1-\sum_{\ell=1}^{L}\left(s_{\ell}^{jt}\right)^{2})\sum_{k=1}^{L}\sum_{\ell\ne k}\frac{s_{k}^{jt}s_{\ell}^{jt}}{1-\sum_{m}^L\left(s_{m}^{jt}\right)^{2}}\sum_{i=1}^I s_{i}^{kjt} s_{i}^{\ell jt}\\
     &=\sum_{k=1}^{L}\sum_{\ell\ne k}s_{k}^{jt}s_{\ell}^{jt}\sum_{i=1}^{I}s_{i}^{kjt}s_{i}^{j \ell t}.
\end{align*}
This calculation is the same in the results for product category because $1-\sum_{m}^L\left(s_{m}^{jt}\right)^{2}$ cancels in the calculation of the collocation term.

As mentioned in the main text, the collocation term determines how much purely local concentration (without cross-market linkages) can affect national concentration terms. 
Conversely, the collocation term determines how much can be learned about local competitive environments using national information. 
If it were large enough, national concentration numbers can be informative about local markets. %
However, the collocation term for the U.S. retail sector is low and remarkably stable across time and products. 
This implies that local concentration can only have a limited effect on national trends, making changes in national concentration mostly informative about cross-market concentration. %

\begin{figure}[tb!]
\begin{center}
    \caption{Share of Local Concentration Term in National Concentration}
    \includegraphics[]{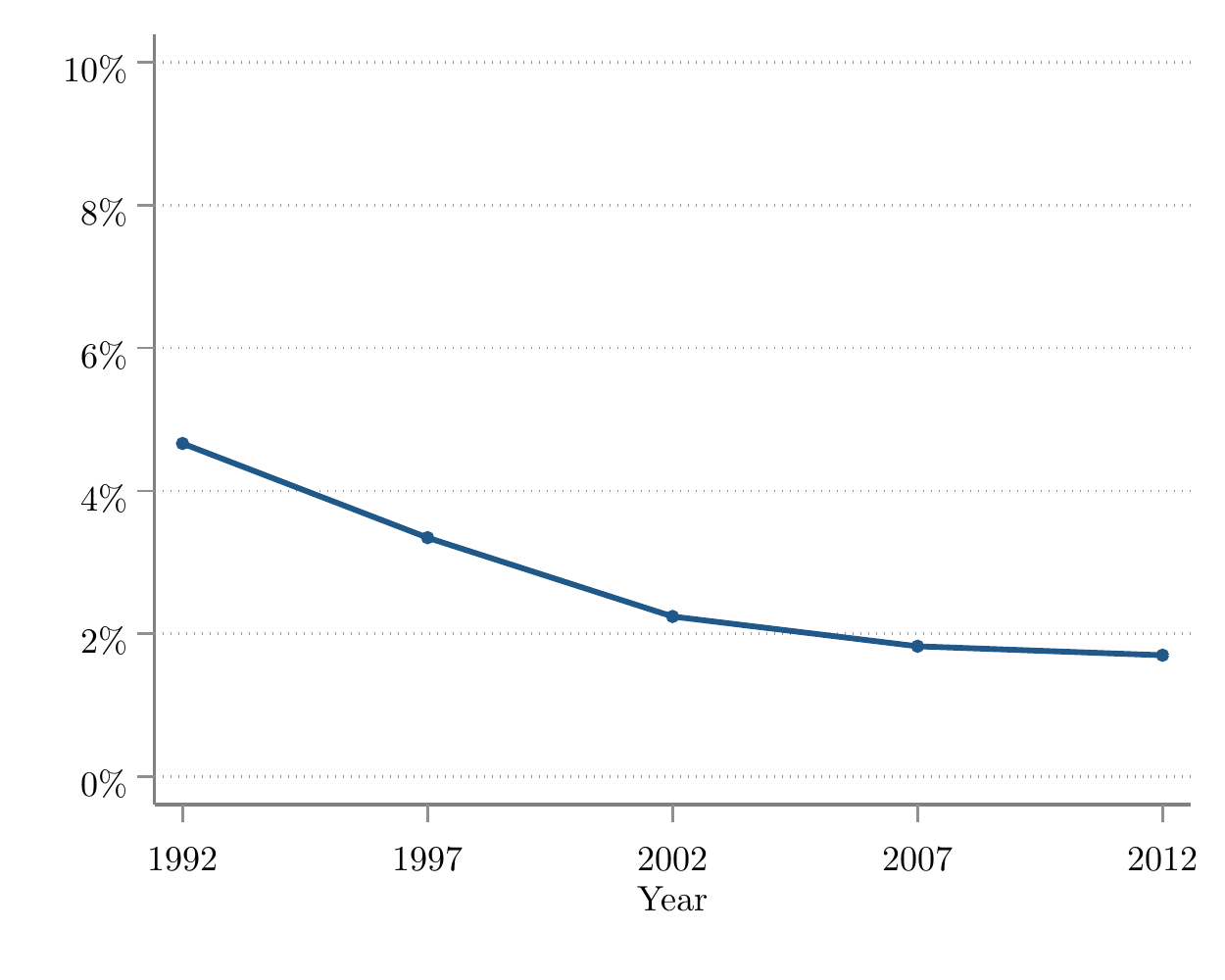}
    \caption*{\footnotesize \textit{Notes:} The numbers are based on calculations from the Census of Retail Trade.  The share of local concentration is measured as the ratio of the local concentration term in equation \eqref{eqn:decomp} to the national Herfindahl-Hirschman Index (HHI). The local concentration term is the product of the collocation term and local HHI. (CBDRB-FY20-P1975-R8604)}
    \label{fig:decomp}
\end{center}
\end{figure}

Figure \ref{fig:decomp} shows the contribution of local concentration to the level national concentration by year, defined as the ratio of the local term in the decomposition to the national HHI, $\nicefrac{P(\ell_x = \ell_y)P(i_x=i_y | \ell_x = \ell_y)}{P(i_x=i_y)}$.
The contribution of local concentration to national concentration is small---never above 5 percent.
Moreover, the contribution of local concentration to national concentration has been falling over time as national concentration has been increasing.  
By 2012, local concentration accounted for  just 1.7 percent of the national concentration level.

\begin{figure}[tbh!]
    \begin{center}
    \caption{Collocation across Product Categories}\label{fig:coll_prod}
    
        \includegraphics[scale=1.05]{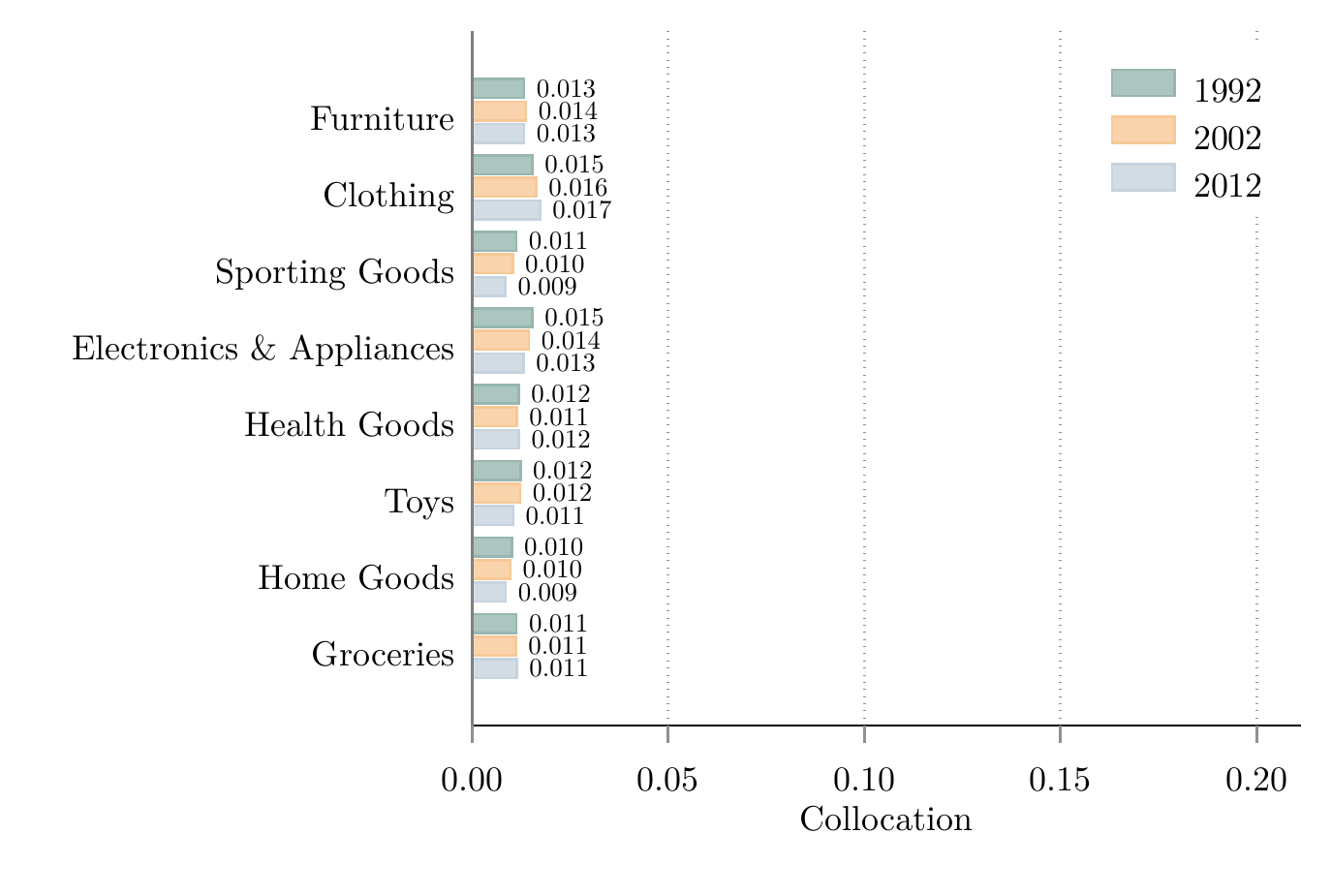}

    \caption*{\footnotesize \textit{Notes:} The numbers are collocation indexes by product weighted by market size from the Census of Retail Trade. 
    (CBDRB-FY20-P1975-R8604)}
    \label{fig:nat_cross_prod}
    \end{center}
\end{figure}

The pattern of low collocation terms and a prominent role of cross-market concentration applies across all product categories. 
Figure \ref{fig:coll_prod} shows that the collocation term is always low, less than 2 percent, and stable over time.  
By the early 1990s, only furniture and groceries have contributions of over 10 percent, with the local contribution in all other products being no higher than 5.5 percent, and as low as 2 percent.

\setlength{\abovedisplayskip}{10pt}
\setlength{\belowdisplayskip}{10pt}
\newpage
\FloatBarrier

\section{Cleaning and Aggregating Product Lines Data} \label{app:lines} 

\begin{figure}[ht]
\begin{center}
\caption{Sample Product Lines Form}
\includegraphics[scale=0.7]{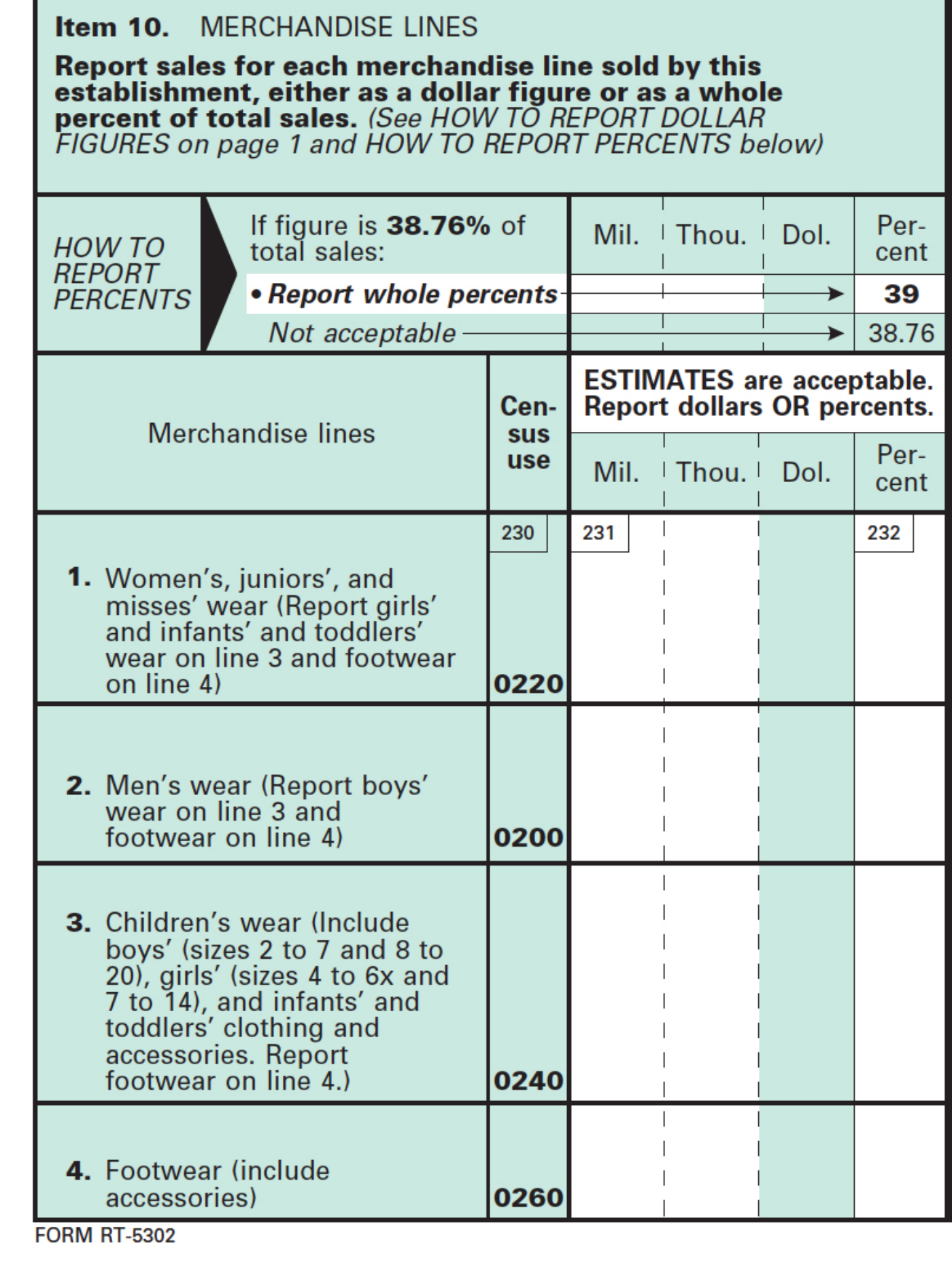}
\label{fig:sample_form}
\end{center}
\end{figure}

The Economic Census collects data on establishment-level sales in a number of product categories (Figure \ref{fig:sample_form} provides an example form).  Many establishments have missing product line sales either due to them not responding to questions or because they do not receive a form.\footnote{Establishments of large firms are always mailed a form, but small firms are sampled.}  In total, reported product lines data account for about 80 percent of sales. %
We develop an algorithm to impute data for missing establishments, which involves aggregating product line codes into categories such that we can accurately infer each establishment's sales by category with available information.  For example, we aggregate lines for women's clothes, men's clothes, children's clothes, and footwear into a product category called clothing.  

We then establish 18 product categories detailed in Table \ref{tab:Product_Category_List}. Of these 18 product categories, 8 categories that we label ``Main'' account for over 80 percent of store sales in the sample. The other 10 product categories are specialty categories that account for a small fraction of aggregate sales and are sold primarily by establishments in one specific industry.  For example, glasses are sold almost exclusively by establishments in 446130 (optical goods stores).  We create these categories so that establishments that sell these products are not included in concentration measures for the 8 main product categories.

\subsection{Aggregating Product Lines}

The first step of cleaning the data is to aggregate reported broad and detailed product line codes into categories.  Some codes reported by retailers do not correspond to valid product line codes, and we allocate those sales to a miscellaneous category.  The Census analyzes reported product line codes to check for issues and flags observations as usable if they pass this check.  We include only observations that are usable and then map these codes to categories.  We use the reported percentage of total sales accounted for by each product line instead of the dollar value because the dollar value is often missing.  Typically an establishment either reports product line data for 100 percent of its sales or does not report any data.  For the small number of establishments that report product lines data summing to a number other than 100 percent, we rescale the percentages so that they sum to one.\footnote{This procedure has a minimal effect on aggregate retail sales in each category.}  After this procedure, we have sales by product category for all establishments that reported lines data. The resulting categories are listed in Table \ref{tab:Product_Category_List}.

\renewcommand{\tabcolsep}{7pt}
\begin{table}[ht]
\caption{List of Product Categories}\label{tab:Product_Category_List}
\begin{center}
\begin{threeparttable}
\begin{tabular}{lccl}
\hline \hline \Tstrut
 Product Category & Main & Corresponding Industry & Example Firm\\
 \hline \Tstrut
Automotive Goods & N & 441 & Ford Dealer \\
Clothing & Y & 448 & Old Navy \\
Electronics and Appliances & Y & 443 & Best Buy \\
Furniture & Y & 442 & Ikea \\
Services & N & N/A  \\
Other Retail Goods & N & N/A  \\
Groceries & Y & 445 & Trader Joe's  \\
Health Products & Y & 446 & CVS \\
Fuel & N & 447 & Shell Gasoline  \\
Sporting Goods & Y & 451 & Dick's Sporting Goods \\
Toys & Y & 451 & Toys ``R'' Us \\
Home \& Garden & Y & 444 & Home Depot \\
Paper Products & N &  453210 \\
Jewelry & N & 423940 & Jared \\
Luggage & N & 448320 & Samsonite \\
Optical Goods & N & 446130 & Lenscrafters \\
Non-Retail Goods & N & N/A  \\
Books & N & 451211 & Borders \\
\hline
\end{tabular}
\begin{tablenotes}
\item {\footnotesize \textit{Notes:} Authors' created list of product categories.  The Main column indicates that a product category is included in concentration calculations. Firm names were created for illustrative purposes based on industries reported to the Securities and Exchange Commission and do not imply that the firm is in the analytical sample.}
\end{tablenotes}
\end{threeparttable}
\end{center}

\end{table}

\subsection{Imputing Missing Data} \label{app:dept_list}

For the remaining establishments, we impute data using the NAICS code of the establishment, reported sales of other establishments of the same firm in the same industry, and reported activity of the same establishment in other census years.\footnote{Reported product line sales are very similar across establishments of the same firm and the same establishment over time.}  Most establishments are part of single-unit firms, and many do not appear in multiple census years; thus their sales are imputed using only industry information.

Using this aggregation method, almost all establishments have significant sales in only two product categories, which increases confidence in the imputation.  Additionally, we have compared the aggregate sales in our data to the Consumer Expenditure Survey (an independent Bureau of Labor Statistics program), and they are in line with the numbers from that source.\footnote{Retail sales include some sales to companies, so it is expected that retail sales in a product category exceed consumer spending on that category.} 

Where relevant, all sales are deflated using consumer price indexes \citep{CPI}.  We use the food deflator for Groceries, clothing and apparel deflator for Clothing, and the deflator for all goods excluding food and fuel for all other categories.

We find that this procedure predicts sales accurately for most establishments, but a small number of stores in each industry report selling very different products than all other stores in that industry.  In these cases, the prediction can produce substantial error.

\section{Product vs Industry in the CRT} \label{app:crt}

In this section, we provide more information about the retail industries in our sample and how they compare to the products that we study. 
At the highest level of detail there are 61 NAICS industries that are comparable to our products. 
However, eight of the 61 industries account for two-thirds of sales, implying that retail sales are concentrated in small number of NAICS codes. 
Moreover, many of the largest industries sell the same sets of products such as clothing and groceries.

\subsection{Number of six-digit NAICS Industries}

Since the introduction of NAICS in 1997, there have consistently been about 70 NAICS codes with minor changes in each NAICS revision. 
We exclude the 10 codes related to automotive industries and fuel dealers (see Section \ref{sec:data}). 
That leaves about 60 industries which primarily sell goods covered by the product categories we use. 
In 2002, specifically, there were 61 six-digit NAICS codes belonging to 22 four-digit industry groups. 
Retail industry groups and subsectors (three-digit NAICS) have not been changed since NAICS was introduced.

\begin{table}[tbh]
    \caption{NAICS codes by level of Detail } \label{tab: NAICS_Definitions}
    \begin{center}
    \begin{threeparttable}
    \begin{tabular}{l cc}
    \hline \hline
    \Tstrut
     Name & Number of digits  & Number of codes  \\
     \hline
     \Tstrut
    Subsector & 3 & 5 \\
    Industry Group & 4 & 22 \\
    Industry & 6 & 61 \\
    \hline
    \end{tabular}
    \begin{tablenotes}
    \item {\footnotesize \textit{Notes:} 
    The table indicates the number of NAICS codes in the retail sector (NAICS 44-45) using 2002 NAICS codes. 
    Name is the official title of the grouping with a given number of digits. 
    Number of codes is the number of unique codes of a given length included in this papers' sample. 
    Codes in subsectors 441 and 447 are not included in any results in this paper except table \ref{tab:rst}.
    }
    \end{tablenotes}
    \end{threeparttable}
    \end{center}
\end{table}

\subsection{Concentration in Products and Industries}
Although there are many industries in retail, a small number of them account for the majority of sales as is shown in Table \ref{tab: Share_Top_Industry_Agg}. 
In 2002, the largest five industries accounted for 54 percent of sales within the 61 industries that we focus on. 
The largest eight industries accounted for 65 percent of sales which is lower than the 94 percent of sales accounted for by the eight main product categories in our study, but still large. 
The largest 18 industries account for 81 percent of sales leaving only 19 percent of sales for the remaining 43 industries. 
In fact, Table \ref{app:industry_share} shows that only 24 industries have a sales-share of more than one percent of retail sales and only 32 industries have a sales share of more than half of one percent. 
The remaining 29 retail industries are individually small and together account for only 5.5 percent of retail sales.

\begin{table}[tbh]
\caption{Share of Sales by Largest Industries} \label{tab: Share_Top_Industry_Agg}
\begin{center}
\begin{threeparttable}
\begin{tabular}{c cc}
\hline \hline
\Tstrut
 Number of Codes & Industry Share & Product Share  \\
 \hline
 \Tstrut
 5 & 54 & N/A \\
 8 & 65 & 94  \\
10 & 70 & N/A \\
18 & 81 & 100  \\
25 & 89 & N/A \\
\hline
\end{tabular}
\begin{tablenotes}
\item {\footnotesize \textit{Notes:} The table shows the sales share in retail for different groups of 6-digit NAICS industries or product categories.}
\end{tablenotes}
\end{threeparttable}
\end{center}
\end{table}

\begin{table}[tbh]
\caption{Share of Sales by NAICS Industry} \label{app:industry_share}
\begin{center}
\begin{threeparttable}
\begin{tabular}{lll SS}
\hline \hline
\Tstrut
Rank & Industry & Name & {Share} & {Cumulative} \\
\hline
\Tstrut
  1 &     445110&          Supermarkets and other grocery  stores                      & 22.0 &  22.0 \\  
  2 &     452910&                                    Warehouse clubs and supercenters  & 10.6 &  32.5 \\  
  3 &     446110&                                          Pharmacies and drug stores  &  8.7 &  41.2 \\  
  4 &     452112&                                        Discount department stores    &  7.5 &  48.7 \\  
  5 &     454113&                                                 Mail-order houses    &  5.3 &  54.0 \\  
  6 &     452111&                     Department stores (except discount dept stores)  &  5.0 &  59.0 \\  
  7 &     448140&                                            Family clothing stores    &  3.5 &  62.5 \\  
  8 &     442110&                                                  Furniture stores    &  2.8 &  65.2 \\  
  9 &     443112&                     Radio, television, and other electronics stores  &  2.7 &  68.0 \\  
 10 &     448120&                                           Women's clothing stores    &  1.7 &  69.7 \\  
 11 &     452990&                              All other general merchandise stores    &  1.7 &  71.4 \\  
 12 &     445310&                                       Beer, wine, and liquor stores  &  1.5 &  73.0 \\  
 13 &     444220&                      Nursery, garden center, and farm supply stores  &  1.5 &  74.5 \\  
 14 &     454111&                                               Electronic shopping    &  1.5 &  75.9 \\  
 15 &     451110&                                             Sporting goods stores    &  1.4 &  77.3 \\  
 16 &     448310&                                                    Jewelry stores    &  1.3 &  78.6 \\  
 17 &     448210&                                                       Shoe stores    &  1.3 &  79.9 \\  
 18 &     454390&                               Other direct selling establishments    &  1.2 &  81.1 \\  
 19 &     442299&                                 All other home furnishings stores    &  1.2 &  82.4 \\  
 20 &     445120&                                                Convenience stores    &  1.2 &  83.5 \\  
 21 &     453210&                               Office supplies and stationery stores  &  1.1 &  84.7 \\  
 22 &     451120&                                         Hobby, toy, game stores      &  1.0 &  85.7 \\  
 23 &     442210&                                             Floor covering stores    &  1.0 &  86.7 \\  
 24 &     443120&                                        Computer and software stores  &  1.0 &  87.7 \\  
 25 &     453220&                                  Gift, novelty, souvenir stores      &  0.9 &  88.5 \\  
 26 &     454311&                                               Heating oil dealers    &  0.8 &  89.4 \\  
 27 &     451211&                                                       Book stores    &  0.8 &  90.2 \\  
 28 &     443111&                                        Household appliance stores    &  0.8 &  91.0 \\  
 29 &     453998&   Other miscellaneous store retailers                                &  0.6 &  91.6 \\  
 30 &     453930&                                Manufactured (mobile) home dealers    &  0.5 &  92.1 \\  
 31 &     454312&                     Liquefied petroleum gas (bottled gas) dealers    &  0.5 &  92.7 \\  
 32 &     448190&                                             Other clothing stores    &  0.5 &  93.2 \\  
 33 &     448110&                                             Men's clothing stores    &  0.4 &  93.6 \\  
 34 &     453310&                                           Used merchandise stores    &  0.4 &  94.0 \\  
 35 &     453910&                                         Pet and pet supplies stores  &  0.4 &  94.5 \\  
 \hline
\end{tabular}

\begin{tablenotes}
\item {\footnotesize \textit{Notes:} The table shows the share of sales in retail (excluding 441 and 447) that are accounted for by each industry. Cumulative shows the fraction of sales accounted for by the X largest industries. Some industry names have been edited for space. Authors' calculations using the 2002 Economic Census Table ec0244i1 \citep{CRTPublic}. The share and cummulative columns may not sum due to rounding.}
\end{tablenotes}
\end{threeparttable}
\end{center}
\end{table}

\subsection{Industries sell many products: Clothing Example}
Having argued that industries and product categories have similar levels of concentration we now show the overlap between industry sales and product sales. 
We focus on the four product lines the comprise the Clothing product category. 
In Table \ref{app:product_by_ind}, we show which industries account for sales in each of the four product lines. 
We include all industries that are one of the top five sellers of any of the four product lines. 
In the first three categories; men's clothes, women's clothes, and children's clothes no industry accounts for more than 30 percent of sales of each line. 
These products are sold by stores in four different subsectors (three-digit NAICS codes) showing that aggregating industries would still miss the competition between retailers in the sales of these product lines. 

\begin{table}[tbh]
\caption{Share of Product Sales Accounted for by Each Industry - Clothing} \label{app:product_by_ind}
\begin{center}
\begin{threeparttable}
\begin{tabular}{ll SSSS}
\hline \hline
\Tstrut
          &        & {Men's}  & {Women's} & {Children's} &     \\
\multicolumn{2}{l}{Industry Name}    & {Clothing} & {Clothing} & {Clothing} & {Footwear}    \\
\hline \Tstrut
 448110 & Men's Clothing Stores      & 12.0 &  0.1 &  0.3 &  0.9  \\
 448120 & Women's Clothing Stores    &  1.6 & 22.1 &  0.5 &  2.3  \\
 448130 & Children's Clothing Stores &  0.1 &  0.1 & 18.4 &  0.4  \\
 448140 & Family Clothing Stores     & 29.2 & 23.3 & 19.5 &  8.6  \\
 448210 & Footwear Stores            &  2.0 &  0.5 &  0.6 & 50.0  \\
 451110 & Sporting Goods Stores      &  4.1 &  1.0 &  0.9 &  6.6  \\
 452111 & Department Stores          & 21.2 & 21.5 & 16.2 & 14.1  \\
 452112 & Discount Department Stores & 11.8 & 10.9 & 23.4 &  6.1  \\
 452910 &  Warehouse Clubs \& Supercenters & 7.3 &  4.7    & 11.4 &  3.6  \\
 454113 & Mail-order houses          &  3.3 &  5.6 &  1.9 &  3.6  \\  
 \multicolumn{2}{l}{All Other Industries}  & 7.4     &  10.2     & 6.9      & 3.8  \\
 \hline
 \Tstrut
\end{tabular}
\begin{tablenotes}
\item {\footnotesize \textit{Notes:} Numbers represent 100 multiplied by the share of sales in a product line accounted for in each industry from Economic Census Table ec0244slls2 \citep{CRTPublic} Numbers come from variable kbpct for the entire U.S.. These numbers are based on public tables and do not include imputed sales as described in Appendix \ref{app:lines}.}
\end{tablenotes}
\end{threeparttable}
\end{center}
\end{table}

Table \ref{app:ind_by_product} shows how sales within an industry are distributed across product lines. 
There is one industry that specializes in each product. 
For example, 90.3 percent of the sales of 448110 (Men's Clothing Stores) come from men's clothing. 
However, there are many industries that sell multiple product lines. 
For example, 448140 (Family Clothing Stores) receives a significant amount of revenue from each of the four lines. 
This fact motivates our decision to combine the four lines into a clothing product category because it makes our imputation procedure more reliable. 
Finally, there are the industries in the general merchandising subsectors that each receive a significant amount of revenue from outside of these four product lines.
Similar patterns arise in the other product categories.

\begin{table}[tbh]
\caption{Share of Industry Sales Accounted for by Each Product - Clothing} \label{app:ind_by_product}
\begin{center}
\begin{threeparttable}
\begin{tabular}{l SSSSS}
\hline \hline
\Tstrut
  & {Men's} & {Women's} & {Children's} &  &    \\
    & {Clothing} & {Clothing} & {Clothing} & {Footwear} & {Other}   \\
 448110 & 90.3 &  2.2 &  1.3 &  4.4 &  1.8 \\
 448120 &  3.0 & 87.9 &  0.6 &  3.1 &  5.4 \\
 448130 &  0.7 &  1.0 & 93.7 &  2.5 &  2.1 \\
 448140 & 27.1 & 44.5 & 11.0 &  5.5 & 11.9 \\ 
 448210 &  5.1 &  2.5 &  0.9 & 89.8 &  1.7 \\ 
 451110 &  9.6 &  4.8 &  1.2 & 10.8 & 73.6 \\
 452111 & 14.5 & 30.2 &  6.7 &  6.7 & 41.9 \\
 452112 &  5.2 &  9.9 &  6.3 &  1.9 & 76.7 \\
 452910 &  2.3 &  3.0 &  2.2 &  0.8 & 91.7 \\
 454113 &  2.1 &  7.1 &  0.7 &  1.5 & 88.6 \\
 \hline
 \Tstrut
\end{tabular}
\begin{tablenotes}
\item {\footnotesize \textit{Notes:} Numbers represent 100 multiplied by the share of sales in a product line accounted for in each industry from Economic Census Table ec0244slls1 \citep{CRTPublic}. Numbers come from variable lnpctal for the entire U.S.. These numbers are based on public tables and do not include imputed sales as described in Appendix \ref{app:lines}.}
\end{tablenotes}
\end{threeparttable}
\end{center}
\end{table}  
\FloatBarrier
\newpage
\section{Additional Tables and Figures}\label{app:Tab_and_Fig}

\subsection{The Role of Multi-Product Retailers}
Table \ref{tab:multi_prod_9212a} shows how sales for each main product category are distributed across sets of industries. This informs us of which type of establishment accounts for the sales of each product. The main subsector column refers to the NAICS subsector that most closely corresponds to the product category. The NAICS code of the subsector is indicated next to each product category. The main subsector accounts for just over half of sales on average, but this figure varies depending on the product. A larger fraction of sales of Furniture, Home Goods, and Groceries comes from establishments in their respective NAICS subsectors, while Electronics and Toys are more commonly sold by establishments in other subsectors. Over time, the share of sales accounted by the product's own subsector has decreased for most products, with the difference captured by establishments outside of the general merchandise subsector.

\renewcommand{\tabcolsep}{5pt}
\begin{table}[tbh!]
    \caption{Share of Product Category Sales by Establishment Subsector}
    \begin{center}
    \begin{threeparttable}
    \begin{tabular}{l ccc | ccc | ccc }
    \hline \hline
    \Tstrut
      &  \multicolumn{3}{c}{Main Subsector} & \multicolumn{3}{c}{GM} & \multicolumn{3}{c}{Other}   \\
    
     & 	1992 &	2002 &	2012 &	1992 &	2002 &	2012 &	1992 &	2002 &	2012 \\
     \hline
     \Tstrut
    {\small Furniture (442)}                 &	76.3 &	73.1 &	64.4 &	16.9 &	13.3 & 	11.2 &	6.8 &	13.6 &	24.4 \\
    {\small Clothing (448)}                  &	50.9 &	51.8 &	51.1 &	41.4 &	37.7 & 	27.4 &	7.7 &	10.5 &	21.5 \\
    {\small Sporting Goods (451)}            &	55.4 &	52.3 &	54.2 &	30.7 &	29.1 & 	21.2 &	14.0 &	18.7 &	24.6 \\
    {\small Electronic \& Appliances (443)}  &	30.3 &	31.0 &	29.5 &	34.1 &	27.1 &	24.9 &	35.6 &	41.9 &	45.6 \\
    {\small Health Goods (446)}              &	49.0 &	50.0 &	46.8 &	19.0 &	21.3 & 	20.5 &	32.0 &	28.7 &	32.6 \\
    {\small Toys (451)}                      &	40.7 &	27.6 &	22.0 &	45.2 &	47.7 & 	46.9 &	14.1 &	24.7 &	31.1 \\
    {\small Home Goods (444)}                &	63.9 &	72.8 &	72.4 &	17.2 &	11.6 & 	10.9 &	18.9 &	15.6 &	16.6 \\
    {\small Groceries (445)}                 &	79.8 &	67.2 &	59.7 &	6.6	 &16.2	 & 22.8	& 13.6 &	16.6 &	17.5 \\
    \hline
    Average                         & 	55.8 &	53.2 &	50.0 &	26.4 &	25.5 &	23.2 &	17.8 &	21.3 &	26.8 \\
    \hline
    \end{tabular}
    \begin{tablenotes}
    \item {\footnotesize\textit{Notes:} The numbers come from the Census of Retail Trade data. GM includes stores in subsector 452. Other includes sales outside of the main subsector (indicated in parenthesis) and GM. Average is the arithmetic mean of the numbers in the column. (CBDRB-FY20-P1975-R8604)}
    \end{tablenotes}
    \end{threeparttable}
    \end{center}
    \label{tab:multi_prod_9212a}
\end{table}

\newpage
\subsection{Extended Sample}\label{app:1982_Sample}

We now present results with an extended sample that covers the period 1982 to 2012. 
The 1982 and 1987 Censuses of Retail Trade do not include product-level sales for all the categories we consider in our main sample (1992-2012).  
The affected product categories, Toys and Sporting Goods, account for a relatively small share of total retail sales. Therefore, we focus on results for the retail sector as a whole which we believe are reliable for this time period.

\begin{figure}[b!]
\begin{center}
\caption{National and Local Concentration}
\includegraphics[]{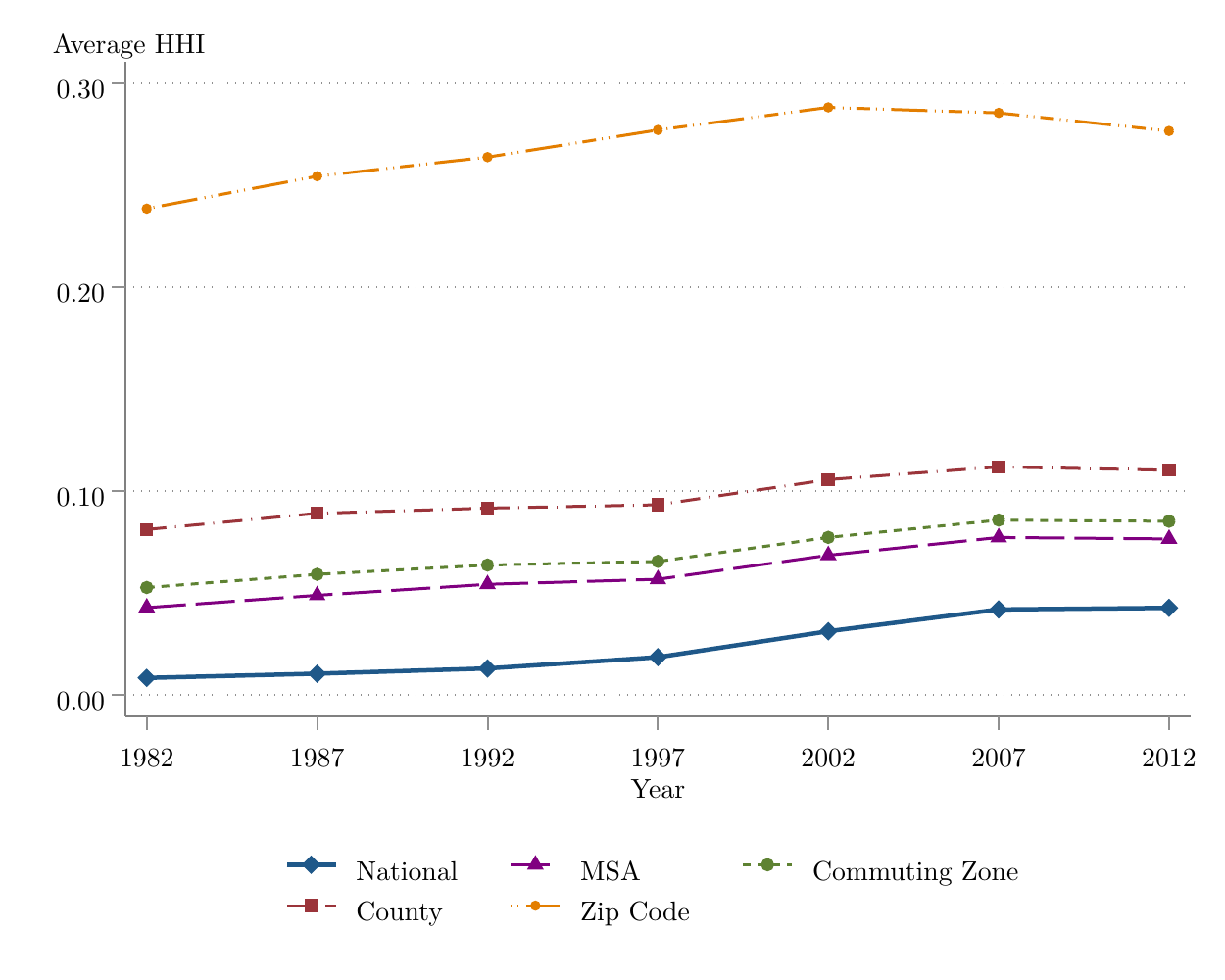}
\caption*{\footnotesize \textit{Notes:} The data are from the Census of Retail Trade.  The Herfindahl-Hirschman Index (HHI) for four different geographic definitions of local markets and national concentration are plotted. The local HHI is aggregated using each location's share of national sales within a product category. The numbers are sales weighted averages of the corresponding HHI in the product categories. (CBDRB-FY20-P1975-R8604)}
\label{fig:local_concen_1982}
\end{center}
\end{figure}

Figure \ref{fig:local_concen_1982} presents measured concentration indexes for different definitions of local markets and the retail sector as a whole going back to 1982. 
We use the store-level NAICS codes imputed by %
\citet{Fort2016} to identify retail establishments prior to 1992.
Relative to Figure \ref{fig:agg_concen} we also include a measure of local concentration where markets are defined by Metropolitan Statistical Areas (MSA). There are more MSAs than commuting zones (about 900 vs 722) and MSAs do not partition the U.S., omitting rural areas.
In practice, the measured concentration level for MSAs is similar to that of commuting zones.

Extending our sample to 1982 does not change the main result of increasing national and local concentration.
All measures show sustained increases between 1982 and 2002.
Looking at the full sample highlights the change in the rate of increase of national concentration after 1997 which contrasts with the slow increase during the 1980s.

\begin{figure}[b!]
\begin{center}
\caption{Share of Local Concentration Term in National Concentration}
\includegraphics[scale=0.9]{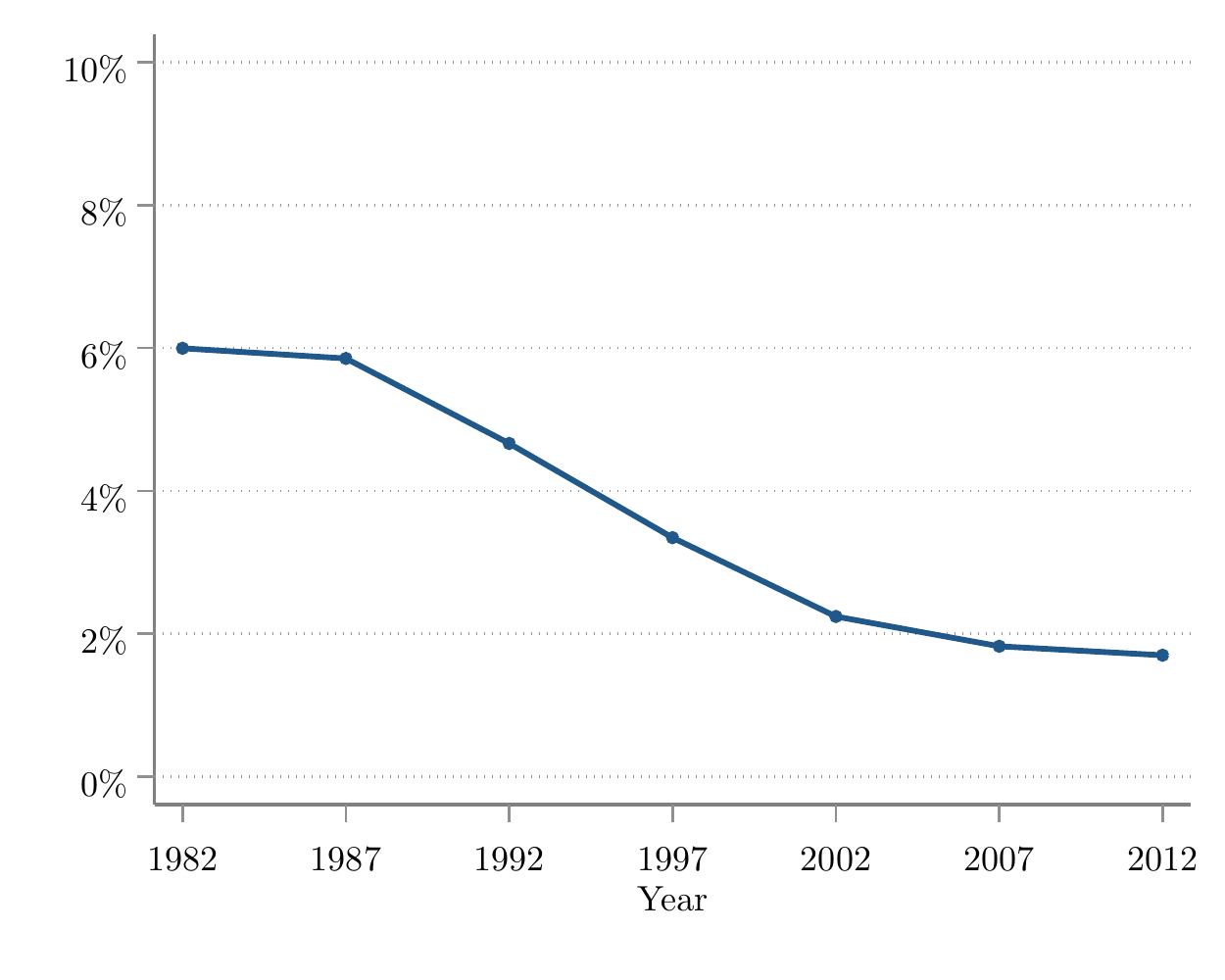}
\caption*{\footnotesize \textit{Notes:} The numbers are from the Census of Retail Trade.  The share of local concentration is measured as the ratio of the local concentration term in equation \eqref{eqn:decomp} to the national Herfindahl-Hirschman Index (HHI). 
We aggregate the local concentration terms across the product categories using their sales shares. (CBDRB-FY20-P1975-R8604)}
\label{fig:decomp_1982}
\end{center}
\end{figure}

Finally, we extend the decomposition exercise of Figure \ref{fig:decomp} to 1982.
The results, shown in Figure \ref{fig:decomp_1982}, show a stark decrease in the contribution of local concentration to national concentration. 
Even though the role of local concentration was never large (always below 6 percent), the share of national concentration attributed to local concentration fell sharply during the 1990s, ending at roughly 2 percent in 2002. 

\newpage
\subsection{Non-Store Retailer Additional Details} \label{app:nonstore_share}

\begin{figure}[b!]
\begin{center}
    \caption{Non-Store Retailers Share across Product Categories}
    \includegraphics[scale=1.05]{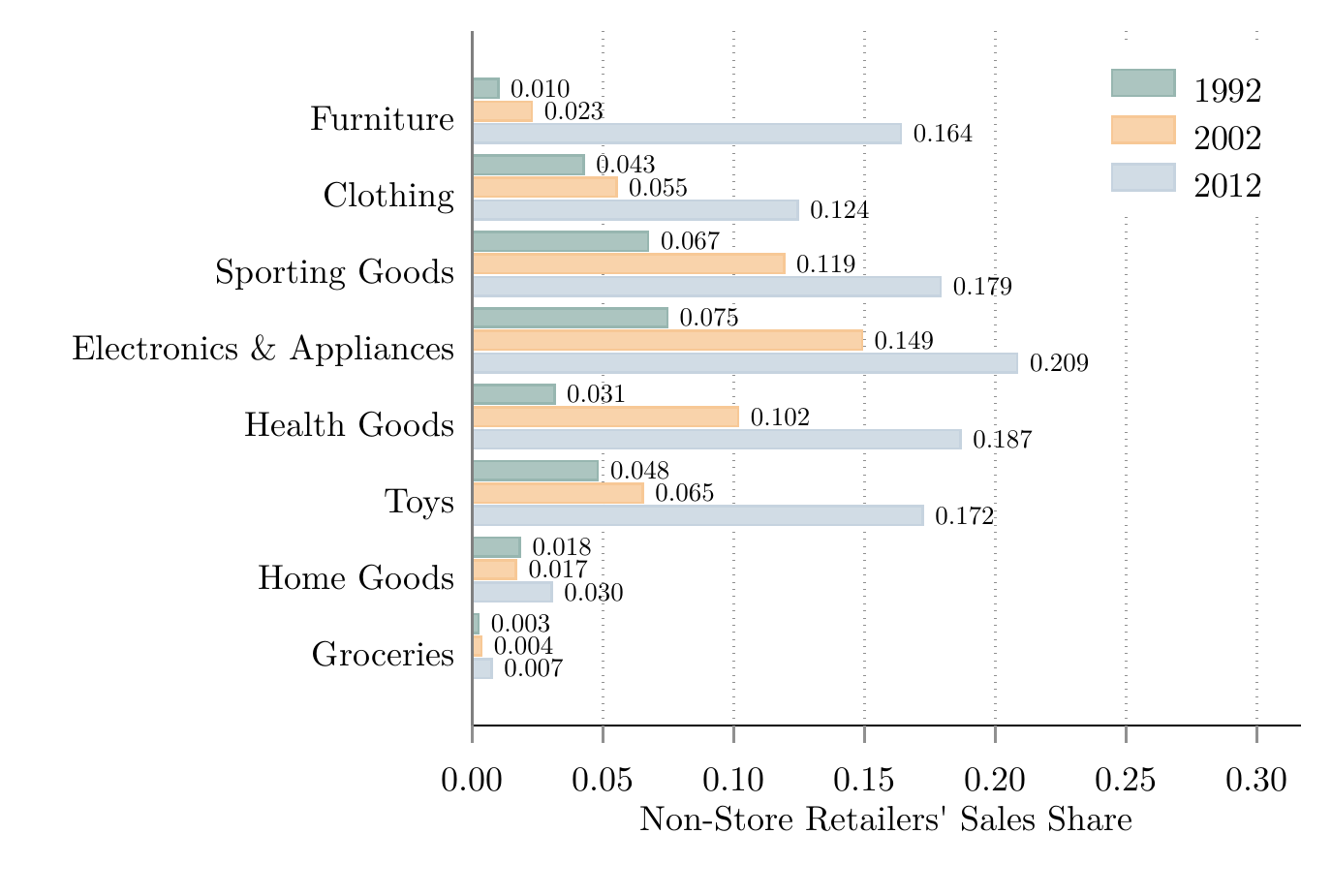}
    \caption*{\footnotesize \textit{Notes:} The numbers are the national sales shares of non-store retailers by product category from the Census of Retail Trade microdata. 
    (CBDRB-FY20-P1975-R8604)}
    \label{fig:online_prod}
\end{center}
\end{figure}

The penetration of non-store retailers varies widely across products.
As Figure \ref{fig:online_prod} shows, the sales share of non-store retailers is highest in Electronics and Appliances, with an initial share of 7.5 percent in 1992 and a share of 20.9 percent in 2012. 
The initial differences were large, with only two categories (Electronics and Sporting Goods) having a share of more than 5 percent. 
By 2012, non-store retailers accounted for more than 15 percent of sales in five of the eight major categories.
Despite this widespread increase, not all products are sold online. 
By 2012, only 0.7 percent of Groceries sales and 3 percent of Home Goods sales were accounted for by non-store retailers.
These two categories account for almost half of all retail sales, which explains the overall low sales share of non-store retailers.

Because the share of national sales going to non-store retailers is quite small these retailers have almost no impact on national concentration numbers until the 2000s. 
In Table \ref{tab:nonstore_nat}, we show national concentration numbers both with and without non-store retailers. 
The trends are similar with non-store retailers slightly lowering the increases in national concentration. 

\begin{table}[tbh!]
    \caption{National HHI with and without Non-Store Retaliers}
    \begin{center}
    \begin{threeparttable}
    \begin{tabular}{l ccccccc}
    \hline \hline
    \Tstrut
     & 	1982 &	1987 &	1992 &	1997 &	2002 &	2007 &	2012 \\
     \hline
     \Tstrut
    Excluding Non-store & 0.0082 & 0.0103 & 0.0128 & 0.0184 & 0.0311 & 0.0418 & 0.0426 \\
    Including Non-store & 0.0083 & 0.0103 & 0.0124 & 0.0175 & 0.0286 & 0.0373 & 0.0391 \\
    \hline
    \end{tabular}
    \begin{tablenotes}
    \item {\footnotesize\textit{Notes:} The numbers come from the Census of Retail Trade data. The numbers in the first row are the same numbers as in Figure \ref{fig:agg_concen} and represent the National HHI excluding non-store retailers. The numbers in the second row include establishments in NAICS 4541. 
    (CBDRB-FY20-P1975-R8604 and CBDRB-FY23-P1975-R10585)}
    \end{tablenotes}
    \end{threeparttable}
    \end{center}
    \label{tab:nonstore_nat}
\end{table}

\newpage 
\subsection{Top 4 Firm Shares}

Here we show that the average market share of the four largest firms in each product and local market increases steadily during our sample for all geographies larger than a zip code. We find the largest increases in commuting zone and metropolitan statistical areas. The average share of the four largest firms for zip codes increases until 2002 and then decrease over the next 10 years.

\begin{table}[htb]
\caption{Average Top 4 Firm Share}
\begin{center}
\begin{threeparttable}
\begin{tabular}{l  SSSSSSS}
\hline
\hline 
 & \multicolumn{1}{c}{1982} & \multicolumn{1}{c}{1987} & \multicolumn{1}{c}{1992} & \multicolumn{1}{c}{1997} & \multicolumn{1}{c}{2002} & \multicolumn{1}{c}{2007} & \multicolumn{1}{c}{2012}  \\
\hline
\Tstrut
Commuting Zone &	0.35 &	0.37 &	0.38 &	0.38 &	0.41 &	0.42 &	0.42 \\
MSA              &	0.31 &	0.33 &	0.34 &	0.35 &	0.38 &	0.39 &	0.39 \\
County          &	0.43 &	0.45 &	0.45 &	0.45 &	0.47 &	0.47 &	0.46 \\
Zip	 & 0.70	& 0.71 &	0.72 &	0.72 &	0.72 &	0.70 &	0.68 \\
\hline 
\end{tabular}
\begin{tablenotes}
\item {\footnotesize\textit{Notes:} Results come from the Census of Retail Trade. The market share of the 4 firms with the greatest sales in each product category and location in each year are summed. These results are then aggregated using a weighted average of the sales share of each product and location in a year. (CBDRB-FY20-P1975-R8604)}
\end{tablenotes}
\end{threeparttable}
\end{center}
\label{tab:Top4_Shares}
\end{table}

\subsection{10 Year Concentration Changes}\label{app: 10y_Concentration_Change}

\begin{figure}[tbh!]
    \begin{center}
    \caption{Changes in Concentration across Markets}
    
    \begin{subfigure}[]{0.49\textwidth}
        \begin{center}
        \caption{Unweighted 1992--2002}\label{fig:distr_2002}
        \includegraphics[scale=0.825]{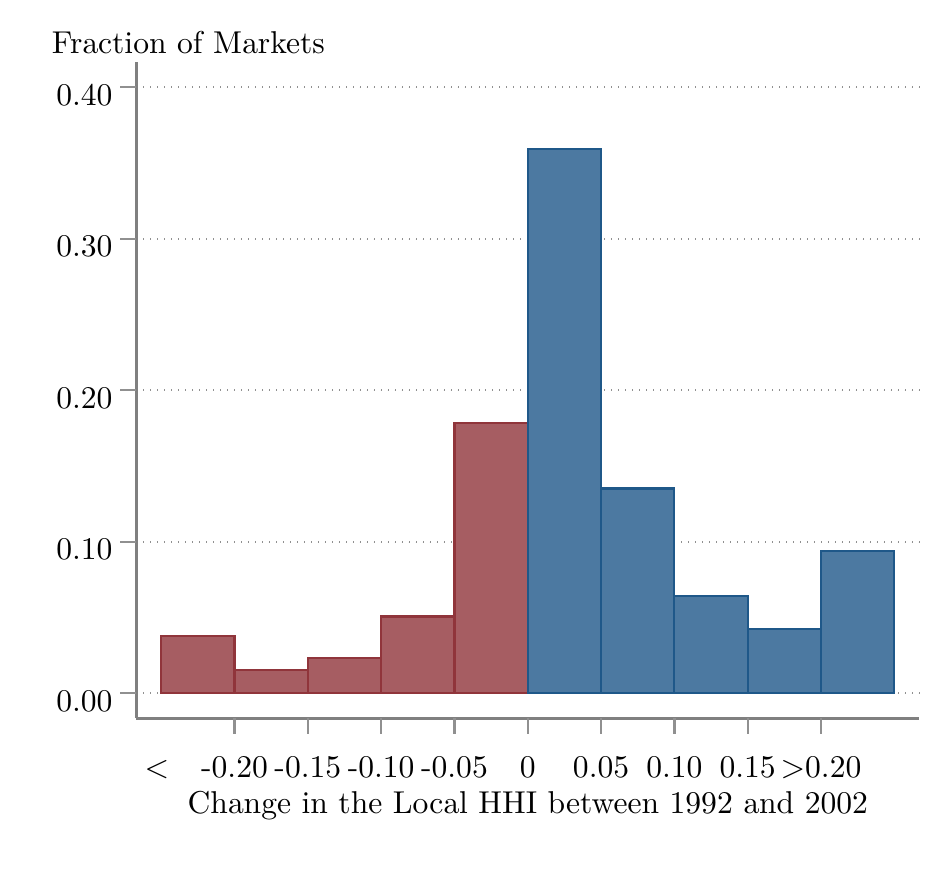}
        \end{center}
    \end{subfigure}%
    ~
    \begin{subfigure}[]{0.49\textwidth}
        \begin{center}
        \caption{Unweighted 2002--2012}\label{fig:distr_2012}
        \includegraphics[scale=0.825]{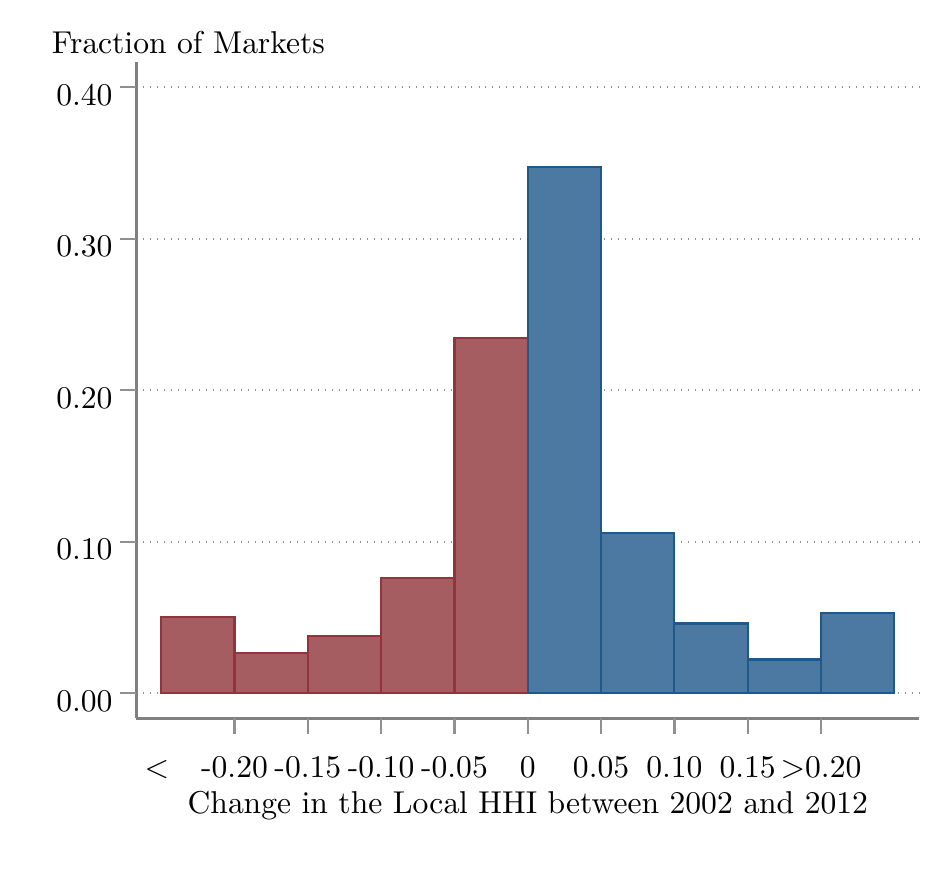}
        \end{center}
    \end{subfigure}
    
    \begin{subfigure}[]{0.49\textwidth}
        \begin{center}
        \caption{Weighted 1992--2002}\label{fig:distr_W_2002}
        \includegraphics[scale=0.825]{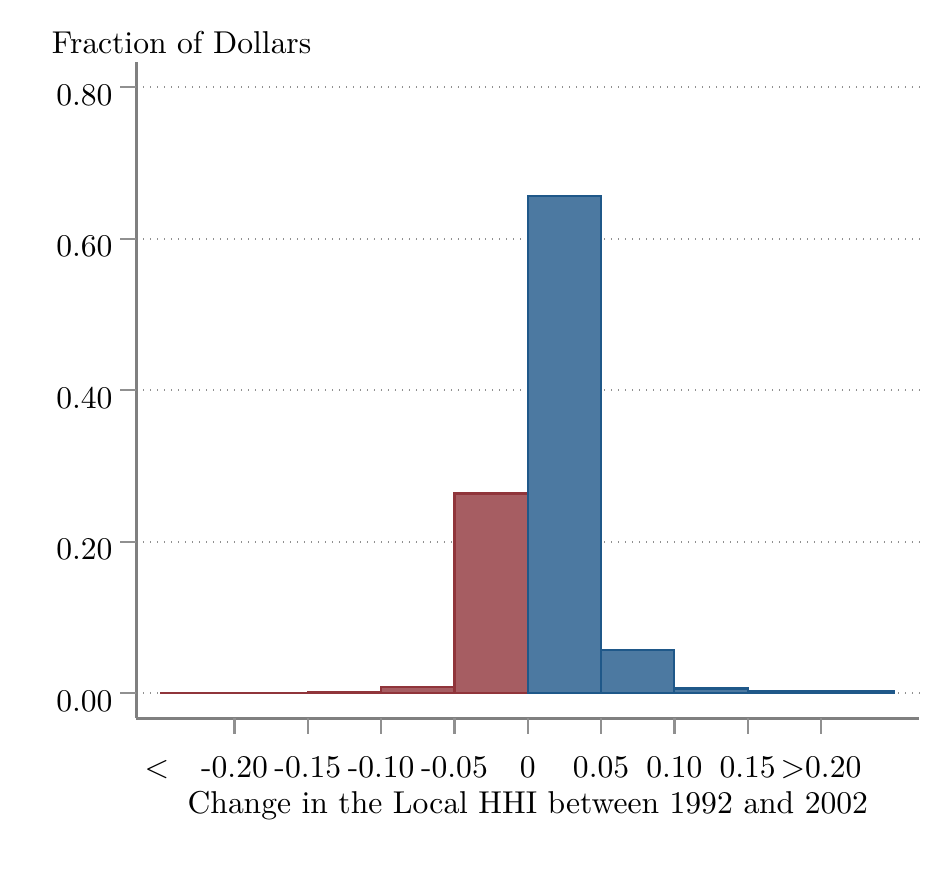}
        \end{center}
    \end{subfigure}%
    ~
    \begin{subfigure}[]{0.49\textwidth}
        \begin{center}
        \caption{Weighted 2002--2012}\label{fig:distr_W_2012}
        \includegraphics[scale=0.825]{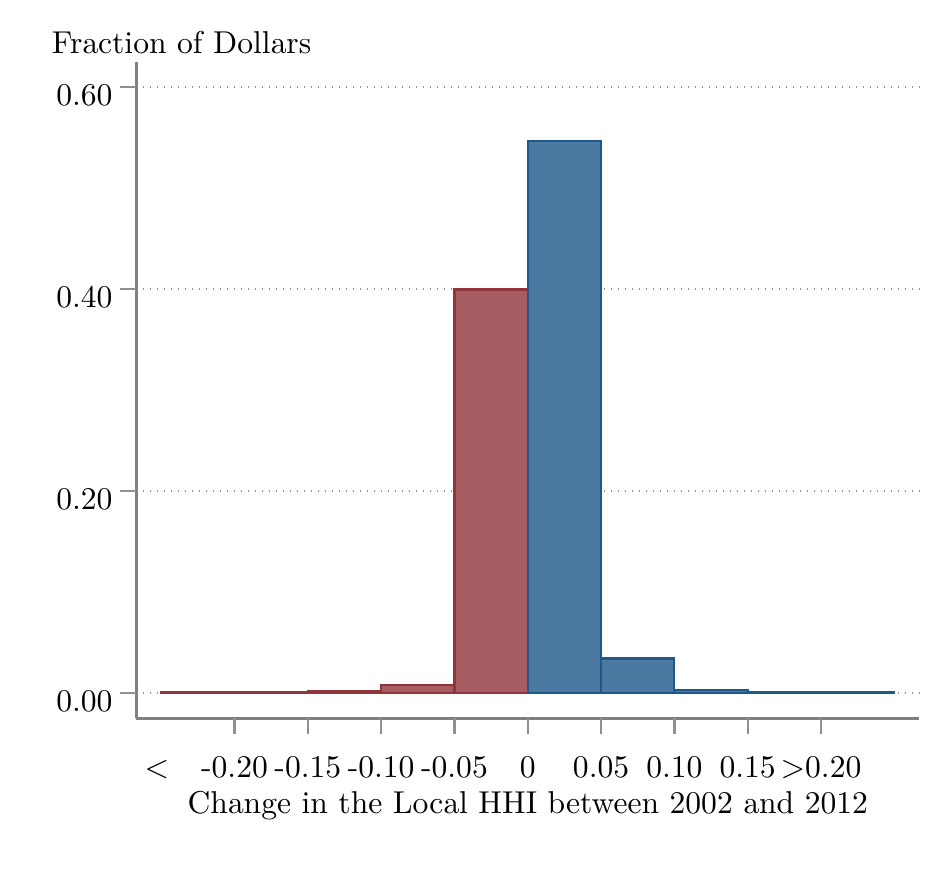}
        \end{center}
    \end{subfigure}%
    
    \caption*{\footnotesize \textit{Notes:} The numbers are based on calculations from the Census of Retail Trade.  The top panels show the fraction of markets, commuting zone/product category pairs, with changes in concentration of a given size.  The bottom panels weight markets by the value of sales in the product category. The columns report changes for the decades 1992 to 2002 and 2002 to 2012. 
    (CBDRB-FY20-P1975-R8604)}
    \label{fig:distr_10y}
    \end{center}
\end{figure}

We now turn to the distribution of changes in concentration across markets.
We find that the increases in concentration have been broad based. 
Almost \Paste{pct_local_increase_2002} percent of dollars spent in 2012 are spent in markets that have increased concentration since 2002 (Figure \ref{fig:distr_W_2012}). 
In just 10 years, 23 percent of markets have increases in concentration of over 5 percentage points (Figure  \ref{fig:distr_2012}).   
These changes are significant.  
For comparison, the Department of Justice considers a 2 percentage point increase in the local HHI potential grounds for challenging a proposed merger \citep{Justice2010}.

Figures \ref{fig:distr_2002} and \ref{fig:distr_W_2002} show that the changes in concentration were even more widespread between 1992 to 2002. 
Over \Paste{pct_local_increase_1992} percent of markets, accounting for \Paste{pct_local_increase_1992_weighted} percent of retail sales, increased their concentration. 
In both the 1992--2002 and 2002--2012 decades, the majority of retail sales occurred in markets with relatively small increases in concentration (between 0 and 5 percentage point increases in the market's HHI).
These markets account for 66 percent of retail sales in 2002 and 55 percent in 2012.

\newpage
\subsection{Industry-Based Results}\label{app:ind}

\begin{figure}[th]
    \begin{center}
    \caption{National and Local Concentration Across Industries}
    \begin{subfigure}[th!]{1\textwidth}
        \vspace{-0.11cm}
        \caption{National Concentration}\label{fig:nat_ind}
        \vspace{-0.28cm}
        \includegraphics[scale=1.05]{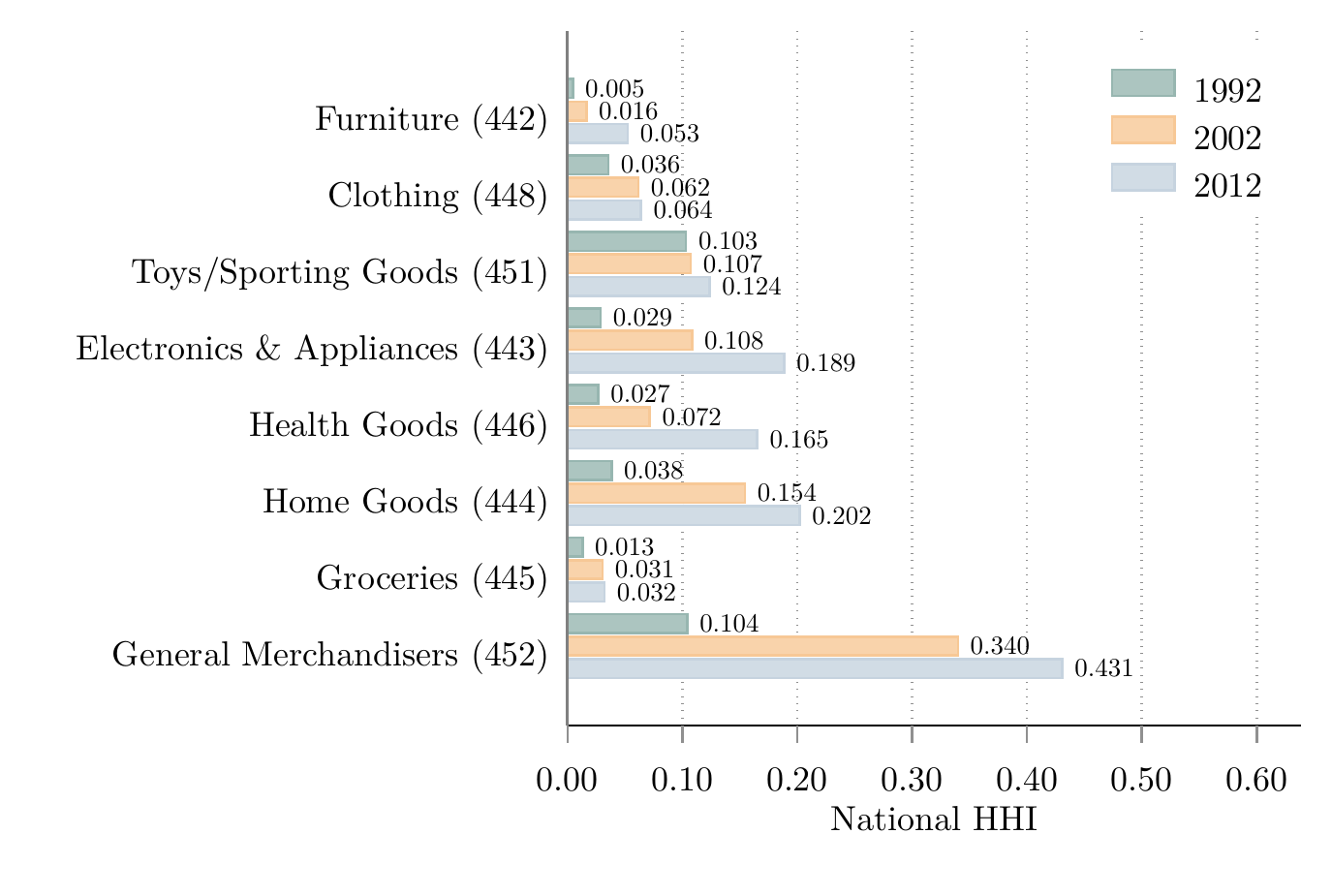}
        \vspace{-0.52cm}
    \end{subfigure}%
    
    \begin{subfigure}[th!]{1\textwidth}
        \caption{Local Concentration}\label{fig:local_ind}
        \vspace{-0.24cm}
        \includegraphics[scale=1.05]{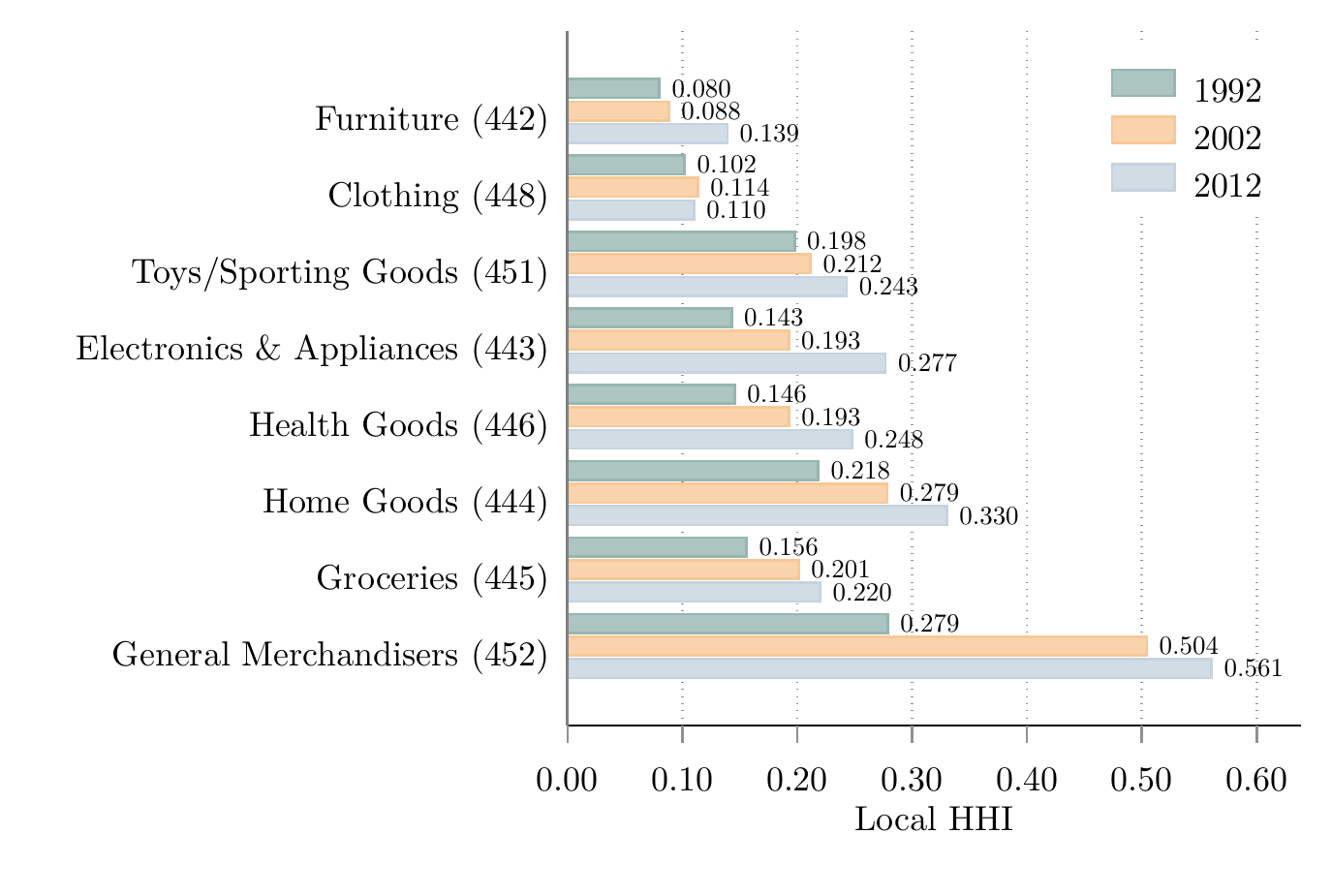}
        \vspace{-0.30cm}
    \end{subfigure}%
    \vspace{-0.45cm}
    
    \caption*{\footnotesize \textit{Notes:} The data are from the Census of Retail Trade. Numbers are the national and local (commuting zone) Herfindahl-Hirschman Index (HHI) for various industries weighted by market size. Concentration is calculated using 6-digit NAICS codes and aggregated to the 3-digit NAICS using each industry's share of sales. (CBDRB-FY20-P1975-R8604)}
    \label{fig:nat_local_ind}
    \end{center}
\end{figure}

This subsection provides more details on our industry based results.
Figure \ref{fig:nat_local_ind} shows national and local concentration for eight retail subsectors (3-digit NAICS). 
Local concentration is defined at the commuting zone level. 
The increasing trends we documented for national concentration in the retail sector are present in all subsectors, but the increase is particularly strong for general merchandisers (NAICS 452) at both the national and at the local level. 
The general merchandise subsector includes department stores, discount general merchandisers, and supercenters. 
Over time a small number of firms have come to dominate this format. 
Similar patterns arise in local concentration. Figure \ref{fig:local_ind} shows local concentration for the major subsectors, calculated as a weighted average of the industries comprising each subsector. 
Local industry concentration levels are higher than national and they also increase.

\newpage
\FloatBarrier
\subsection{The Relationship Between National and Local HHIs}\label{app:counterfactuals}

This subsection presents the levels of the national HHI in the counterfactual economy described in Section \ref{sec:decomposition}.
Table \ref{tab:counterfactual_hhi} presents the underlying numbers for Figure \ref{fig:counterfactual_hhi}.

\begin{table}[htb!]
    \caption{Multi-Market Firms Counterfactual}
    \begin{center}
    \begin{threeparttable}
    \begin{tabular}{l c ccccc}
        \hline \hline
        \Tstrut
        & & \multicolumn{5}{c}{National HHI} \\
         & & 1992 & 1997 & 2002 & 2007 & 2012 \\
        \hline
        Actual HHI & & 0.0128 & 0.0184 & 0.0311 & 0.0418 & 0.0426 \\
        Counterfactual HHI 1992  & & 0.0128 &  0.0135 & 0.0158 & 0.0171 & 0.0161 \\
        Counterfactual HHI 1997  & & --- & 0.0184 & 0.0232 & 0.0249 & 0.0239 \\
        Counterfactual HHI 2002  & & --- & --- & 0.0311 & 0.0354 & 0.0340 \\
        Counterfactual HHI 2007  & & --- & --- & --- & 0.0418 & 0.0404 \\
        \hline
    \end{tabular}
    \begin{tablenotes}
        \item {\footnotesize \textit{Notes:} 
        The numbers are based on calculations from the Census of Retail Trade.  
        The table reports the level of the national HHI and its counterfactual levels preserving the market structure of each Census year. 
        The counterfactual values are computed by preserving the rank of firms according to the initial market structure, assigning to the firm with the $r^{th}$ largest shares in the initial year the market share of the $r^{th}$ firm in each subsequent year.
        (CBDRB-FY23-P1975-R10585)
        }
    \end{tablenotes}
    \end{threeparttable}
    \end{center}
    \label{tab:counterfactual_hhi}
\end{table}

\renewcommand{\tabcolsep}{20pt}
\newpage
\FloatBarrier

\section{Comparison to Rossi-Hansberg, Sarte, and Trachter (2020)}\label{app:rh}

This section compares our results to those in \citet*{Rossi-Hansberg2018} (hereafter RST) for the retail sector and explains the factors contributing to the differences between our papers. Unlike us, they find a reduction in the local HHI for the retail sector between 1990 and 2014. RST present results for many sectors of the economy.  In what follows we discuss only their results in the retail sector.  However, our discussion of aggregation methods is relevant for all sectors.

There are three key differences between our paper and RST's that each partially explains the opposite results regarding local concentration.  First, we use different data sources: while RST use the National Establishment Time Series (NETS), this paper uses confidential data from the CRT and the LBD.  Second, we have different definitions of markets: this paper defines markets by product based on NAICS-6 classification of establishments, while RST define markets by industry based on SIC-8 or SIC-4 classification of establishments. Third, we differ in the methodology used to aggregate markets. This paper aggregates market-level concentration using contemporaneous weights, and we report the change in this (aggregate) index of local concentration. In contrast,  RST aggregate the change in market-level concentration using end-of-period weights and report this (aggregate) change. 

We argue that the CRT is likely to provide better data for the study of concentration in local markets, and we show that changing from NETS to CRT data alone explains a third of the discrepancy in the change of local concentration (while controlling for market definition and aggregation methodology). Another third of the difference in estimates is explained by the definition of product markets (by changing detailed SIC-8 industries to more aggregated SIC-4 industries). The proper definition of a product market (SIC-8, SIC-4, NAICS-6, product category) can depend on the question being asked. We argue in Section \ref{sec:Department-Level_Revenue} that product categories are the proper way to study retail markets. The final third of the difference in estimates is explained by the aggregation methodology. We argue that the method used by RST is biased toward finding decreasing local concentration, and we show that their method could find evidence of decreasing concentration in a time series, even when concentration is not changing in the cross-section. This occurs when markets become less concentrated as they grow.
Below we expand upon these differences and their implications for the measurement of local concentration.

\paragraph{Data sources} The baseline results in RST are based on the NETS, a data product from Walls and Associates that contains information on industry, employment, and sales by establishments. 
These data have been shown to match county-level employment counts relatively closely \citep*{Barnatchez2017}, but the data do not match the dynamics of businesses \cite{Crane2020}.  
The results in this paper are based on the CRT, a data set assembled and maintained by the U.S. Census Bureau covering all employer retail establishments. 

Both the NETS data and the CRT use the establishment's reported industry and sales when available and both have some degree of imputation for establishments that do not report. 
However, the CRT can often impute using administrative records from the IRS.\footnote{Response to the CRT is required by law.  Single-unit establishments are randomly sampled for sales in the CRT, while the non-sampled units have their sales imputed. See \url{http://dominic-smith.com/data/CRT/crt_sample.html} for more details.} 
Beyond this, the two data sets differ in other two relevant aspects. First, the CRT contains sales by product category for the majority of sales, while the NETS contains only industry, allowing us to define markets by product categories and account for cross-industry competition by general merchandisers (see Section \ref{sec:Department-Level_Revenue}). 
Second, the NETS includes non-employer establishments, while the CRT does not.  According to official estimates, non-employer establishments account for about 2 percent of retail sales in 2012 (Economy-Wide Key Statistics: 2012 Economic Census of the United States).\footnote{\url{https://factfinder.census.gov/faces/tableservices/jsf/pages/productview.xhtml?pid=ECN_2012_US_00A1&prodType=table}} 
On the whole, the CRT provides a more accurate picture of activity in the retail sector.

\paragraph{Definition of product markets} 
We adopt a different definition than RST for what constitutes a product market. Each definition of product market has its own pros and cons, and researchers may choose one over the other depending on the specific context. 
We define markets by a combination of a geographical location and a product category that we construct using the detailed data on sales provided by the CRT, along with the (NAICS-6) industry classification of establishments (see Section \ref{sec:Department-Level_Revenue}). 
As we mentioned above, doing this treats multi-product retailers as separate firms, ignoring economies of scope, in favor of putting all sales in a product category in the same market. 
However, we also present results defining markets by industry and find that the same patterns of higher national and local concentration arise, but with stronger magnitudes. See Section \ref{sec:industry}.

In contrast, RST define markets by the establishment's industry, using both SIC-8 and SIC-4 codes. Some examples of SIC-8 codes are department stores, discount (53119901); eggs and poultry (54999902); and Thai restaurants (58120115).\footnote{NETS allows for 914 retail SIC-8 codes. A full list is available at \url{https://www.dnb.com/content/dam/english/dnb-solutions/sales-and-marketing/sic_8_digit_codes.xls}. 
RST indicate that many SIC-8 codes are rarely used (data appendix), but without access to the NETS data, we cannot assess the relative significance of each code for economic activity.}  
SIC-8 codes may be overly detailed for retail product markets, to the point that many retailers will sell multiple types of goods.  For example, calculating concentration in eggs and poultry (54999902) would miss the fact that many eggs and poultry are sold by chain grocery stores (54119904) and discount department stores (53119901).  This suggests that aggregating to less detailed codes may provide a better definition of product markets.  
To that end, RST present results for SIC-4 codes.  
When concentration is calculated using SIC-4 codes, the decrease in local concentration is much smaller, a 8 percentage point decrease instead of a 17 percentage point decrease.\footnote{The change from SIC-8 to SIC-4 has little effect on concentration outside of retail (RST Data Appendix). The numbers are read off graphs for the change in retail sector concentration for zip codes between 1990 and 2012.}

Incidentally, the SIC-4 codes are quite similar to the NAICS-6 codes available in the CRT, except restaurants are included in the SIC definition of retail but not in NAICS.\footnote{In the results in the main text, we exclude automotive dealers, gas stations, and non-store retailers because of concerns related to ownership data and defining which markets they serve (see Section \ref{sec:data} for further discussion).  
This has little impact on the estimates for local concentration.} 
This makes the concentration measures based on each classification more closely comparable. Yet, even in this setting (NETS SIC-4 versus CRT NAICS-6) there are still significant differences between our studies. We will go back to this comparison when we discuss Figure \ref{fig:rst_comparison} and Table \ref{tab:rst} below.

\paragraph{Aggregation methodology} 
The final difference comes from how we aggregate the market-level changes in concentration into an aggregate index of local concentration.
We compute the local HHI index by first computing the HHI for each pair of product category $(j)$ and location $(\ell)$. 
Then we aggregate across locations, weighting each market (location-product) HHI by the market's share of the product's national sales.
Doing this provides a measure of the average local HHI for each product. 
Finally, we aggregate across products, weighting by the product's share of national retail sales, to obtain an average local HHI.
Every step in the aggregation maintains the interpretation of the HHI as a probability, which also makes the levels of the HHI comparable across time.
We do this for each period $(t)$ and  report the time series for this index. 
The average local HHI is then given by
\begin{align} \label{eqn:agg_local_HHI}
    HHI_t = \underbrace{\sum_{j} s_{j}^{t}}_{Products} \underbrace{\sum_{\ell} s_{\ell}^{jt}}_{Locations} \cdot HHI_{j \ell t}, \quad \text{where } HHI_{j \ell t}=\sum_{i}\left(s_i^{j \ell t}\right)^2.
\end{align}

RST use a different methodology. Instead of computing concentration in the cross-section, they calculate the change in concentration between $t$ and some initial period and then aggregate these changes weighting by the period \textit{$t$} share of employment of each industry $(j)$ in total retail employment.
Their index for the change in concentration is given by\footnote{Equation \ref{eqn:rst} is taken from RST, with notation adjusted to match the notation in this paper.}
\begin{align} \label{eqn:rst}
    \Delta HHI_t^{RST} = \sum_{j\ell} s_{j\ell}^{t} \Delta HHI_{j \ell t},
\end{align}
where $s_{j\ell}^t$ is the sales share of industry $j$ and location $\ell$ in the country at time $t$\footnote{RST weight markets by their employment share $\left(e_{j\ell}^t\right)$ instead of their sales share $\left(s_i^{j \ell t}\right)$. However, their data appendix shows  this has no effect on the results.} and $\Delta HHI_{j \ell t}$ is the change in the revenue-based HHI in industry $j$ and location $\ell$ between the base period and time $t$.  

The key difference between the methodologies is that RST do not account for the size of a market in the initial period. 
This is shown in equation \ref{eqn:rst_diff}, which subtracts the two measures of concentration from each other.  
After canceling terms, the difference between the two measures is 
\begin{align}
    \Delta HHI -\Delta HHI^{RST} &=  \sum_{j\ell} \underbrace{(s_{j\ell}^{t} - s_{j\ell}^{0})}_{\Delta s_{j\ell}^t}\cdot HHI_{m \ell 0}. \label{eqn:rst_diff}
\end{align}

RST will weight markets that increase in size over time by more in the initial period, while those that decrease will be weighted less relative to our measure.  
As markets grow, they typically become less concentrated resulting in RST weighting markets with decreasing concentration more than markets with increasing concentration.\footnote{A similar point is made in Appendix E of \citet{Ganapati2018a} using LBD data.}

Figure \ref{fig:rst_weights} shows that this methodology can find decreasing concentration in a time series, even when concentration is not changing in the cross-section. 
Consider three firms (A, B, and C) that operate in two markets and have the same size. 
In the first period $(t-1)$, firms A and B operate in market 1 and firm C operates in market 2. Consequently, the HHI is 0.5 and 1 for each market, respectively, and the aggregate (cross-sectional) HHI is $\nicefrac{2}{3}$. 
In period $t$, market 1 shrinks and market 2 grows, with firm B changing markets. 
This change does not affect the cross-sectional distribution of local (market-specific) concentration, but it does imply an increase in concentration in market 1 and a decrease in market 2. 
Despite there being no changes in the cross-sectional HHI, RST's methodology would report a decrease in local concentration $\left(\Delta HHI=\nicefrac{-1}{6}\right)$, driven by the decrease in market 2's HHI (which happens to be the largest market in period $t$).

\begin{figure}[htb!] 
    \centering 
    \caption{Example of RST Methodology}\label{fig:rst_weights}
    \begin{tikzpicture}
            \draw[line width=0.5mm,blue] (1,6) circle (2cm);
            \draw[line width=0.5mm,red] (1,1) circle (1.4142cm);
        
            \node[line width=0.5mm] at (1,9.5) {\textbf{Period t-1}};
            \node[line width=0.5mm] at (1,8.5) {\textbf{Market 1 - HHI=1/2}};
            \node[line width=0.5mm] at (1,3.5) {\textbf{Market 2 - HHI=1.0}};
            \draw[line width=0.5mm,blue] (-1,6) -- (3,6);
            \node at (1,7) {Firm A};
            \node at (1,5) {Firm B};
            \node at (1,1) {Firm C};
            \draw[line width=0.5mm,blue] (7.5,6) circle (1.4142cm);
            \draw[line width=0.5mm,red] (7.5,1) circle (2cm);
            
            \node[line width=0.5mm] at (7.5,9.5) {\textbf{Period t}};
            \node[line width=0.5mm] at (7.5,8.5) {\textbf{Market 1 - HHI=1.0}};
            \node[line width=0.5mm] at (7.5,3.5) {\textbf{Market 2 - HHI=1/2}};
            \node at (7.5,6) {Firm A};
            \draw[line width=0.5mm,red] (5.5,1) -- (9.5,1);
            \node at (7.5,2) {Firm B};
            \node at (7.5,0) {Firm C};
            \draw[->] (3.2,6) -- (5.8,6) node(x)[midway,above]{$\Delta HHI=1/2$} ;
            \draw[->] (2.7,1) -- (5.3,1) node(x)[midway,above]{$\Delta HHI=-1/2$};
            \draw[line width=0.2mm] (-1.5,-1.5) -- (3.5,-1.5);
            \draw[line width=0.2mm] ( 5.0,-1.5) -- (10.0,-1.5);
            \draw[line width=0.2mm] ( 1.0,-2.5) -- (7.5,-2.5);
            \node[line width=0.5mm] at (1.0,-2) {\textbf{Cross-Section HHI=2/3}};
            \node[line width=0.5mm] at (7.5,-2) {\textbf{Cross-Section HHI=2/3}};
            \node[line width=0.5mm] at (4.25,-3) {\textbf{RST Weighted} $\boldsymbol{\Delta}$\textbf{HHI=-1/6}};
            
    \end{tikzpicture}
    \vspace{0.2cm}
    \caption*{\footnotesize \textit{Notes:} The figure shows how market and cross-sectional concentration indices are computed under our methodology (difference in cross-section Herfindahl-Hirschman Index (HHI)) and that of \cite{Rossi-Hansberg2018}. The economy has two markets and three firms. Firms are of the same size. Markets change size from period $t-1$ to period $t$, but the cross-sectional distribution of markets and concentration does not change. The weighting methodology used by \cite{Rossi-Hansberg2018} puts more weight on market 2, which increases size between $t-1$ and $t$ and has a reduction in concentration. The result is a decrease in aggregate concentration when changes are measured according to this methodology, while cross-section HHI does not change.}
\end{figure}

\paragraph{Quantifying differences} 
Figure \ref{fig:rst_comparison} quantifies the role of each of the differences highlighted above for the change in local concentration between 1992 and 2012.\footnote{RST use 1990 as the base year instead of 1992.  This is unlikely to matter as RST find small changes in concentration between 1990 and 1992.} 
To make the comparison clear, we define markets by industry throughout the exercise.\footnote{To be precise, we define a market either by an SIC-8, an SIC-4, or a NAICS-6 industry in a given location. Our preferred definition of markets by product categories implies a change in the level of the HHI that makes the comparison with the results in RST less transparent.} 
Overall, Figure \ref{fig:rst_comparison} shows that the difference in the estimated change  of local HHI is explained in roughly equal parts by the three differences highlighted above: data source (CRT versus NETS), industry definition (NAICS-6 versus SIC-8), and aggregation methodology.
We discuss each step in more detail below.

\begin{figure}[tb]
    \centering
    \caption{RST Comparison}
    \includegraphics[scale=0.9]{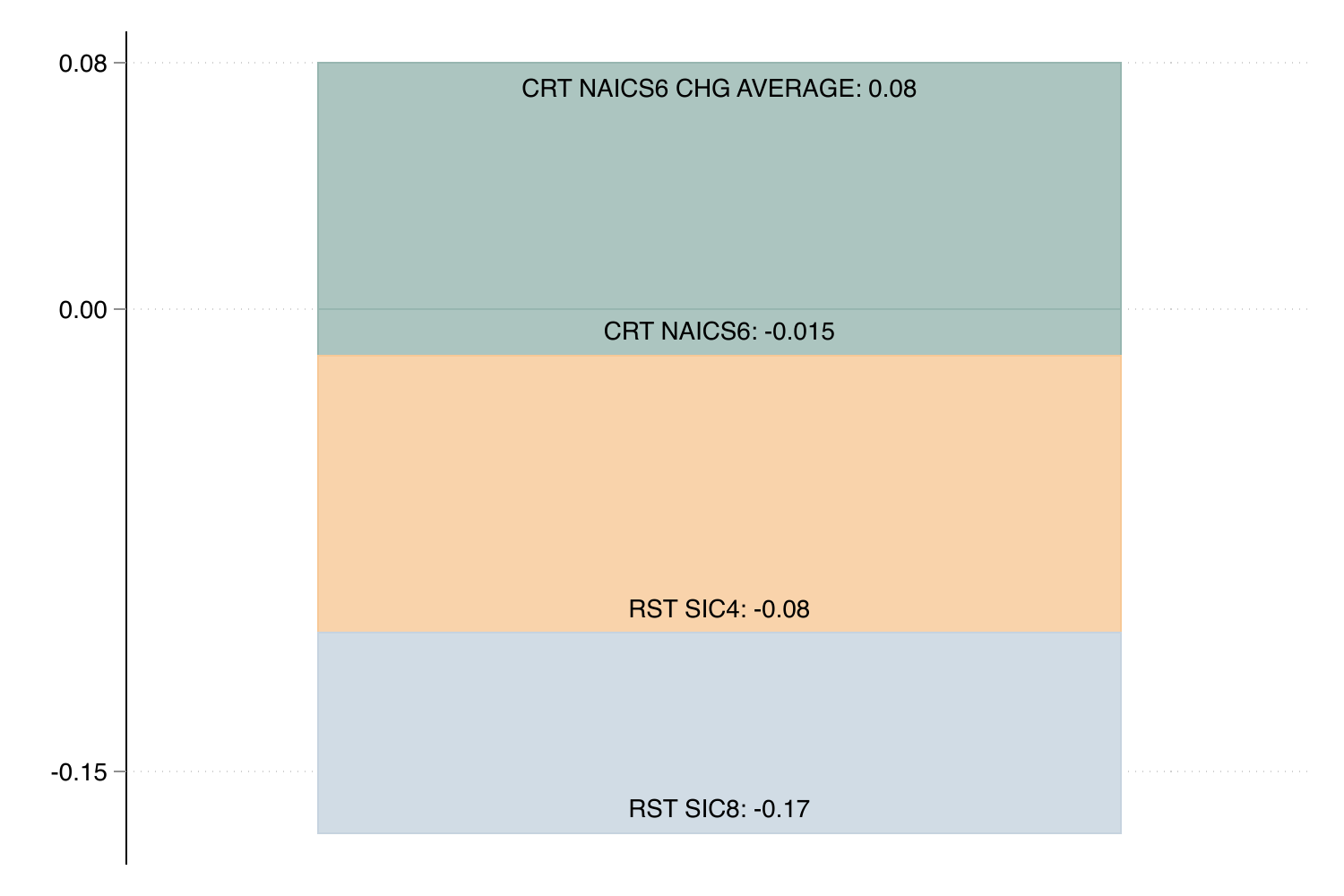}
    \label{fig:rst_comparison}
        \vspace{0.2cm}
    \caption*{\footnotesize \textit{Notes:} The figure shows various estimates for the change in local HHI between 1992 and 2012. The estimates vary according to the data source, industry definition, and aggregation methodology. The lowest estimate corresponds to \cite{Rossi-Hansberg2018}'s estimate using SIC-8 industries, and the second lowest estimate corresponds to using SIC-4 industries. The second highest estimate corresponds to using Census of Retail Trade  microdata and  NAICS-6 industries (which are similar to SIC-4 industries), and the highest estimate computes indices under our aggregation methodology instead of that of \cite{Rossi-Hansberg2018}. (CBDRB-FY20-P1975-R8604)}
\end{figure}

The lowest estimate for the change in local concentration (a decrease of 0.17 points in local HHI) corresponds to RST's baseline estimate using NETS data and SIC-8 for industry classification. 
Once industries are aggregated to the SIC-4 level (to improve comparability across establishments), the estimate increases by 9 percentage points, still implying a reduction of 8 percentage points in the local HHI.
The next estimate reproduces RST's methodology using microdata from the CRT. 
Changing from NETS to CRT data implies a further increase in the estimate of 6.5 percentage points, with the overall change suggesting a minor decease of local HHI of 1.5 percentage points.\footnote{Part of this difference could be explained in theory by the inclusion of restaurants in SIC-4; however, the industry by industry results in RST's Figure 7 suggest that this is not the case because they find diverging trends in most retail industries.}
Next we change the weighting methodology to ours (as explained above). 
Doing so increases the estimated change of local concentration again (by 9.5 percentage points), implying an overall increase of local HHI of 8 percentage points.\footnote{These numbers use all retail firms, including those that were dropped for the main sample in the paper. Concentration numbers are calculated for zip codes and aggregated according to each zip code's share of employment.}

Table \ref{tab:rst} provides a more detailed account of the estimates presented in Figure \ref{fig:rst_comparison} and also includes estimates of changes in local concentration for intermediate census years (1997, 2002, and 2007). 
In the first panel, national concentration, we compare the numbers in RST (Figure 1b) to numbers calculated for NAICS-based measures (including all 6-digit industries in NAICS) and product-based measures.  
In all three cases, national concentration is increasing significantly. 
Despite differences in the initial levels of concentration (column 1), the national HHI increases by two to three times.\footnote{The level of concentration is not provided in RST.}

The second panel of Table \ref{tab:rst} compares concentration measured at the zip code level using RST's weighting methodology as described above. 
We also provide results for the set of establishments that are included in the product-based results in the paper. 
Using their methodology, we find evidence for slight decreases in local concentration of 1 to 2 percentage points whether markets are aggregated using sales or employment weights.
These decreases are much less severe than the 17 percentage point decrease in RST.

The final panel of Table \ref{tab:rst} compares concentration measured at the zip code level using our aggregation method.  
This method finds significant increases in local concentration across both NAICS samples. Local HHI increased between 7.1 and 8.5 percentage points; that is, the average dollar in 2012 is spent in a more concentrated market than the average dollar in 1992.

\renewcommand{\tabcolsep}{12pt}
\begin{table}[htb]
\caption{Comparison of Concentration to RST}
\begin{center}
\begin{threeparttable}
\begin{tabular}{ll  lllll}
\hline
\hline 
\multicolumn{6}{c}{National Concentration \Tstrut}  \\
 
\hline
\Tstrut
& \multirow{2}{*}{Weight} & Level & \multicolumn{4}{c}{Change from 1992} \\
  &  & 1992 & 1997 & 2002 & 2007 & 2012 \\
  \cline{2-7} \Tstrut
RST & Emp.  & N/A & 0.020 & 0.030 & 0.050 & 0.055  \\
NAICS Based & Sales & 0.029 & 0.017 & 0.056 & 0.076 & 0.087 \\
Product Based & Sales & 0.013 & 0.006 & 0.018 & 0.029 & 0.030 \\
 \hline 
 \multicolumn{7}{c}{Zip Code Concentration: End-of-Period Weights \Tstrut}  \\
\hline
\Tstrut
& Weight & Level & \multicolumn{4}{c}{Change from 1992} \\
\cline{2-7} \Tstrut
RST & Emp. &  & -0.070 & -0.100 & -0.140 & -0.170 \\
\multirow{2}{*}{NAICS Based} & Emp. & & -0.022 & -0.016 & -0.019 & -0.015  \\
                             & Sales & N/A & -0.023 & -0.015 & -0.017 & -0.011 \\
\multirow{2}{*}{Paper Sample} & Emp. & & -0.020 & -0.013 & -0.018 & -0.017  \\
                             & Sales & & -0.024 & -0.009 & -0.013 & -0.011 \\
Product based & N/A &  & N/A & N/A & N/A & N/A \\
\hline 
 \multicolumn{7}{c}{Zip Code Concentration: Current Period Weights \Tstrut}  \\
\hline \Tstrut
& Weight & Level & \multicolumn{4}{c}{Change from 1992} \\
\cline{2-7} 
RST & N/A & N/A & N/A & N/A & N/A & N/A  \\
\multirow{2}{*}{NAICS Based} & Emp.  & 0.507 & 0.025 & 0.060 & 0.068 & 0.080  \\
                             & Sales & 0.498 & 0.018 & 0.052 & 0.062 & 0.071 \\
\multirow{2}{*}{Paper Sample} & Emp. & 0.524  & 0.029 & 0.069 & 0.075 & 0.083  \\
                             & Sales & 0.530 & 0.022 & 0.073 & 0.081 & 0.085 \\
Product Based & Sales & 0.2637 & 0.013 & 0.024 & 0.022 & 0.013 \\
\hline 
\end{tabular}
\begin{tablenotes}
\item {\footnotesize\textit{Notes:} The numbers come from the Census of Retail Trade and  \citet{Rossi-Hansberg2018} (RST).  Numbers from RST are taken from retail series in  Figure 2.   The level column contains the 1992 level of concentration. The formula for changes in concentration using end-of-period weights does not depend on the initial 1992 level as shown in RST, and consequently the level column does not apply to these calculations. NAICS-based measures concentration calculated including all NAICS industries.  Paper sample uses only establishments included in the sample for the product-based results.  Retail in RST is defined using SIC codes that include restaurants. (CBDRB-FY19-P1179-R7207, CBDRB-FY20-P1975-R8604)}
\end{tablenotes}
\end{threeparttable}
\end{center}
\label{tab:rst}
\end{table}
\note{We also have the data for 1982 and 1987 for this table. We also have the data by commuting zone fro NAICS and Select NAICS.}

\setlength{\abovedisplayskip}{5pt}
\setlength{\belowdisplayskip}{5pt}
\newpage
\section{Concentration and markups}\label{app: HHI_Markups_General}

In this appendix, we reproduce standard results relating the HHI with firms' profitability when firms compete à la Cournot. 
We then extend those results to competition with differentiated goods and discuss aggregation across markets.
Finally, we present implied changes in retail markups using the relationship between them and the HHI.

\paragraph{Cournot competition with a homogeneous good.}
Consider a market with $N$ producers of a homogeneous good competing à la Cournot.
Firms are heterogeneous in their marginal cost, $c_i$. 
The (inverse) demand for the good is captured by a price function $P\left(Q\right)$, where $Q\equiv\sum_i q_i$ is the total amount of the good being produced. 
The problem for each producer is
\begin{align}
    \max_{q_i} P\left(Q\right)q_i - c_i q_i
\end{align}
The producer's optimal choice is characterized by a markup over marginal cost rule,
\begin{align}
    P\left(Q\right) = \left[1 - \frac{s_i}{\varepsilon}\right]^{-1} c_i,
\end{align}
where $\varepsilon^{-1}\equiv-\nicefrac{Q}{P}\nicefrac{\partial P}{\partial Q}$ is the elasticity of demand and $ s_i \equiv \nicefrac{P q_i}{P Q}$ the producer's market share.

The producer's rule implies a relationship between aggregate markups (or gross margins) and the HHI.
To see this, consider the market's aggregate profits:
\begin{align}
    \Pi & \equiv \sum_{i=1}^{N} \left( P\left(Q\right)q_i - c_i q_i\right) \nonumber \\
        & = P Q  \sum_{i=1}^{N} \left( s_i + \left[ 1 - \frac{s_i}{\varepsilon}\right] s_i \right) \nonumber \\
        & = P Q \frac{\text{HHI}}{\varepsilon}
\end{align}
Then we can get the expression linking markups to the HHI.
The markups or gross-margins are defined as the ratio of total (variable) cost to revenue,  $\mu\equiv\nicefrac{\sum_i c_i q_i}{PQ}$.
\begin{align}
    P Q - \sum_{i=1}^{N} c_i q_i  & = P Q \frac{\text{HHI}}{\varepsilon} \nonumber \\
    \mu & = \left[ 1 - \frac{\text{HHI}}{\varepsilon} \right]^{-1}. 
\end{align}

\paragraph{Cournot competition with differentiated goods.}
Consider a market with $N$ producers of a differentiated goods competing à la Cournot.
Firms are heterogeneous in their marginal cost, $c_i$. 
The demand for the goods comes from homothetic preferences described by a homogeneous-of-degree-1 function, so that we write the aggregate quantity as $Q=F\left(q_1,\ldots,q_N\right)$.
The demand for an individual good $\left(q_i\right)$ is characterized by $\nicefrac{p_i}{P}=F_i\left(\nicefrac{q_i}{Q}\right)$, where $p_i$ is good $i$'s price, $P$ is the ideal price index (i.e., $PQ=\sum_i p_i q_i$), and $F_i\left(\cdot\right)$ is the $i^{th}$ partial derivative of $F$.

The problem for each producer is
\begin{align}
    \max_{q_i} p_i\left(q_i,Q,P\right)q_i - c_i q_i
\end{align}
The producer's optimal choice is characterized by
\begin{align}
    \left[ \frac{\partial p_i}{\partial q_i} + \frac{\partial p_i}{\partial Q} \frac{\partial Q}{\partial q_i} +\frac{\partial p_i}{\partial P} \frac{\partial P}{\partial Q} \frac{\partial Q}{\partial q_i} \right]q_i + p_i - c_i & = 0  \nonumber\\
    p_i \left[ \left(\frac{q_i}{p_i}\frac{\partial p_i}{\partial q_i}\right)  +  \left(\frac{Q}{p_i}\frac{\partial p_i}{\partial Q}\right)  \left(\frac{Q}{q_i}\frac{\partial Q}{\partial q_i}\right) + \left(\frac{P}{p_i}\frac{\partial p_i}{\partial P}\right) \left(\frac{Q}{P}\frac{\partial P}{\partial Q}\right) \left(\frac{q_i}{Q}\frac{\partial Q}{\partial q_i}\right) + 1\right]  & = c_i . 
\end{align}
In the above expressions we assume that the producer ignores second order effects of their choices on the price through the change in other firm's prices $\left(\nicefrac{\partial p_i}{\partial P} \nicefrac{\partial P}{\partial p_j} \nicefrac{\partial p_j}{\partial q_i} \right)$.
This effect is zero when the aggregate quantity is a CES aggregator. 

We can express the optimal pricing rule in terms of the elasticity of demand  $\left(\varepsilon_i^{-1}\equiv-\nicefrac{q_i}{p_i}\nicefrac{\partial p_i}{\partial q_i} \right)$ and the producer's market share $\left( s_i \equiv \nicefrac{p_i q_i}{P Q} \right)$.
To do this we first establish two results from the producer's demand,$\nicefrac{p_i}{P}=F_i\left(\nicefrac{q_i}{Q}\right)$:
\begin{align}
    \frac{Q}{p_i}\frac{\partial p_i}{\partial Q} = - \frac{q_i}{p_i}\frac{\partial p_i}{\partial q_i} = \frac{1}{\varepsilon_i} \quad \& \quad \frac{P}{p_i}\frac{\partial p_i}{\partial P} = 1. 
\end{align}
With these results we can simplify the expression to
\begin{align}
    p_i \left[ \frac{\varepsilon_i-1}{\varepsilon_i}  + \left(  \frac{1}{\varepsilon_i} + \frac{Q}{P}\frac{\partial P}{\partial Q}\right) \left(\frac{q_i}{Q}\frac{\partial Q}{\partial q_i}\right) \right]  & = c_i. 
\end{align}
Finally, we assume that the expenditure in the market is fixed so that $\nicefrac{Q}{P}\nicefrac{\partial P}{\partial Q}=1$ and we derive a final result relating the elasticity of $Q$ with the producer's market share. 
We start from the definition of $Q$ and consider the total differential, focusing on $dq_i\neq0$,
\begin{align}
    1&=F\left(\frac{q_1}{Q},\ldots,\frac{q_N}{Q}\right) \nonumber\\
    0&=F_i\left(\frac{q_i}{Q}\right)\frac{dq_i}{Q}-\sum_{j=1}^{N}F_j\left(\frac{q_j}{Q}\right)\frac{q_j}{Q^2}dQ\nonumber\\
    0&=F_i\left(\frac{q_i}{Q}\right)dq_i-\left(\sum_{j=1}^{N}F_j\left(\frac{q_j}{Q}\right)\frac{q_j}{Q}\right)dQ\nonumber\\
    \frac{dQ}{dq_i}&=\frac{p_i}{P},
\end{align}
where the sum in the second to last line is equal to 1 because of Euler's theorem and we replace for the relative price form the producer's demand.
With this result in hand, it follows that $\nicefrac{q_i}{Q}\nicefrac{\partial Q}{\partial q_i}=s_i$.
Replacing in the producer's optimal pricing rule we obtain
\begin{align}
    p_i  & = \frac{\varepsilon_i}{\varepsilon_i-1} \left[ 1  - s_i \right]^{-1} c_i. \label{eq: Optimal_Price_Cournot_Differentiated}
\end{align}
This is the same result as in \citet{Atkeson2008Pricing-to-MarketPrices} and \citet{Grassi2017} under CES preferences, when all firms face the same (constant) elasticity of demand $(\varepsilon_i=\varepsilon)$.

We can aggregate to a relationship between profitability and the HHI at the market level as in the homogeneous good case when the elasticity is common across firms. 
\begin{align}
    \Pi \equiv \sum_{i=1}^{N} \left( p_i q_i - c_i q_i\right) %
        & = P Q  \sum_{i=1}^{N} \left( s_i - \frac{\varepsilon-1}{\varepsilon} \left[ 1  - s_i \right] s_i \right) \nonumber \\
        & = P Q \left[\frac{1}{\varepsilon}+\frac{\varepsilon-1}{\varepsilon}\text{HHI}\right].
\end{align}
Then we can get the expression linking markups to the HHI.
The markups or gross-margins are defined as the ratio of total (variable) cost to revenue,  $\mu\equiv\nicefrac{\sum_i c_i q_i}{PQ}$.
\begin{align}
    P Q - \sum_{i=1}^{N} c_i q_i  & = P Q \left[\frac{1}{\varepsilon}+\frac{\varepsilon-1}{\varepsilon}\text{HHI}\right] \nonumber \\
    \mu & = \frac{\varepsilon}{\varepsilon-1}\left[1-\text{HHI}\right]^{-1}. \label{eq: Agg_Markup_HHI_Cournot}
\end{align}

\paragraph{Aggregation across markets.}
We also compute the average markup across markets $(\ell)$ for a given product of industry $(j)$.  
We can compare this measure to gross margins by product or industry obtained from the ARTS. 
We define the average markup as the ratio between product $j's$ total sales and total labor costs of the product across markets $\left(\ell=1,\ldots,L\right)$, 
\begin{align}
    \mu_{j} & \equiv \frac{\sum_{\ell=1}^{L} P_{j}^{\ell}Q_{j}^{\ell}}{\sum_{\ell=1}^{L} C_{j}^{\ell}Q_{j}^{\ell}}  
    = \frac{\sum_{\ell=1}^{L} P_{j}^{\ell}Q_{j}^{\ell}}{\sum_{\ell=1}^{L} \frac{1}{\mu_{j}^{\ell}}P_{j}^{\ell}Q_{j}^{\ell}} 
    = \left[\sum_{\ell=1}^{L} \left(\mu_{j}^{\ell}\right)^{-1}\theta_{j}^{\ell}\right]^{-1},
\end{align}
where $\mu_{j}^{\ell}=\nicefrac{P_{j}^{\ell}}{C_{j}^{\ell}}$ is the average markup (gross margin) of product/industry $j$ in market $\ell$, with $C_{j}^{\ell}\equiv\nicefrac{\left(\sum_{i} c_i^{j,\ell} q_i^{j,\ell}\right)}{Q_{j}^{\ell}}$ the average cost in market $\ell$, and  $\theta_{j}^{\ell}\equiv\nicefrac{p_{j}^{\ell}y_{j}^{\ell}}{\sum_{\ell=1}^{L} p_{j}^{\ell}y_{j}^{\ell}}
$ is the share of product/industry $j$ sales accounted for by market $\ell$.  

Using the result in \eqref{eq: Agg_Markup_HHI_Cournot} we express the markup in terms of market concentration.
\begin{align}
\mu_{j}=\left[\sum_{\ell=1}^{L} \left(\frac{\epsilon_{j}^{\ell}}{\epsilon_{j}^{\ell}-1}\right)^{-1}\left[1-\text{HHI}_{j}^{\ell}\right]\theta_{j}^{\ell}\right]^{-1}
\end{align}
If the elasticity of demand of good $j$ is common across markets the expression simplifies to:
\begin{align}
    \mu_{j}=\frac{\epsilon_{j}}{\epsilon_{j}-1}\left[1-\text{HHI}_{j}\right]^{-1}, \label{eq: Agg_Markup_HHI_Cournot_Product}
\end{align}
where $\text{HHI}_{j}\equiv\sum_{\ell=1}^{L} \text{HHI}_{j}^{\ell}\theta_{j}^{\ell}$
is the sales weighted HHI of product $j$ across markets.
\clearpage
\section{Model of Product Competition in Retail}\label{app:Model}

In this Appendix, we present a model of the retail sector where retailers compete in local product markets.
Consumers have traditionally chosen between nearby stores selling a given product when purchasing goods. 
Accordingly, we focus on a setup where competition is exclusively local.
The model provides us with an explicit link between the local HHI and average retailers' margins for each product which in the model are captured as markups.

In Appendix \ref{app: HHI_Markups_General}, we showed that Cournot competition implies a relationship between markups and local HHIs. 
Now, we build upon this relationship and show how that equation fits into a fully specified model of oligopolistic competition in local product markets. 
We then estimate the key parameters of the model, namely elasticities of substitution for each product, to match the 1993 level of retailers' gross margins using data from the ARTS. 

We use the model to find the implied change in product margins (markups) coming from changes in local concentration.
We find that increases in local concentration imply a \Paste{markup_prod_increase} percentage point increase in markups between 1992 and 2012, roughly a third of the observed increase in gross margins in the ARTS during that period.
The results are similar in magnitude to the ones we obtained in the case of competition in homogeneous goods described in Appendix \ref{app: HHI_Markups_General}, see Table \ref{tab:Cournot_Margins}.

The model follows \citet{Grassi2017} and \citet{Atkeson2008Pricing-to-MarketPrices}.
The model economy has $L$ locations, in each of them there are $J$ products being transacted in local product markets. 
Each market has $N_{j\ell}$ retail firms that compete with one another in product $j$. 
Competition takes place at the location-product level. 
A perfectly competitive sector aggregates goods across firms for each product and location, aggregates products by location into location-specific retail goods, and aggregates each location's retail output into a final consumption good. 
A single representative consumer demands the final consumption good and supplies labor in each location.

\subsection{Technology}

A retailer $i$ selling product $j$ in location $\ell$ produces using a constant-returns-to-scale technology that combines labor $\left(n\right)$ and potentially other inputs $\left\{x_k\right\}_{k=1}^{K}$:
\begin{align}
    y_i^{j\ell} = z_i^{j\ell} F\left(x_1,\ldots,x_K,n_i^{j\ell}\right),
\end{align} 
where $z_i^{j\ell}$ represents the productivity of the retailer and $F$ is homogeneous of degree 1. 

The homogeneity of $F$ implies that the retailer has a constant marginal cost of production that we denote $c_i^{j\ell}$. 
Retailers differ in their marginal costs, reflecting productivity differences between small and large (multi-market) retailers, or differential pricing from suppliers.
because of differences in productivity and in the prices of the inputs they require for production. In this way the model captures the cost advantages associated with large multi-market retail firms which is the most relevant aspect of large firms when thinking about local competition.
Retailers maximize profits independently in each market:
\begin{align}
    \pi_i^{j\ell} = p_i^{j\ell} y_i^{j\ell} - c_i^{j\ell} y_i^{j\ell},
\end{align}

The demand faced by the individual retailer comes from the aggregation sector that serves the consumer. 
Aggregation takes place in three levels. 
First, a local aggregator firm that combines the output of the $N_{j\ell}$ retail firms selling product $j$ in location $\ell$. 
The firm operates competitively using the following technology: 
\begin{align}
    y_j^\ell = \left( \sum_{i=1}^{N_{j\ell}} \left( y_i^{j\ell} \right)^{\frac{\epsilon_j - 1}{\epsilon_j}}\right)^{\frac{\epsilon_j}{\epsilon_j -1}}; \qquad \epsilon_j > 1. \label{eqn:Y_j^m Aggregator}
\end{align}

Then, the combined product bundles, $y_j^\ell$, are themselves aggregated into local retail output, $y_\ell$, through the following technology:
\begin{align}
    y_\ell = \prod_{j=1}^J \left( y_j^\ell\right)^{\gamma_j^\ell}; \qquad \sum_{j=1}^J \gamma_j^\ell = 1,
\end{align}
where $\gamma_j^\ell$ is the share of product $j$ in retail sales in location $\ell$

Finally, the national retail output is created by combining local output, $y_\ell$, from the $L$ locations in the country: 
\begin{align}
    y = \prod_{\ell=1}^L \left(y_\ell\right)^{\beta_\ell}; \qquad \sum_{\ell=1}^L \beta_\ell =1,
\end{align}
where $\beta_\ell$ corresponds to the share of location $\ell$ in national retail sales.  

The aggregation process implies the following demand and prices:
\begin{align}
    y_{\ell}&=\beta_{\ell}\frac{P}{p_{\ell}}\cdot y \quad & P&=\prod_{\ell=1}^{L}\left(\frac{p_{\ell}}{\beta_{\ell}}\right)^{\beta_{\ell}} \label{eq:location_demand}\\    
    y_{j}^{\ell}&=\gamma_{j}^{\ell}\frac{p_{\ell}}{p_{j}^{\ell}}y_{\ell} \quad  & p_{\ell}&=\prod_{j=1}^{J}\left(\frac{p_{j}^{\ell}}{\gamma_{j}^{\ell}}\right)^{\gamma_{j}^{\ell}} \label{eq:product_demand}\\
    y_{i}^{j\ell}&=\left(\frac{p_{i}^{j\ell}}{p_{j}^{\ell}}\right)^{-\epsilon_j}y_{j}^{\ell} \quad  & p_{j}^{\ell}&=\left(\sum_{i=1}^{N}\left(p_{i}^{j\ell}\right)^{1-\epsilon_j}\right)^{\frac{1}{1-\epsilon_j}}
    \label{eq:firm_demand}
\end{align}

\subsection{Pricing to market and average markups}\label{app:competition}

Firms compete directly in the sales of each product in a given location. 
Firms compete à la Cournot, choosing the quantity $\left(y_{i}^{j\ell}\right)$ in a non-cooperative fashion, taking as given the choices of other firms.
Firms are aware of the effect of their choices $\left(p_{i}^{j\ell},y_{i}^{j\ell}\right)$ on the price and quantity of the product in the market they operate in $\left(p_{j}^{\ell},y_{j}^{\ell}\right)$.
The choice of quantity implies a pricing policy for the firms according to their residual demand.

The solution to the pricing problem is the same as in Appendix \ref{app: HHI_Markups_General} for the differentiated goods case. 
The optimal price satisfies equation \eqref{eq: Optimal_Price_Cournot_Differentiated} and depends on the firm's share of sales in their local product market, $s_i^{j\ell}$ and the elasticity of demand they face, $\epsilon_j$. 
\begin{align}
    \mu_{i}^{j\ell}=
        \frac{\epsilon_j}{\epsilon_j-1}\left[1-s_{i}^{j\ell}\right]^{-1} 
    \label{eqn:Markups_Prop}
\end{align}
In the model, retailers with lower costs can charge lower prices and increase their market share, making it so that high-markup retailers are low-cost. 

Average markups in a market satisfy equation \eqref{eq: Agg_Markup_HHI_Cournot} and average product markups (across markets) satisfy equation \eqref{eq: Agg_Markup_HHI_Cournot_Product} from Appendix \ref{app: HHI_Markups_General}.
\begin{align}
     \mu_j   &= \frac{\epsilon_j}{\epsilon_j-1} \left[1- \sum_{\ell=1}^{L} s_\ell^j \text{HHI}_j^\ell \right]^{-1},
     \label{eqn:prod_markup}
\end{align}
where $s_\ell^j$ is the share of market $\ell$ in product's $j$ national sales.
As markets become more concentrated, average markups increase. 
The sensitivity of markups to increases in concentration is larger for products with a lower elasticity of demand. 
Retail markups (averaging across products) are
\begin{align}
    \mu=\left[\sum_{j=1}^{J}\left(\mu_{j}\right)^{-1}s_{j}\right]^{-1},
\end{align}
where $s_j$ is the share of product $j$ in national retail sales.

\subsection{Estimation and Data}

The two key ingredients for analyzing markups are firms' market shares by product in each location, $s_i^{j\ell}$, and the elasticity of substitution for each product, $\epsilon_j$. 
We obtain the shares directly from the CRT and estimate the elasticities using equation \eqref{eqn:prod_markup}.  
Specifically, we use the product HHIs calculated in Section \ref{sec:Product_Concentration} and gross margins from the Annual Retail Trade Survey (ARTS) which are available beginning in 1993.

The ARTS provides the best source to compare our results to because it computes markups using cost of goods sold, which are the most direct data analogue to markups in the model. 
The ARTS samples firms with activity in retail, collecting data on sales and costs for each firm. 
The firm-level markups collected by ARTS represent an average markup across the products that the firm sells.
The information available is similar to information in Compustat, but the ARTS includes activity of non-public firms that account for a significant share of retail sales.

The ARTS also provides us with margins for detailed industries, which we convert to product margins using the CRT.\footnote{
    Appendix \ref{app:markups} presents additional results based on industry-level markups as well as robustness exercises with alternative measures of product markups.
    } 
We do this in three steps. 
First, we compute margins for the industries, $k(j)\in K$ most closely related to each of the eight major product categories, $j$.
We also compute margins for general merchandisers (NAICS 452).\footnote{
    For instance, we relate clothing to NAICS 448 and groceries to NAICS 445; see Appendix \ref{app:ind} for a complete list.
    } 
As before, the sales of each specialized industry are all assigned to its own product category, while the sales of general merchandisers are divided across products using the share of each product category's sales that come from general merchandisers in the CRT.
Second, we estimate a scaling factor $\lambda=0.82$ that measures how large general merchandise margins are relative to what would be implied by other industries' margins:
\begin{align}
    \mu_{GM}^{ARTS}=\lambda\sum_{j}\omega_j^{GM}\mu_{k(j)}^{ARTS},
\end{align}
where $\mu_{k(j)}^{ARTS}$ is the measured margins of industry $k(j)$ in the ARTS and $\omega_j^{GM}$ is the share of sales of product $j$ in general merchandising from the CRT.
We use the scaling factor $\lambda$ to construct product-specific margins for general merchandisers while being consistent with the measured margins from the ARTS. 
The margin of general merchandisers in product $j$ is then $\mu_{GM}^j=\lambda\mu_{k(j)}^{ARTS}$. 
Finally, we compute product-level margins in a model-consistent way as 
\begin{align}
    \mu_j=\left(\frac{1-\omega_{GM}^{j}}{\mu_{k(j)}^{ARTS}}+\frac{\omega_{GM}^{j}}{\mu_{GM}^{j}}\right)^{-1}%
    ,
\end{align}
where $\omega_{GM}^j$ is the share of general merchandisers in product $j$'s sales.
In this way, product-level margins incorporate the effect of competition from general merchandisers.

\subsection{Changes in Concentration and Markups}\label{app:Changes_Markups_Model}

We conduct two exercises with the model. 
First, we fit the model to match product markups in 1993 given the observed levels of local concentration in 1992, which provides us with estimates of the elasticities of substitution for each product category. 
Holding these estimates fixed, we can extend the model through 2012 and obtain the change in markups implied by the observed increase in local concentration. 
This exercise explains one-third of the increase in markups observed in the ARTS.
Second, we can fit the model to match observed markups for each economic census year by allowing the elasticities of substitution to be time varying.
To match the increase in markups, the model implies a decrease in the elasticity of substitution for most products.

\begin{figure}[tb!]
\begin{center}
    \caption{Local Concentration and Markups}
    \includegraphics[scale=1]{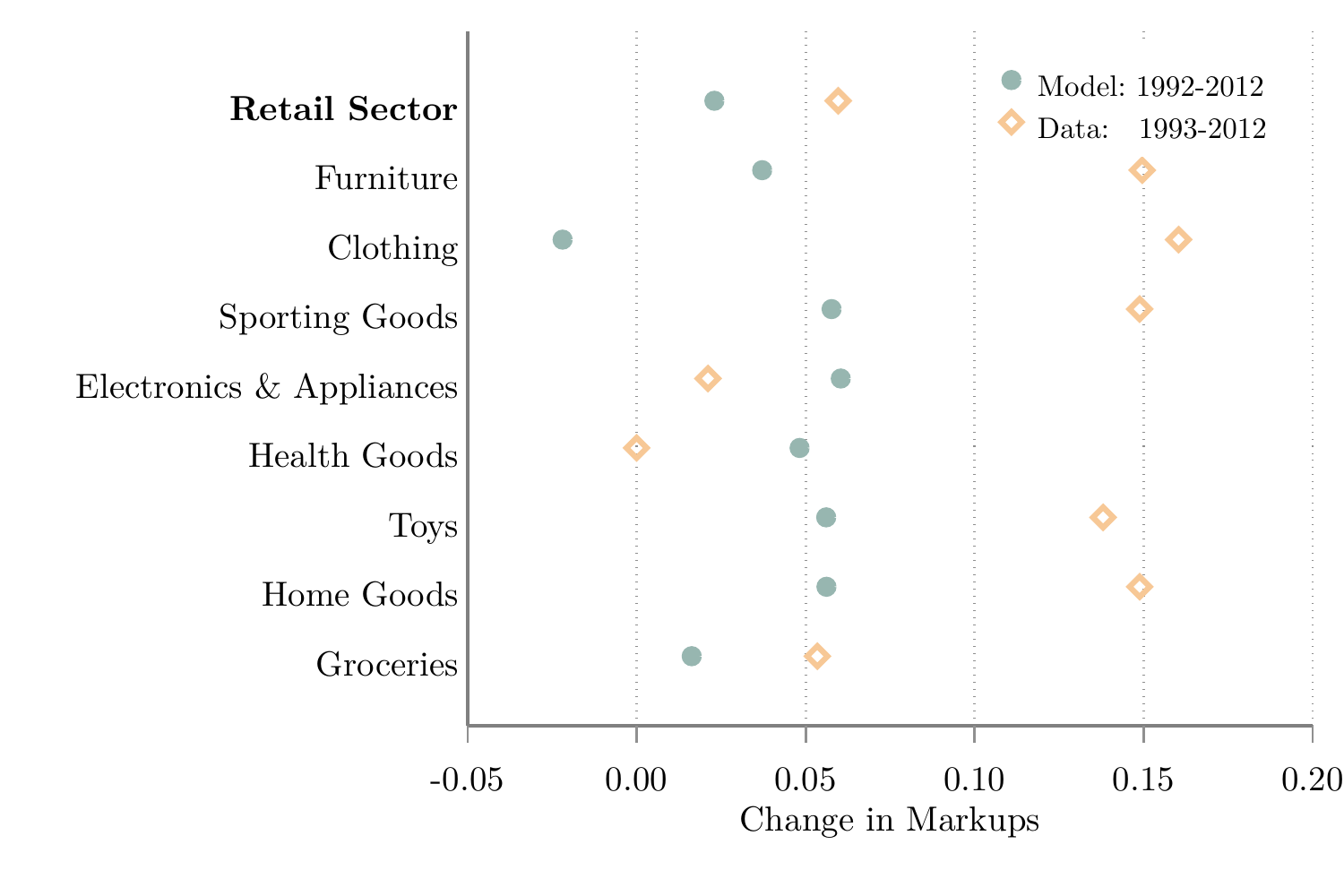}
    \caption*{\footnotesize \textit{Notes:} Diamonds mark the change in product markups between 1992 and 2012 from the Annual Retail Trade Survey and Census of Retail Trade data and a weighted average across products for the retail sector. Circles mark the change in markups implied by the change in local concentration given the model estimates for 1992. }
    \label{fig:model_aggregates_gm}
\end{center}
\end{figure}

The increase in local concentration implies an increase in retail markups of \Paste{markup_prod_increase} percentage points between 1992 and 2012, about one-third of the 6 percentage point increase in product-level markups. 
Figure \ref{fig:model_aggregates_gm} shows that in all but two product categories, the observed increase in markups is higher than what is implied by the rise in product-level HHI.
These results are robust to alternative measures of product markups (see Appendix \ref{app:markups}).
The changes in model markups in Figure \ref{fig:model_aggregates_gm} assume that the elasticity of demand faced by firms are constant over time and vary only because of changes in local HHI. 
However, many changes in the competitive environment of retail can be reflected in changes in these elasticities rather than changes in market concentration. 

\renewcommand{\tabcolsep}{10pt}
\begin{table}[t]
    \caption{Estimated Elasticities of Substitution}
    \begin{center}
    \begin{threeparttable}
    \begin{tabular}{l ccc }
        \hline \hline
        \Tstrut
        \multirow{2}{*}{Product Category} & \multicolumn{3}{c}{$\epsilon_j$} \\
         & 1992 & 2002 & 2012 \\
        \hline
        Furniture & 2.70 & 2.43 & 2.43 \\
        Clothing & 3.07 & 2.83 & 2.48 \\
        Sporting Goods & 3.73 & 3.77 & 3.20 \\
        Electronics \& Appliances & 4.48 & 5.74 & 4.95 \\
        Health Goods & 4.38 & 5.30 & 5.09 \\
        Toys & 5.55 & 5.91 & 4.91 \\
        Home Goods & 4.85 & 4.13 & 3.92 \\
        Groceries & 5.82 & 5.39 & 6.40 \\
        \hline
    \end{tabular}
    \begin{tablenotes}
        \item {\footnotesize \textit{Notes:} The data are authors' estimates of product elasticities of substitution using industry markups from the Annual Retail Trade Survey and product-level local Herfindahl-Hirschman Indexes calculated from the Census of Retail Trade. The elasticities are the solution to equation \eqref{eqn:prod_markup}.}
    \end{tablenotes}
    \end{threeparttable}
    \end{center}
    \label{tab:epsilon}
\end{table} 

Table \ref{tab:epsilon} shows the value of the elasticity of substitution needed to match the level of markups in each year.
We find the lowest elasticities of substitution in Clothing and Furniture. 
These are categories that feature many different brands only available from a small set of retail firms, leaving more room for differentiation than in products such as Toys and Groceries, where different firms carry similar or even identical physical products.

To match the observed increase in markups, most product categories require a decrease in their elasticity of substitution. 
The magnitude of the decrease depends on the initial level of the elasticities as markups respond more to changes for lower elasticities. The decreasing trend for the elasticities of substitution is consistent with the findings of \cite{Bornstein2018}, \cite{Brand2020}, and \cite{Neiman2019}, who link the decrease to the rise of store and brand loyalty/inertia. 
The exception to the trend of decreasing elasticities of substitution are Electronics \& Appliances and Health Goods, which instead require an increase in their elasticities. 
Health Goods had almost no change in markups in the data, but based on the change in concentration, markups should have increased by about 5 percentage points.

\subsection{Additional Markup Results}\label{app:markups}

We perform the same exercises as above using various assumptions regarding the behavior of markups. 
Our first set of results changes only the measure of markups we use, keeping the change in local concentration constant as measured by product-level local (commuting-zone) HHI. 
We find that the changes implied by the change in local product-level concentration on markups are robust to changes in the level of markups. 
The second set of results uses changes in industry-level concentration instead of product-level concentration.
We find that these changes imply larger increases in markups for the retail sector as a whole. 
This follows from the fact that industry-level measures of concentration ignore the competition between general merchandisers and other retailers.

\paragraph{Product-based results}
We consider four alternative measures of markups. 
Our baseline measure constructs product-level markups combining information from the ARTS and the CRT. 
Our second measure assigns to each product category the markup of its main NAICS industry without adjusting for the role of general merchandisers.
Our last two measures consider the possibility that markups are much lower or higher than we estimate, respectively decreasing markups by fifty percent or doubling the product-level markups we constructed in our baseline. 
Table \ref{tab:epsilon_robustness} reports the level of markups in 1992 under each of our alternative measures.

We estimate the implied elasticity of substitution for each product category using equation \eqref{eqn:prod_markup} from the model. 
The implied elasticities are reported in Table \ref{tab:epsilon_robustness}. 
The level of the elasticities varies to match the level of markups under each alternative specification but the general rank stays largely unchanged between the product- and industry-based exercises.

\renewcommand{\tabcolsep}{8pt}
\begin{table}[tb!]
    \caption{Markups Robustness: Estimated Elasticities of Substitution}
    \begin{center}
    \begin{threeparttable}

    \begin{tabular}{lSS|SS|SS|SS}
        \hline \hline
        \Tstrut
         & \multicolumn{2}{c|}{Product $\mu$} & \multicolumn{2}{c|}{Industry  $\mu$} & \multicolumn{2}{c|}{Low $\mu$} & \multicolumn{2}{c}{High  $\mu$}\tabularnewline
         & $\mu_{j}^{92}$ & $\epsilon_{j}$ & $\mu_{j}^{92}$ & $\epsilon_{j}$ & $\mu_{j}^{92}$ & $\epsilon_{j}$ & $\mu_{j}^{92}$ & $\epsilon_{j}$ \\
        \hline 
        {\small Furniture} & 1.67 & 2.7 & 1.73 & 2.5 & 1.33 & 4.7 & 2.33 & 1.8 \\
        {\small Clothing} & 1.55 & 3.1 & 1.69 & 2.6 & 1.28 & 5.6 & 2.10 & 2.0 \\
        {\small Sporting Goods} & 1.47 & 3.7 & 1.57 & 3.2 & 1.24 & 7.8 & 1.94 & 2.2 \\
        {\small Electronics \& Appliances} & 1.34 & 4.5 & 1.44 & 3.6 & 1.17 & 9.1 & 1.68 & 2.6 \\
        {\small Health Goods} & 1.38 & 4.4 & 1.44 & 3.8 & 1.19 & 9.6 & 1.77 & 2.5 \\
        Toys & 1.43 & 5.6 & 1.57 & 3.9 & 1.21 & 27.8 & 1.85 & 2.7 \\
        {\small Home Goods} & 1.32 & 4.9 & 1.37 & 4.2 & 1.16 & 10.2 & 1.63 & 2.8 \\
        {\small Groceries} & 1.31 & 5.8 & 1.33 & 5.4 & 1.16 & 16.7 & 1.62 & 3.0 \\
        \hline 
    \end{tabular}
    \vspace{0.5cm}
    \begin{tablenotes}
        \item {\footnotesize \textit{Notes:} The data are authors' estimates of product elasticities of substitution using different measures of markups for each product category in 1992. Our baseline measures correspond to product-level markups. In industry $\mu$ we assign to each product category the markup of its main NAICS industry. In low $\mu$ we half the product-level markup.  In low $\mu$ we double the product-level markup. Markup information comes from the Annual Retail Trade Survey. The elasticities are the solution to equation \eqref{eqn:prod_markup} using the measured product-level local Herfindahl-Hirschman Indexes from the Census of Retail Trade.}
    \end{tablenotes}
    \end{threeparttable}
    \end{center}
    \label{tab:epsilon_robustness}
\end{table}

Finally, we use the changes in product-based local HHI computed in Section \ref{sec:concen} along with equation  \eqref{eqn:prod_markup} to compute the change in markups implied by the model and the changes in local concentration. 
Table \ref{tab:markup_change_robustness} presents the implied change in markups under our four alternative measures.
It is clear that the choice of of the level of markups does not affect our main result regarding the effect of local concentration on markups.

\renewcommand{\tabcolsep}{10pt}
\begin{table}[tb!]
    \caption{Markups Robustness: Implied changes in markups}
    \begin{center}
    \begin{threeparttable}

    \begin{tabular}{lSSSS}
        \hline \hline
        \Tstrut
         & \multicolumn{1}{c}{Product $\mu$} & \multicolumn{1}{c}{Industry  $\mu$} & \multicolumn{1}{c}{Low $\mu$} & \multicolumn{1}{c}{High $\mu$} \\
        \hline 
        Furniture & 0.04 & 0.04 & 0.03 & 0.05 \\
        Clothing & -0.02 & -0.02 & -0.02 & -0.03 \\
        Sporting Goods & 0.06 & 0.06 & 0.05 & 0.08 \\
        Electronics \& Appliances & 0.05 & 0.05 & 0.04 & 0.06 \\
        Health Goods & 0.06 & 0.06 & 0.05 & 0.08 \\
        Toys & 0.06 & 0.06 & 0.05 & 0.07 \\
        Home Goods & 0.06 & 0.06 & 0.05 & 0.07 \\
        Groceries & 0.02 & 0.02 & 0.01 & 0.02 \\
        \textbf{Retail Sector} & 0.02 & 0.02 & 0.02 & 0.03 \\
        \hline 
    \end{tabular}
    \vspace{0.5cm}
    \begin{tablenotes}
        \item {\footnotesize \textit{Notes:} The data are authors' estimates of the changes in markups implied by the change in product-level local concentration. Our baseline measures uses product-level markups. In ``Industry $\mu$'' we assign to each product category the markup of its main NAICS subsector. In ``Low $\mu$'' we half the product-level markup.  In ``High $\mu$'' we double the product-level markup. Markup information comes from the Annual Retail Trade Survey and product-level local Herfindahl-Hirschman Indexes from the Census of Retail Trade.}
    \end{tablenotes}
    \end{threeparttable}
    \end{center}
    \label{tab:markup_change_robustness}
\end{table}

\renewcommand{\tabcolsep}{8pt}
\begin{table}[tb!]
    \caption{Markups Robustness: Industry Estimates}
    \begin{center}
    \begin{threeparttable}
    
    \begin{tabular}{lSSSSSS}
    \hline 
     & \multicolumn{1}{c}{$\text{HHI}_{i}^{92}$} & \multicolumn{1}{c}{$\text{HHI}_{i}^{12}$} & $\epsilon_{i}$ & $\mu_{i}^{92}$ & $\Delta\mu_{i}^{\text{ARTS}}$& $\Delta\mu_{i}^{\text{Model}}$ \\
    \hline 
    {\small Furniture (442)} & 0.08 & 0.14 & 2.7 & 1.73 & 0.15 & 0.12 \\
    {\small Clothing (448)} & 0.10 & 0.11 & 2.9 & 1.69 & 0.16 & 0.02 \\
    {\small Electronics \& Appliances (443)} & 0.14 & 0.28 & 5.2 & 1.44 & 0.00 & 0.27 \\
    {\small Health Goods (446)} & 0.15 & 0.25 & 5.3 & 1.44 & 0.02 & 0.20 \\
    {\small Toys and Sporting Goods (451)} & 0.20 & 0.24 & 4.8 & 1.57 & 0.15 & 0.09 \\
    {\small Home Goods (444)} & 0.22 & 0.30 & 15.4 & 1.37 & 0.14 & 0.23 \\
    {\small Groceries (445)} & 0.16 & 0.22 & 9.1 & 1.33 & 0.05 & 0.11 \\
    {\small General Merchandisers (452)} & 0.28 & 0.56 & \multicolumn{1}{c}{3606} & 1.39 & -0.03 & 0.89 \\
    Retail Sector  &  &  &  & 1.42 & 0.05 & 0.30 \\
    \hline 
    \end{tabular}
    \vspace{0.5cm}
    \begin{tablenotes}
        \item {\footnotesize \textit{Notes:} The data are industry-level local Herfindahl-Hirschman Indexes from the Census of Retail Trade and markups from  the Annual Retail Trade Survey. The elasticities are the solution to equation \eqref{eqn:prod_markup}. Decimal places are not shown on the $\epsilon_i$ for General Merchandisers due to its magnitude. High levels of $\epsilon$ imply essentially the same markups. For instance an $\epsilon_{GM}=165.7$ implies the same level of markups up to the second decimal as the estimate we report.}
    \end{tablenotes}
    \end{threeparttable}
    \end{center}
    \label{tab:markup_indsutry_robustness}
\end{table}

\paragraph{Industry-based results}
We also consider how our results would change if used industry-level measures of concentration. 
In Section \ref{sec:industry} we show that this leads to larger measured changes in local concentration at the industry level.
Consequently, using industry-level measures of concentration would have led to a higher implied change in markups.

Table \ref{tab:markup_indsutry_robustness} presents the results of our exercise using industry-level markups from the ARTS and industry-level local concentration from the CRT. 
The change in industry-level concentration are much larger than those in product-based measures.
This is particularly true in the general merchandise subsector (NAICS 452), where the change in local concentration implies a change in markups of 89 percentage points.
During this period general merchandiser markups were almost unchanged in ARTS.
These two facts can be reconciled by the fact that general merchandisers face significant competition from retailers outside of their industry.
When aggregated, these changes imply an increase in retail markups of 30 percentage points which significantly exceeds the change in margins observed in the ARTS (\Paste{ARTS_Markup_9312} percentage points).

\subsection{Uniform prices across locations}\label{app:pricing}

In this subsection, we show that retail markups are lower on average if multi-market firms engage in uniform pricing. Consider the problem of firm $i$ that sales product $j$ across various locations $\ell\in{\cal L}_{i}$ and sets a uniform price (in a Bertrand fashion). %
The firm's problem is:
\begin{align}
    \max_{p_{i}^{j}}\,\sum_{\ell\in{\cal L}_{i}}\left[p_{i}^{j}y_{i}^{j\ell}-c_{i}^{j\ell}y_{i}^{j\ell}\right]
\end{align}
\[
\text{s.t. }y_{i}^{j\ell}=\left(\frac{p_{i}^{j}}{p_{j}^{\ell}}\right)^{-\epsilon_j}y_{j}^{\ell}\qquad y_{j}^{\ell}=\gamma_{j}^{\ell}\frac{p_{\ell}y_{\ell}}{p_{j}^{\ell}}\qquad p_{j}^{\ell}=\left(\sum_{i=1}^{N}\left(p_{i}^{j\ell}\right)^{1-\epsilon_j}\right)^{\frac{1}{1-\epsilon_j}}
\]
Replacing the constraints:
\begin{align}
    \max_{p_{i}^{j}}\,\sum_{\ell\in{\cal L}_{i}}\left[\left(p_{i}^{j}\right)^{1-\epsilon_j}-c_{i}^{j\ell}\left(p_{i}^{j}\right)^{-\epsilon_j}\right]\left(\sum_{i=1}^{N}\left(p_{i}^{j\ell}\right)^{1-\epsilon_j}\right)^{-1}\gamma_{j}^{\ell}p_{\ell}y_{\ell}
\end{align}
The first order condition is:
\begin{align}
    0 & =\sum_{\ell\in{\cal L}_{i}}\left[\left[\left(1-\epsilon_j\right)p_{i}^{j}+\epsilon_jc_{i}^{j\ell}\right]\frac{y_{i}^{j\ell}}{p_{i}^{j}}-\left(1-\epsilon_j\right)\left[p_{i}^{j}-c_{i}^{j\ell}\right]s_{i}^{j\ell}\frac{y_{i}^{j\ell}}{p_{i}^{j}}\right] \nonumber \\
    0 & =\sum_{\ell\in{\cal L}_{i}}\left[-\left(\epsilon_j-1\right)\left(1-s_{i}^{j\ell}\right)y_{i}^{j\ell}p_{i}^{j}+\left(\epsilon_j-\left(\epsilon_j-1\right)s_{i}^{j\ell}\right)c_{i}^{j\ell}y_{i}^{j\ell}\right]
\end{align}
Rearranging:
\begin{align}
    p_{i}^{j}=\frac{\sum_{\ell}\left(\epsilon_j-\left(\epsilon_j-1\right)s_{i}^{j\ell}\right)c_{i}^{j\ell}y_{i}^{j\ell}}{\sum_{\ell}\left(\epsilon_j-1\right)\left(1-s_{i}^{j\ell}\right)y_{i}^{j\ell}}
\end{align}
If marginal cost is constant across markets then we define the markup:
\begin{align}
    p_{i}^{j}=\mu_{i}^{j}c_{i}^{j}\qquad\mu_{i}^{j}=\frac{\sum_{\ell}\left(\epsilon_j-\left(\epsilon_j-1\right)s_{i}^{j\ell}\right)y_{i}^{j\ell}}{\sum_{\ell}\left(\epsilon_j-1\right)\left(1-s_{i}^{j\ell}\right)y_{i}^{j\ell}}
\end{align}

The firm's markup reflects its market power across different markets, captured by the firm's output-weighted average share, $\hat{s}_i^j$.
Define $\hat{y}_{i}^{j\ell}\equiv\frac{y_{i}^{j\ell}}{\sum_{\ell}y_{i}^{j\ell}/L_i}$ and $\hat{s}_{i}^{j}\equiv\sum_{\ell}s_{i}^{j\ell}\hat{y}_{i}^{j\ell}$, then:
\begin{align}
\mu_{i}^{j} 
    =\frac{\sum_{\ell}\left(\epsilon_j-\left(\epsilon_j-1\right)s_{i}^{j\ell}\right)\hat{y}_{i}^{j\ell}}{\sum_{\ell}\left(\epsilon_j-1\right)\left(1-s_{i}^{j\ell}\right)\hat{y}_{i}^{j\ell}}
    = \frac{\epsilon_j-\left(\epsilon_j-1\right)\sum_{\ell}s_{i}^{j\ell}\hat{y}_{i}^{j\ell}}{\left(\epsilon_j-1\right)\left(1-\sum_{\ell}s_{i}^{j\ell}\hat{y}_{i}^{j\ell}\right)} 
    = \frac{\epsilon_j-\left(\epsilon_j-1\right)\hat{s}_{i}^{j}}{\left(\epsilon_j-1\right)\left(1-\hat{s}_{i}^{j}\right)} 
    \label{eq: Unifrom_Markup}
\end{align}
The firm's uniform markup is lower than the average markup if the firm chooses prices in each market separately.
To see this, define the firm's average price in product $j$ such that: $p_i^j y_i^j = \sum_\ell p_i^{j\ell}y_i^{j\ell}$, where $y_i^j\equiv\sum_\ell y_i^{j\ell}$. 
It follows that $p_i^j = \sum_\ell p_i^{j\ell}\hat{y}_i^{j\ell}$. 
The average markup would then be: 
\begin{align}
    \bar{\mu}_i^j\equiv\frac{p_i^j}{c_i^j}=\sum_\ell \frac{p_i^{j\ell}}{c_i^j}\hat{y}_i^{j\ell}=\sum_\ell \mu_i^{j\ell}\hat{y}_i^{j\ell},
    \label{eq: Average_Firm_Markup}
\end{align} 
which is the output-weighted average of the individual market (Bertrand) markups
\begin{align}
    \mu_i^{j\ell} = \frac{\epsilon_j-\left(\epsilon_j-1\right){s}_{i}^{j\ell}}{\left(\epsilon_j-1\right)\left(1-{s}_{i}^{j\ell}\right)}. 
    \label{eq: Bertrand_Markup}
\end{align}
The average markup $\bar{\mu}_i^j$ in \eqref{eq: Average_Firm_Markup} is higher than the uniform markup $\mu_i^j$ in \eqref{eq: Unifrom_Markup}. 
The result follows from Jensen's inequality as the Bertrand markup in \eqref{eq: Bertrand_Markup} is convex in the firm's sales share.

\renewcommand{\tabcolsep}{5pt}

\end{document}